\documentclass{article}
\usepackage{graphicx} 
\usepackage{multirow}
\usepackage{jheppub}
\usepackage{amsmath,latexsym}
\usepackage{amsmath}
\usepackage{slashed}
\usepackage{graphicx}
\usepackage{bm}
\usepackage{booktabs}
\usepackage{tikz-feynman}
\tikzfeynmanset{/tikzfeynman/momentum/arrow shorten=0.30}
\tikzfeynmanset{/tikzfeynman/warn luatex=false}
\usepackage{comment}
\usepackage{subcaption}
\usepackage{xcolor}
\usepackage[normalem]{ulem}
\usepackage{amsfonts}
\usepackage{float}
\usepackage{cancel}

\title{Anatomy and Phenomenology of Minimal Flavor Deconstruction in the Lepton Sector}
\author[a,b]{Antonio Masiero,}
\author[a,b]{Paride Paradisi,}
\author[c]{Daniel Quieroz,}
\author[a,b,d]{Andrea Sainaghi,}
\author[c]{Nicola Valori,}
\author[c]{Oscar Vives,}

\affiliation[a]{Dipartimento di Fisica e Astronomia ``Galileo Galilei'', Universit\`a degli Studi di Padova, \\Via F. Marzolo 8, 35131 Padova, Italy}
\affiliation[b]{Istituto Nazionale di Fisica Nucleare (INFN), Sezione di Padova,
\\Via F. Marzolo 8, 35131 Padova, Italy}
\affiliation[c]{Institut de F\'isica Corpuscular (CSIC-Universitat de València), Departament de F\'isica Teòrica, Dr. Moliner 50, E-46100 Burjassot (València), Spain}
\affiliation[d]{Physik-Institut, Universit\"at Z\"urich, Winterthurerstrasse 190, CH-8057 Z\"urich, Switzerland}

\emailAdd{antonio.masiero@pd.infn.it}
\emailAdd{paride.paradisi@pd.infn.it}
\emailAdd{daniel.queiroz@uv.es}
\emailAdd{andrea.sainaghi@phd.unipd.it}
\emailAdd{nicola.valori@uv.es}
\emailAdd{oscar.vives@uv.es}

\newcommand{\HH}{\Theta}

\abstract{
We investigate the low-energy phenomenology of a minimal flavor-deconstructed framework in the lepton sector within an effective field theory approach, focusing on the interplay between flavor and CP violation. Starting from the ultraviolet completion of the model, we derive the effective Yukawa structure through a systematic spurion expansion beyond leading order and identify the dominant sources of flavor and CP violation. 
We show that, while leading-order effects to dipole operators are approximately aligned with the Yukawa matrices, next-to-leading order contributions generically induce physical CP-violating phases and flavor misalignment, leading to potentially observable low-energy signals.
After constructing the corresponding low-energy effective theory, we analyze the phenomenological implications for charged lepton flavor violating observables, lepton flavor universality tests, and electric dipole moments (EDMs). We find that future searches for $\mu-e$ conversion and the electron EDM can probe scales in the multi-10~TeV range under natural assumptions on the flavor structure and CP phases.
Our results highlight the complementarity between flavor-violating and CP-violating observables and demonstrate that precision measurements in the lepton sector provide a powerful probe of flavor-deconstructed scenarios beyond the direct reach of collider experiments.
}

\begin{document}

\maketitle
\flushbottom

\section{Introduction}
\label{sec:intro}

After the most significant achievement of the Large Hadron Collider (LHC), namely the discovery of the Higgs boson, a cornerstone of the Standard Model (SM), no unambiguous evidence for physics beyond the Standard Model (BSM) has so far emerged at the TeV scale associated with electroweak symmetry breaking. Nevertheless, a number of well-established experimental and observational facts—such as the existence of dark matter, non-zero neutrino masses, and the baryon asymmetry of the Universe—provide compelling evidence for BSM physics, whose underlying nature and characteristic energy scale(s) remain unknown. It is therefore becoming increasingly clear that progress in the exploration of BSM scenarios will rely on a comprehensive exploitation of the complementarity among the three main frontiers of investigation: high-energy, high-intensity, and astroparticle physics. 

In view of our present ignorance of BSM physics and recalling historical precedents where new particles were first inferred via virtual loop effects, precision physics plays an increasingly central role. The comparison between high-precision theoretical predictions and experimental data allows one to probe deviations signalling the incompleteness of the SM. Observables associated with SM symmetry violations are particularly sensitive to new physics (NP) beyond the reach of current colliders, with flavor-changing processes and CP violation providing prominent examples.

Although the peculiar structure of the Yukawa couplings and the origin of CP violation do not represent a “technical” problem for the SM, they remain among the most puzzling features of Nature for which no satisfactory explanation is provided within the SM framework. Remarkably, CP violation has so far been observed only in the Yukawa sector, suggesting a common dynamical origin with the mechanism responsible for flavor generation. Several extensions of the SM have been proposed in an attempt to address the so-called “flavor puzzle”, in particular scenarios in which NP couples preferentially to third-generation fermions~\cite{Isidori:2012ts,Barbieri:2012uh,Allwicher:2023shc,Barbieri:2011ci}. Such constructions arise naturally in a variety of ultraviolet completions, including supersymmetric models~\cite{Barbieri:2011ci,Papucci:2011wy,Larsen:2012rq}, composite Higgs scenarios~\cite{Barbieri:2012tu,Matsedonskyi:2014iha,Panico:2016ull}, and, more recently, flavor non-universal gauge theories~\cite{Allwicher:2020esa,Fuentes-Martin:2020bnh,Fuentes-Martin:2022xnb,Fuentes-Martin:2024fpx,Greljo:2018tuh,Barbieri:2023qpf,Barbieri:2024zkh,Bordone:2017bld,Davighi:2023evx,Davighi:2023iks,Covone:2024elw,Lizana:2024jby,FernandezNavarro:2023rhv,FernandezNavarro:2024hnv,FernandezNavarro:2025zmb,Greljo:2024ovt,Fabri:2025fsc,Isidori:2025rci,Beneito:2025fzf,Greljo:2025mwj,Arkani-Hamed:2026wwy,Gosnay:2026dye,Blennow:2026psy,Pesut:2026fyl}, often referred to as flavor deconstruction (FD). Notably, Yukawa couplings provide no information on the scale of NP, which may lie anywhere between the multi-TeV range and the Planck scale. However, a relatively low symmetry-breaking scale could induce sizeable contributions to precision observables, such as leptonic and hadronic dipole transitions and rare decays.

Specifically, within the FD framework, the flavor symmetry is realised as a direct product of flavor-dependent gauge groups, which are spontaneously broken down to the SM gauge group at the electroweak (EW) scale. This setup can reproduce the observed hierarchical pattern of fermion masses and mixings through multiple symmetry-breaking scales, some of which may lie in the few-TeV range, together with the associated new field content~\cite{Davighi:2023iks}. While flavor mixing has been extensively investigated in a variety of FD realizations, the origin and implications of CP violation in the Cabibbo--Kobayashi--Maskawa (CKM) matrix have received comparatively limited attention in the literature.

In this work we assess the impact of CP violation in FD models, with particular emphasis on low-energy observables in the lepton sector. This is motivated by the exceptional sensitivity currently achieved in precision experiments, as well as by expected future improvements. In particular, the electron electric dipole moment (EDM) is experimentally constrained to $d_{e} < 4.1 \times 10^{-30}\, e\,\mathrm{cm}$~\cite{Roussy:2022cmp}, with ongoing efforts aimed at further strengthening this bound~\cite{Hiramoto:2022fyg,Fitch:2020jil,Vutha:2018tsz}.
These limits imply an indirect sensitivity to NP scales reaching the multi-$100\,\mathrm{TeV}$ regime, thereby establishing the electron EDM as one of the most sensitive probes of BSM physics~\cite{Kley:2021yhn,Panico:2018hal,Aebischer:2021uvt,Ardu:2025rqy}.

We focus on a specific setup, referred to as Minimal Flavor Deconstruction, in which a set of flavor-specific abelian gauge symmetries is broken down to the SM hypercharge, thereby reproducing the SM gauge group at low energies~\cite{Barbieri:2023qpf,Barbieri:2024zkh}. To realise this structure, the SM matter content is extended by additional fermionic and scalar fields, such that the Yukawa sector acquires new physical phases and thus provides additional sources of CP violation.

To properly account for the leading effects contributing to loop-induced CP-violating observables, a complete computation quickly becomes intractable, even within the simplified setup considered here. We therefore employ a spurion analysis~\cite{Georgi:1984zwz,tHooft:1979rat,Chivukula:1987py,Hall:1990ac,DAmbrosio:2002vsn,Feruglio:2015gka} to systematically capture the dominant contributions to both mass and dipole operators. 
This approach is essential, since at leading order (LO) the dipole operators are approximately aligned with the Yukawa matrices, whereas next-to-leading order (NLO) effects generically induce flavor misalignment and physical CP-violating phases.
Such phases are not only generically expected, but in fact required in order to reproduce the observed CKM phase. This has important phenomenological implications, indicating EDMs as one of the most sensitive probes of the framework. Notably, our analysis can be straightforwardly extended to a broad class of models addressing the flavor puzzle.

The paper is organised as follows. In Section~\ref{sec:minimal flav dec} we introduce the benchmark model and discuss in detail the lepton Yukawa sector using a spurion-based analysis. In Section~\ref{sec:SMEFT} we perform the matching onto the Standard Model Effective Field Theory (SMEFT) after integrating out the heavy degrees of freedom. In Section~\ref{sec: pheno FD lepton sector} we study the resulting phenomenological implications, with particular emphasis on CP-violating observables, while Section~\ref{sec:conclusions} is devoted to our conclusions.
Additional technical details relevant to the analysis are provided in the appendices.

\section{Minimal flavor deconstruction}\label{sec:minimal flav dec}
\subsection{Review of the model}\label{sec:review}

Flavor deconstruction (FD) refers to a framework in which gauge interactions are generally non-universal across the three fermion generations of the SM~\cite{Allwicher:2020esa,Fuentes-Martin:2020bnh,Fuentes-Martin:2022xnb,Fuentes-Martin:2024fpx,Greljo:2018tuh,Barbieri:2023qpf,Barbieri:2024zkh,Bordone:2017bld,Davighi:2023evx,Davighi:2023iks,Covone:2024elw,Lizana:2024jby,FernandezNavarro:2023rhv,FernandezNavarro:2024hnv,FernandezNavarro:2025zmb,Isidori:2025rci,Greljo:2025mwj,Arkani-Hamed:2026wwy}.
Following~\cite{Barbieri:2024zkh, Barbieri:2023qpf}, we consider a UV embedding 
of the SM that realizes a FD of hypercharge, while keeping the non-abelian interactions universal across the different fermion generations. 
The full UV gauge group is given by the following structure
\begin{equation}
    \mathrm{SU}(3)_C\times \mathrm{SU}(2)_L\times \mathrm{U}(1)^{[3]}_{Y} \times \mathrm{U}(1)^{[12]}_{\mathrm{B-L}} \times \mathrm{U}(1)^{[2]}_{T_{3R}} \times \mathrm{U}(1)^{[1]}_{T_{3R}}\,,
\end{equation}
where $[i]$ means that the gauge interaction is non-trivial for the $i$-th generation of fermion. The abelian part of this UV gauge group is broken down to the SM hypercharge by means of a chain of spontaneous symmetry breakings (SBB) given by 
\begin{equation}\label{eq:SSB}
    \mathrm{U}(1)^{[3]}_{Y} \times \mathrm{U}(1)^{[12]}_{\mathrm{B-L}} \times \mathrm{U}(1)^{[2]}_{T_{3R}} \times \mathrm{U}(1)^{[1]}_{T_{3R}}
\;\xrightarrow{}\;
\mathrm{U}(1)^{[3]}_{Y} \times \mathrm{U}(1)^{[12]}_{\mathrm{B-L}} \times \mathrm{U}(1)^{[12]}_{T_{3R}}
\;\xrightarrow{}\;
\mathrm{U}(1)_{Y}\,.
\end{equation}

In addition to all matter SM fields, the particle content is given by four new scalars $\phi,\chi^{\ell},\chi^{q}$, $\sigma$, and nine vector-like fermions $F_{1},F_2,F_3$ with $F=U,D,E$. The cancellation of gauge anomalies involving the extra $\mathrm{U}(1)$ factors requires the presence of at least two right-handed neutrinos. In the following, we specialize to the lepton sector, while the gauge quantum number of the up and down quark sectors can be chosen in a similar manner. The full particle content relevant to the lepton sector is shown in Tab.~\ref{Table:ContModA}. The SSB is due to the vacuum expectation values of the extra scalars. In particular, the first symmetry breaking is triggered by $\langle \sigma\rangle$, while the final step of the breaking pattern is due to $\langle \phi \rangle \sim \langle \chi^{\ell,q} \rangle$. The mass hierarchy in this model reads
\begin{equation}\label{eq:masshierarchy}
    M_{E_{3}} \gg \langle \sigma \rangle \gg M_{E_{1,2}} \gg \langle \chi^{\ell,q}\rangle,\langle\phi \rangle \gg \langle H \rangle\,.
\end{equation}

\begin{table}[h]
\centering
\begin{tabular}{cc|c|c|c|c|c}
 &  & $\mathrm{U}(1)_Y^{[3]}$ & $\mathrm{U}(1)_{\mathrm{B-L}}^{[12]}$ & $\mathrm{U}(1)_{T_{3R}}^{[2]}$ & $\mathrm{U}(1)_{T_{3R}}^{[1]}$ & $\mathrm{SU}(3)\times \mathrm{SU}(2)$ \\ [0.1cm]\hline

\multirow{7}{*}{$\begin{array}{c} {\rm SM~fermions}\\ (i=1,2) \end{array}$}
& $\ell_i$ & $0$ & $-1$ & $0$ & $0$ & $(\mathbf{1},\mathbf{2})$ \\

& $\ell_3$ & $-1/2$ & $0$ & $0$ & $0$ & $(\mathbf{1},\mathbf{2})$ \\ \cline{2-7}

& $e_1$ & $0$ & $-1$ & $0$ & $-1/2$ & $(\mathbf{1},\mathbf{1})$ \\ 

& $e_2$ & $0$ & $-1$ & $-1/2$ & $0$ & $(\mathbf{1},\mathbf{1})$ \\ 

& $e_3$ & $-1$ & $0$ & $0$ & $0$ & $(\mathbf{1},\mathbf{1})$ \\ \cline{2-7}

& $\nu_1$ & $0$ & $-1$ & $0$ & $1/2$ & $(\mathbf{1},\mathbf{1})$ \\ 

& $\nu_2$ & $0$ & $-1$ & $1/2$ & $0$ & $(\mathbf{1},\mathbf{1})$ \\ 
\hline

\multirow{2}{*}{$\begin{array}{c} {\rm heavy~fermions}\\ (\alpha=1,2) \end{array}$}
& $E_\alpha$ & $-1/2$ & $-1$ & $0$ & $0$ & $(\mathbf{1},\mathbf{1})$ \\ 

& $E_3$ & $0$ & $-1$ & $-1/2$ & $0$ & $(\mathbf{1},\mathbf{1})$ \\ 
\hline

\multirow{5}{*}{${\rm Scalars}$}
& $H$ & $-1/2$ & $0$ & $0$ & $0$ & $(\mathbf{1},\mathbf{2})$ \\ 

& $\chi^\ell$ & $1/2$ & $-1$ & $0$ & $0$ & $(\mathbf{1},\mathbf{1})$ \\ 

& $\chi^q$ & $-1/6$ & $1/3$ & $0$ & $0$ & $(\mathbf{1},\mathbf{1})$ \\ 

& $\phi$ & $1/2$ & $0$ & $-1/2$ & $0$ & $(\mathbf{1},\mathbf{1})$ \\ 

& $\sigma$ & $0$ & $0$ & $1/2$ & $-1/2$ & $(\mathbf{1},\mathbf{1})$ \\

\hline
\end{tabular}
\caption{Fermion and scalar matter content of the minimal flavor deconstruction model, together with their charge assignments under the full gauge group.}
\label{Table:ContModA}
\end{table}
In addition to heavy scalar and fermionic states, this extension of the SM predicts the existence of four additional gauge bosons $X_i$, which accompany the SM gluons $A^A$ and the $\mathrm{SU}(2)_L$ gauge fields $W^I$, and are associated with the abelian gauge sectors of the theory. 
The corresponding Lagrangian can then be decomposed as follows:
\begin{equation} 
\label{eq:lagrangianUVmodel}
    \mathcal{L}_{\mathrm{UV}}= \, \mathcal{L}_{\text{gauge}} \,+ \, \mathcal{L}_{\text{scalars}} \,+\,\sum_{\psi= q,\ell,u,d,e}\sum_{i=1}^3\overline{\psi}^i i\slashed D\psi^i \, + \, \sum_{F=U,D,E}\sum_{\alpha=1}^{3}\overline{F}_\alpha(i\slashed D-M_{F_\alpha})F_\alpha \, + \, \sum_{\psi=u,d,e}\mathcal{L}_{Y}^\psi\,,
\end{equation} 
where $\mathcal{L}_{\text{gauge}}$ contains the kinetic terms of the gauge fields, $\mathcal{L}_{\text{scalars}}$ encodes the kinetic term of the scalar fields as well as the scalar potential, and $\mathcal{L}_{Y}^\psi$ represents the Yukawa sector of the theory. 
Restricting to the abelian gauge groups, the covariant derivative reads
\begin{equation}\label{eq: cov der}
D_{\mu}^{X} = \partial_{\mu}
 - i\Big( g_{1}\,Y^{[3]}X_{1\mu}
        + g_{2}\,\frac{1}{2}(\mathrm{B-L})^{[12]}X_{2\mu}
        + g_{3}\,T_{3R}^{[2]}\,X_{3\mu}
        + g_{4}\,T_{3R}^{[1]}\,X_{4\mu}\Big),
\end{equation}
while the scalar sector is given by
\begin{equation}\label{eq:scalarsector}
    \mathcal{L}_{\text{scalars}}=\sum_{i}\left[(D_\mu\phi_i)^\dag(D^\mu\phi_i) +\mu_i^2\phi_i^\dag \phi_i\right]-\sum_{i,j}\lambda_{ij}(\phi_i^\dag \phi_i)(\phi_j^\dag \phi_j)\,,
\end{equation}
where the sum runs over all scalar fields in the theory. An additional term of the form $\chi^{\ell}(\chi^{q})^{3}$ may also be present, which would introduce new potential sources of CP violation. However, motivated by the existence of a complex phase in the CKM matrix, we neglect this contribution and focus instead on CP-violating effects arising from the Yukawa sector. Accordingly, all parameters in the scalar sector are taken to be real.
The requirement of a well-defined scalar potential, together with the SSB condition, implies $\lambda_{ii} > 0$ and $\mu_i > 0$, while the off-diagonal couplings $\lambda_{ij}$ with $i \neq j$ may also take negative values. Finally, the Yukawa Lagrangian $\mathcal{L}_Y$ includes all allowed renormalizable interactions among heavy and SM fermions and a single scalar field. Restricting to the lepton sector, $\mathcal{L}^e_Y$ reads:
\begin{equation}\label{eq:Yukawa}
    \begin{split}
        -\mathcal{L}^e_Y &=\,  y^e_3\, \bar{\ell}_{3} e_{3} H + (Y^e)_{i\alpha}\,  \bar{\ell}_{i} E_{\alpha}  H
+  (Y^{\chi})_{\alpha}\,   \bar{E}_{\alpha} e_{3}  \chi^\ell + (Y^{\phi}_2)_{\alpha }\,  \bar{E}_{\alpha} e_{2} \phi^{\dag}  
+ (Y^{\phi}_3)_{\alpha}\,    \bar{E}_{L3} E_{R\alpha} \phi^{} \\
&\qquad  +\,  (\hat{Y}^{\phi}_3)_{\alpha}\, \bar{E}_{L\alpha} E_{R3}  \phi^{\dag}  + Y^{\sigma}\,  \bar{E}_{3} e_{1} \sigma+ {\rm h.c.} \,,
    \end{split}
\end{equation}
where $i,\alpha = 1,2$. The Yukawa Lagrangians for the up- and down-type quark sectors are analogous to that of the charged lepton sector, with $\chi^{\ell}$ replaced by $\chi^{q}$.

After SSB as defined in Eq.~\eqref{eq:SSB}, the vector bosons mix into four mass eigenstates, denoted by $B$, $Z_{23}$, $Z_{23}^\prime$, and $Z_H$. The field $B$ corresponds to the massless gauge boson associated with the SM hypercharge, while the remaining three states acquire masses of order $M_{Z_H} \sim \langle \sigma \rangle$ and $M_{Z_{23}}, M_{Z_{23}^\prime} \sim \langle \phi \rangle, \langle \chi^{\ell,q} \rangle$. Owing to the mass hierarchy specified in Eq.~\eqref{eq:masshierarchy}, the state $Z_H$ will be neglected in the following. The covariant derivative relevant to the abelian sector of the model can thus be written as:
\begin{equation}\label{eq: cov derivative broken phase}
D^{\mu\,X} = \partial^{\mu}
 - i\Big( g^{\prime}\,YB^{\mu}
        + g_{23}\, Q_{23}\,Z^{\mu}_{23}
        + g^{\prime}_{23}Q^{\prime}_{23}\,Z^{\prime\mu}_{23}\Big),
\end{equation}
where $g'$ denotes the SM gauge coupling associated with the hypercharge, which follows the decomposition
\begin{equation}
    Y=Y^{[3]}+\frac{1}{2}(\mathrm{B-L})^{[12]}+ T_{3R}^{[12]}\,.
\end{equation}
The remaining  couplings and charges, $g_{23}^{(\prime)}$ and $Q_{23}^{(\prime)}$, are given by suitable combinations of the gauge couplings $g_i$ introduced in Eq.~\eqref{eq: cov der}. For the purposes of this manuscript, the explicit form of $g_{23}^{(\prime)}$ is left unspecified, and the only condition imposed
\begin{equation}\label{eq:universality between first 2 generations}
    Q_{23}^{(\prime)\psi_{1}}=Q_{23}^{(\prime)\psi_{2}}\neq Q_{23}^{(\prime)\psi_{3}}\,,
\end{equation}
which is is dictated by the flavor symmetry of the model.\footnote{This follows from the the quantum numbers in Tab. \ref{Table:ContModA}, once the first stage of SSB has been triggered.}

With regard to the scalar sector of the theory, SSB leaves four heavy scalar states in addition to a pseudoscalar boson. Owing to the presence of quartic couplings $\lambda_{ij}$, scalar mixing is also induced. The detailed structure of such mixing lies beyond the scope of the present analysis; we therefore label the heavy physical degrees of freedom collectively as the scalar states of the broken theory, following the same mass hierarchy pattern as their corresponding VEVs.
Furthermore, SSB induces a shift in the Higgs mass term in the scalar Lagrangian, which can be written as
\begin{equation}\label{eq: scalar lagrangian broken symmetry}
    \mathcal{L}_{\text{scalars}}\supset \left[\mu^2-\sum_{i}\langle\phi_i\rangle^2 \lambda_{i H}\right] H^\dag H\equiv (\mu')^2 H^\dag H\,.
\end{equation}
In order to trigger EWSB, one must require that, at the electroweak scale, $(\mu')^2 \approx (90\,\mathrm{GeV})^2$. This introduces a naturalness tension, since for $\langle \phi_i \rangle = \mathcal{O}(\mathrm{TeV})$, either $\lambda_{iH} \lesssim \mathcal{O}(10^{-2})$ is required, or $\lambda_{iH} = \mathcal{O}(1)$ must be accompanied by a non-trivial cancellation at the percent level.

\subsection{Lepton Yukawa matrix from a spurion analysis} \label{Sec:spurion}
A systematic and simple way to derive the Yukawa matrices after integrating out the 
heavy degrees of freedom is to perform a \emph{spurion analysis}~\cite{Georgi:1984zwz,tHooft:1979rat,Chivukula:1987py,Hall:1990ac,DAmbrosio:2002vsn,Feruglio:2015gka}. In this subsection, we highlight this approach and discuss the possible contributions to the masses (or, more precisely, Yukawa couplings) of the SM charged leptons, including NLO effects in the $\varepsilon_i$ expansion (see below). This technique also allows for an unambiguous identification of potential CP-violating sources originating from the UV completion of the model.

In a spurion analysis a set of parameters, the spurions, are promoted to background fields transforming under the underlying flavor symmetry.
Indeed, by switching off all couplings in Eq.~\eqref{eq:Yukawa}, the theory becomes invariant under independent rotations and rephasings of leptons carrying identical gauge quantum numbers. From Tab.~\ref{Table:ContModA}, the lepton doublets involving the first two generations, together with the two light vector-like leptons, form doublets of the flavor symmetry and therefore transform under $2\times2$ unitary matrices $U$ and $V_{L,R}$. The remaining fields ($\ell_{3}, e_{1}, e_{2}, e_{3}, E_{3,L}, E_{3,R}$) are singlets and transform only under $\mathrm{U}(1)$ phase rotations $(\ell_3, e_1, e_2, e_3, v_{L3}, v_{R3})$.

The Lagrangian in Eq.~\eqref{eq:Yukawa} is then formally invariant under the full flavor symmetry if the Yukawa couplings and vector-like lepton masses are promoted to spurion fields transforming as follows:
\begin{subequations}
\begin{alignat}{3}
\label{eq:Couptrans}
& y_3^e         && \qquad\to\qquad && e^{i \ell_3} y_3^e e^{-i e_3} \,, \\
& Y^e           && \qquad\to\qquad && U \cdot Y^e \cdot V_R^\dagger \,, \\
& Y^\chi        && \qquad\to\qquad && V_L \cdot Y^\chi e^{-i e_3} \,, \\
& Y^\phi_2      && \qquad\to\qquad && V_L \cdot Y^\phi_2 e^{-i e_2} \,, \\
& Y^\phi_3      && \qquad\to\qquad && V_R \cdot Y^\phi_3 e^{-i v_{L3}} \,, \\
& \hat Y^\phi_3 && \qquad\to\qquad && V_L \cdot \hat Y^\phi_3 e^{-i v_{R3}} \,, \\
& Y^\sigma      && \qquad\to\qquad && e^{i v_{L3}} Y^\sigma e^{-i e_1} \,, \\
& M_{E_\alpha}  && \qquad\to\qquad && V_L \cdot M_{E_\alpha} \cdot V_R^\dagger \,, \\
& M_{E_3}       && \qquad\to\qquad && e^{i v_{L3}} M_{E_3} e^{-i v_{R3}} \, .
\end{alignat}
\end{subequations}
Mass terms for the light SM fermions are then generated by connecting right-handed and left-handed fields through multiple spurion insertions, in a flavor-invariant manner. Similarly, being chirality changing, dipole operators feature the same spurion expansion as the mass terms. These operators will be directly linked to LFV or CP-violating observables, such as lepton EDMs or $\ell_{i}\rightarrow\ell_{j}\gamma$.

As we will show in Sec.~\ref{sec:flav structure to smeft}, a consistent treatment of all relevant contributions to these observables requires the spurion expansion to be performed at NLO, since such effects become parametrically of the same order as the LO contributions once the Yukawa matrix is diagonalised. Accordingly, in the following we also expand the leptonic Yukawa matrix up to NLO.

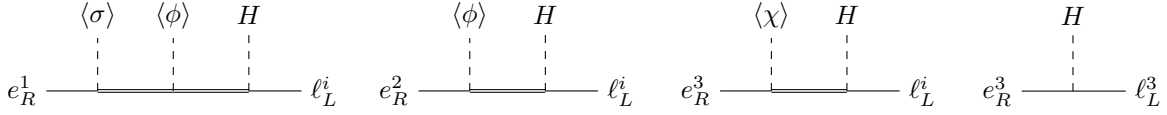
\begin{figure}[ht!]
\centering
\begin{subfigure}[b]{0.28\textwidth}
\centering
\scalebox{1}{
\begin{tikzpicture}
    \begin{feynman}
        \vertex(a) at (0,0) {$e_R^1$};
        \vertex(b) at (1,0);
        \vertex(c) at (2,0);
        \vertex(b1) at (1,1) {$\langle \sigma \rangle$};
        \vertex(c1) at (2,1) {$\langle \phi \rangle$};
        \vertex(d) at (3,0);
        \vertex(d1) at (3,1) {$H$};
        \vertex(e) at (4,0) {$\ell_L^i$};
        \diagram{
        (a) -- (b) -- [double] (c) -- [double] (d) -- (e),
        (b) -- [scalar] (b1),
        (c) -- [scalar] (c1),
        (d) -- [scalar] (d1),
        };
    \end{feynman}
\end{tikzpicture}}
\end{subfigure}
\hfill
\begin{subfigure}[b]{0.22\textwidth}
\centering
\scalebox{1}{
\begin{tikzpicture}
    \begin{feynman}
        \vertex(a) at (0,0) {$e_R^2$};
        \vertex(c) at (1,0);
        \vertex(c1) at (1,1) {$\langle \phi \rangle$};
        \vertex(d) at (2,0);
        \vertex(d1) at (2,1) {$H$};
        \vertex(e) at (3,0) {$\ell_L^i$};
        \diagram{
        (a) -- (c) -- [double] (d) -- (e),
        (c) -- [scalar] (c1),
        (d) -- [scalar] (d1),
        };
    \end{feynman}
\end{tikzpicture}}
\end{subfigure}
\hfill
\begin{subfigure}[b]{0.22\textwidth}
\centering
\scalebox{1}{
\begin{tikzpicture}
    \begin{feynman}
        \vertex(a) at (0,0) {$e_R^3$};
        \vertex(c) at (1,0);
        \vertex(c1) at (1,1) {$\langle \chi \rangle$};
        \vertex(d) at (2,0);
        \vertex(d1) at (2,1) {$H$};
        \vertex(e) at (3,0) {$\ell_L^i$};
        \diagram{
        (a) -- (c) -- [double] (d) -- (e),
        (c) -- [scalar] (c1),
        (d) -- [scalar] (d1),
        };
    \end{feynman}
\end{tikzpicture}}
\end{subfigure}
\hfill
\begin{subfigure}[b]{0.17\textwidth}
\centering
\scalebox{1}{
\begin{tikzpicture}
    \begin{feynman}
        \vertex(a) at (0,0) {$e_R^3$};
        \vertex(c) at (1,0);
        \vertex(c1) at (1,1) {$H$};
        \vertex(e) at (2,0) {$\ell_L^3$};
        \diagram{
        (a) -- (c) -- (e),
        (c) -- [scalar] (c1),
        };
    \end{feynman}
\end{tikzpicture}}
\end{subfigure}

\caption{Feynman diagrams contributing to the lepton Yukawa couplings in minimal flavor deconstruction at LO in the spurion expansion.}
\label{fig:mssLO}
\end{figure}

\paragraph{LO effects.} The lepton Yukawa matrix arises after the new scalar fields $\phi$, $\chi^\ell$ and $\sigma$ acquire their VEVs, denoted generically as $\langle\phi_i\rangle$. At the LO in the \emph{spurion} insertions, one obtains:
\begin{subequations}\label{eq:mass_spurionLO}
    \begin{align}
 \mathcal{Y}^{e}_{i1} & =Y^e\cdot M_{E_\alpha}^{-1}\cdot \hat{Y}_3^{\phi}\cdot M_{E_3}^{-1}\cdot Y^\sigma \langle\phi\rangle \langle\sigma\rangle\, , \\
 \mathcal{Y}^{e}_{i2} & =Y^e\cdot M_{E_\alpha}^{-1}\cdot Y_2^{\phi} \langle\phi\rangle\, ,\\
\mathcal{Y}^{e}_{i3}&=Y^e\cdot M_{E_\alpha}^{-1}\cdot Y^{\chi} \langle\chi^\ell\rangle\,,\\
\quad \mathcal{Y}^{e}_{33} &=y_3^e\,,
\end{align} 
\end{subequations}
where $i=1,2$, and which can be conveniently rewritten in a compact matrix form as follows:
\begin{equation}\label{eq:Yukawa electron after diag}
 \mathcal{Y}^e  \approx  \begin{pmatrix}
  Y^e_{1\alpha} \hat{Y}^{\phi}_{3,\alpha}  Y^{\sigma} \varepsilon_\sigma \varepsilon_{\phi }  &    -Y^e_{1\alpha} Y^{\phi}_{2,\alpha} \varepsilon_\phi  & -Y^{e}_{1\alpha}  Y^{\chi}_{\alpha} \varepsilon_\chi   \\
   Y^e_{2\alpha}  \hat{Y}^{\phi}_{3, \alpha}  Y^{\sigma} \varepsilon_\sigma \varepsilon_{\phi } & -Y^e_{2\alpha}  Y^{\phi}_{2,\alpha} \varepsilon_\phi  &   -Y^{e}_{2\alpha}  Y^{\chi}_{\alpha}  \varepsilon_\chi  \\
 \approx 0 &  \approx 0 & y^e_3
 \end{pmatrix},
 \end{equation}
where we have introduced the small dimensionless parameters defined as\footnote{In order to correctly reproduce the hierarchical structure of the charged-lepton mass matrix, we assume all Yukawa couplings to be of $\mathcal{O}(1)$, with the exception of $Y^{e}_{i\alpha} \sim y_{3}^{e} \sim y_{\tau} \approx 10^{-2}$.
}

\begin{equation}\label{eq:varepsilon definition}
    \varepsilon_\phi\equiv \langle\phi\rangle/M_{E_{1,2}}\,,\qquad \varepsilon_\chi\equiv \langle\chi^\ell\rangle/M_{E_{1,2}}\,,\qquad \varepsilon_\sigma\equiv \langle\sigma\rangle/M_{E_3}\,.
\end{equation}Diagrammatically, the LO contribution is shown in Fig.~\ref{fig:mssLO}. 

\paragraph{NLO effects.} 
The spurion analysis outlined above can be systematically extended to derive NLO corrections to the lepton Yukawa matrix. In general, both Yukawa interactions and dipole operators receive contributions not only from genuine higher-order spurion insertions, but also from wave-function renormalization effects.
Pure mass-term corrections arise from diagrams connecting left- and right-handed leptons through heavy fermion propagators, while additional contributions originate from the LO mass terms in Eq.~\eqref{eq:mass_spurionLO} combined with wave-function renormalization effects~\cite{DAmbrosio:2002vsn,King:2004tx,Espinosa:2004ya,Davidson:2007si}. After electroweak symmetry breaking, the lepton mass term can therefore be written as
\begin{equation}
R_m^\dagger
\bigg(m^{\circ}_\ell 
+ \eta^{\ell}_{m} -{1\over2}\eta_{R}^{\ell} m^{\circ}_{\ell}-{1\over2}m^{\circ}_{\ell} \eta_{L}^{\ell}\bigg)
L_{m}= m_{\ell}\,.
\label{eq:mass_corr}
\end{equation}
where $m_{\ell}^{0}$ and $\eta_{m}^{\ell}$ denote the LO and NLO contributions to the lepton mass matrix, respectively, while $\eta_{L}^{\ell}$ and $\eta_{R}^{\ell}$ encode the wave-function renormalization of the left- and right-handed fields. The unitary matrices $L_{m}$ and $R_{m}$ diagonalize the resulting mass matrix. Additional details, together with a model-independent derivation of this expression, are provided in Appendix~\ref{Appendix:massterms}.

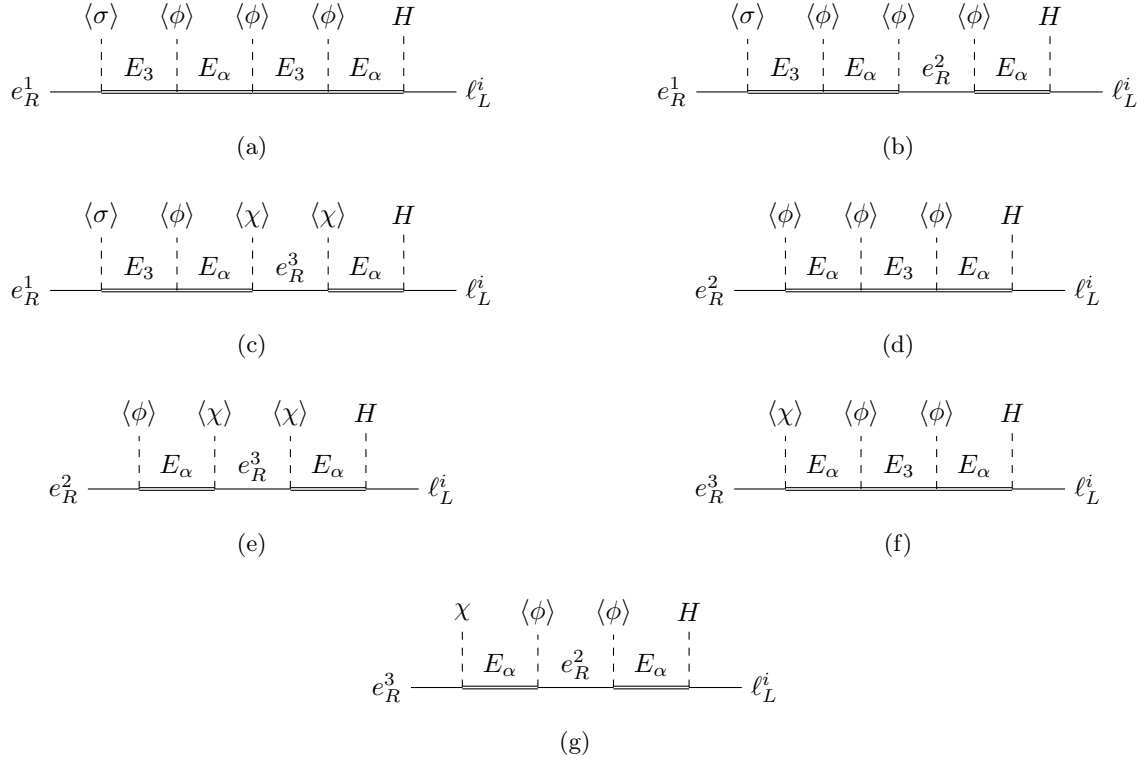
\begin{figure}[ht]
\centering

\begin{subfigure}[b]{0.45\textwidth}
\centering
\begin{tikzpicture}[baseline={(current bounding box.center)}]
    \begin{feynman}
        \vertex(a) at (0,0) {$e_R^1$};
        \vertex(b) at (1,0);
        \vertex(c) at (2,0);
        \vertex(b1) at (1,1) {$\langle\sigma\rangle$};
        \vertex(c1) at (2,1) {$\langle\phi \rangle$};
        \vertex(d) at (3,0);
        \vertex(d1) at (3,1) {$\langle \phi \rangle$};
        \vertex(e) at (4,0);
        \vertex(e1) at (4,1) {$\langle \phi \rangle$};
        \vertex(f) at (5,0);
        \vertex(f1) at (5,1) {$H$};
        \vertex(g) at (6,0) {$\ell_L^i$};
        \vertex(cap1) at (1.5,0.3) {$E_3$};
        \vertex(cap2) at (2.5,0.3) {$E_\alpha$};
        \vertex(cap3) at (3.5,0.3) {$E_3$};
        \vertex(cap4) at (4.5,0.3) {$E_\alpha$};
        \diagram{
            (a) -- [black] (b) -- [double, black] (c) -- [double,black] (d)
                -- [double,black] (e) -- [double,black] (f) -- [black] (g),
            (b) -- [scalar] (b1),
            (c) -- [scalar] (c1),
            (d) -- [scalar] (d1),
            (e) -- [scalar] (e1),
            (f) -- [scalar] (f1),
        };
    \end{feynman}
\end{tikzpicture}
\caption{}
\label{fig:massNLOa}
\end{subfigure}
\hfill
\begin{subfigure}[b]{0.45\textwidth}
\centering
\scalebox{1}{\begin{tikzpicture}[baseline={(current bounding box.center)}]
    \begin{feynman}
        \vertex(a) at (0,0) {$e_R^1$};
        \vertex(b) at (1,0);
        \vertex(c) at (2,0);
        \vertex(b1) at (1,1) {$\langle \sigma \rangle$};
        \vertex(c1) at (2,1) {$\langle \phi \rangle$};
        \vertex(d) at (3,0);
        \vertex(d1) at (3,1) {$\langle \phi \rangle$};
        \vertex(e) at (4,0);
        \vertex(e1) at (4,1) {$\langle \phi \rangle$};
        \vertex(f) at (5,0);
        \vertex(f1) at (5,1) {$H$};
        \vertex(g) at (6,0) {$\ell_L^i$};
        \vertex(cap1) at (1.5,0.3) {$E_3$};
        \vertex(cap2) at (2.5,0.3) {$E_\alpha$};
        \vertex(cap3) at (3.5,0.3) {$e_R^2$};
        \vertex(cap4) at (4.5,0.3) {$E_\alpha$};
        \diagram{
            (a) -- [black] (b) -- [double, black] (c) -- [double,black] (d)
                -- [black] (e) -- [double,black] (f) -- [black] (g),
            (b) -- [scalar] (b1),
            (c) -- [scalar] (c1),
            (d) -- [scalar] (d1),
            (e) -- [scalar] (e1),
            (f) -- [scalar] (f1),
        };
    \end{feynman}
\end{tikzpicture}}
\caption{}
\label{fig:massNLOb}
\end{subfigure}

\vspace{1em}

\begin{subfigure}[b]{0.45\textwidth}
\centering
\begin{tikzpicture}[baseline={(current bounding box.center)}]
    \begin{feynman}
        \vertex(a) at (0,0) {$e_R^1$};
        \vertex(b) at (1,0);
        \vertex(c) at (2,0);
        \vertex(b1) at (1,1) {$\langle \sigma \rangle$};
        \vertex(c1) at (2,1) {$\langle \phi \rangle$};
        \vertex(d) at (3,0);
        \vertex(d1) at (3,1) {$\langle \chi \rangle$};
        \vertex(e) at (4,0);
        \vertex(e1) at (4,1) {$\langle \chi \rangle$};
        \vertex(f) at (5,0);
        \vertex(f1) at (5,1) {$H$};
        \vertex(g) at (6,0) {$\ell_L^i$};
        \vertex(cap1) at (1.5,0.3) {$E_3$};
        \vertex(cap2) at (2.5,0.3) {$E_\alpha$};
        \vertex(cap3) at (3.5,0.3) {$e_R^3$};
        \vertex(cap4) at (4.5,0.3) {$E_\alpha$};
        \diagram{
            (a) -- [black] (b) -- [double, black] (c) -- [double,black] (d)
                -- [black] (e) -- [double,black] (f) -- [black] (g),
            (b) -- [scalar] (b1),
            (c) -- [scalar] (c1),
            (d) -- [scalar] (d1),
            (e) -- [scalar] (e1),
            (f) -- [scalar] (f1),
        };
    \end{feynman}
\end{tikzpicture}
\caption{}
\label{fig:massNLOc}
\end{subfigure}
\hfill
\begin{subfigure}[b]{0.45\textwidth}
\centering
\begin{tikzpicture}[baseline={(current bounding box.center)}]
    \begin{feynman}
        \vertex(a) at (1,0) {$e_R^2$};
        \vertex(c) at (2,0);
        \vertex(c1) at (2,1) {$\langle \phi \rangle$};
        \vertex(d) at (3,0);
        \vertex(d1) at (3,1) {$\langle \phi \rangle$};
        \vertex(e) at (4,0);
        \vertex(e1) at (4,1) {$\langle \phi \rangle$};
        \vertex(f) at (5,0);
        \vertex(f1) at (5,1) {$H$};
        \vertex(g) at (6,0) {$\ell_L^i$};
        \vertex(cap2) at (2.5,0.3) {$E_\alpha$};
        \vertex(cap1) at (3.5,0.3) {$E_3$};
        \vertex(cap3) at (4.5,0.3) {$E_\alpha$};
        \diagram{
            (a) -- [black] (c) -- [double,black] (d) -- [double,black] (e)
                -- [double,black] (f) -- [black] (g),
            (c) -- [scalar] (c1),
            (d) -- [scalar] (d1),
            (e) -- [scalar] (e1),
            (f) -- [scalar] (f1),
        };
    \end{feynman}
\end{tikzpicture}
\caption{}
\label{fig:massNLOd}
\end{subfigure}

\vspace{1em}

\begin{subfigure}[b]{0.45\textwidth}
\centering
\begin{tikzpicture}[baseline={(current bounding box.center)}]
    \begin{feynman}
        \vertex(a) at (1,0) {$e_R^2$};
        \vertex(c) at (2,0);
        \vertex(c1) at (2,1) {$\langle \phi \rangle$};
        \vertex(d) at (3,0);
        \vertex(d1) at (3,1) {$\langle \chi \rangle$};
        \vertex(e) at (4,0);
        \vertex(e1) at (4,1) {$\langle \chi \rangle$};
        \vertex(f) at (5,0);
        \vertex(f1) at (5,1) {$H$};
        \vertex(g) at (6,0) {$\ell_L^i$};
        \vertex(cap2) at (2.5,0.3) {$E_\alpha$};
        \vertex(cap1) at (3.5,0.3) {$e_R^3$};
        \vertex(cap3) at (4.5,0.3) {$E_\alpha$};
        \diagram{
            (a) -- [black] (c) -- [double,black] (d) -- [black] (e)
                -- [double,black] (f) -- [black] (g),
            (c) -- [scalar] (c1),
            (d) -- [scalar] (d1),
            (e) -- [scalar] (e1),
            (f) -- [scalar] (f1),
        };
    \end{feynman}
\end{tikzpicture}
\caption{}
\label{fig:massNLOe}
\end{subfigure}
\hfill
\begin{subfigure}[b]{0.45\textwidth}
\centering
\begin{tikzpicture}[baseline={(current bounding box.center)}]
    \begin{feynman}
        \vertex(a) at (1,0) {$e_R^3$};
        \vertex(c) at (2,0);
        \vertex(c1) at (2,1) {$\langle \chi \rangle$};
        \vertex(d) at (3,0);
        \vertex(d1) at (3,1) {$\langle \phi \rangle$};
        \vertex(e) at (4,0);
        \vertex(e1) at (4,1) {$\langle \phi \rangle$};
        \vertex(f) at (5,0);
        \vertex(f1) at (5,1) {$H$};
        \vertex(g) at (6,0) {$\ell_L^i$};
        \vertex(cap2) at (2.5,0.3) {$E_\alpha$};
        \vertex(cap1) at (3.5,0.3) {$E_3$};
        \vertex(cap3) at (4.5,0.3) {$E_\alpha$};
        \diagram{
            (a) -- [black] (c) -- [double,black] (d) -- [double,black] (e)
                -- [double,black] (f) -- [black] (g),
            (c) -- [scalar] (c1),
            (d) -- [scalar] (d1),
            (e) -- [scalar] (e1),
            (f) -- [scalar] (f1),
        };
    \end{feynman}
\end{tikzpicture}
\caption{}
\label{fig:massNLOf}
\end{subfigure}

\vspace{1em}

\begin{subfigure}[b]{0.7\textwidth}
\centering
\begin{tikzpicture}[baseline={(current bounding box.center)}]
    \begin{feynman}
        \vertex(a) at (1,0) {$e_R^3$};
        \vertex(c) at (2,0);
        \vertex(c1) at (2,1) {$\chi$};
        \vertex(d) at (3,0);
        \vertex(d1) at (3,1) {$\langle \phi \rangle$};
        \vertex(e) at (4,0);
        \vertex(e1) at (4,1) {$\langle \phi \rangle$};
        \vertex(f) at (5,0);
        \vertex(f1) at (5,1) {$H$};
        \vertex(g) at (6,0) {$\ell_L^i$};
        \vertex(cap2) at (2.5,0.3) {$E_\alpha$};
        \vertex(cap1) at (3.5,0.3) {$e_R^2$};
        \vertex(cap3) at (4.5,0.3) {$E_\alpha$};
        \diagram{
            (a) -- [black] (c) -- [double,black] (d) -- [black] (e) -- [double,black] (f) -- [black] (g),
            (c) -- [scalar] (c1),
            (d) -- [scalar] (d1),
            (e) -- [scalar] (e1),
            (f) -- [scalar] (f1),
        };
    \end{feynman}
\end{tikzpicture}
\caption{}
\label{fig:massNLOg}
\end{subfigure}

\caption{Feynman diagrams contributing to the lepton Yukawa couplings in minimal flavor deconstruction at NLO in the spurion expansion.}
\label{fig:massNLO}
\end{figure}

Overall, at NLO, the matrix elements $\mathcal{Y}^{e}_{i1}$ ($i=1,2$) receive contributions from the following terms:
\begin{subequations}\label{eq: spurions NLO i1}
    \begin{align}
    \mathcal{Y}_{i1}^{e(a)}&=Y^e\cdot M_{E_\alpha}^{-1}\cdot\Big[\hat{Y}_3^{\phi}\cdot M_{E_3}^{-1}\cdot (Y_3^{\phi})^\dag\cdot M_{E_\alpha}^{-1} \langle\phi\rangle^2\Big]\cdot \hat{Y}_3^{\phi}\cdot M_{E_3}^{-1}\cdot Y^\sigma \langle\phi\rangle\langle\sigma\rangle \,,\\
    \mathcal{Y}_{i1}^{e(b)} & =-\,Y^e\cdot M_{E_\alpha}^{-1}\cdot\Big[Y_2^{\phi}\cdot (Y_2^\phi)^\dag \langle\phi\rangle^2M_{E_\alpha}^{-2}\Big]\cdot \hat{Y}_3^{\phi}\cdot M_{E_3}^{-1}\cdot Y^\sigma \langle\phi\rangle\langle\sigma\rangle\,,\\
    \mathcal{Y}_{i1}^{e(c)}&= -\,Y^e\cdot M_{E_\alpha}^{-1}\cdot\Big[Y^{\chi}\cdot (Y^\chi)^\dag \langle\chi^{\ell}\rangle^2M_{E_\alpha}^{-2}\Big]\cdot \hat{Y}_3^{\phi}\cdot M_{E_3}^{-1}\cdot Y^\sigma \langle\phi\rangle\langle\sigma\rangle\,,
\end{align}
\end{subequations}
where the corresponding spurion diagrams are shown in Figs.~\ref{fig:massNLOa}, \ref{fig:massNLOb}, and \ref{fig:massNLOc}. The first diagram corresponds to the contribution encoded in $\eta^{\ell}_{m}$ after electroweak symmetry breaking, while the latter two arise from the wave-function renormalization of the right-handed fields, generating terms proportional to $\eta^{\ell}_{R} m_{\ell}^{0}$ in Eq.~\eqref{eq:mass_corr}.
Similarly, for the second and third generations, we obtain
\begin{subequations}\label{eq: spurions NLO i2}
\begin{align}
    \mathcal{Y}^{e(d)}_{i2}&=Y^e\cdot M_{E_\alpha}^{-1}\cdot\Big[\hat{Y}^{\phi}_3\cdot M_{E_3}^{-1}\cdot (Y^\phi_3)^\dag\cdot M_{E_\alpha}^{-1} \langle\phi\rangle^2\Big]\cdot Y_2^{\phi}\langle\phi\rangle\,, \\
    \mathcal{Y}^{e(e)}_{i2}&=  -Y^e\cdot M_{E_\alpha}^{-1}\cdot\Big[Y^{\chi}\cdot (Y^\chi)^\dag \langle\chi^\ell\rangle^2M_{E_\alpha}^{-2}\Big]\cdot Y_2^{\phi}\langle\phi\rangle\,, \\
    \mathcal{Y}^{e(f)}_{i3}&= Y^e\cdot M_{E_\alpha}^{-1}\cdot\Big[\hat{Y}^{\phi}_3\cdot M_{E_3}^{-1}\cdot (Y^\phi_3)^\dag\cdot M_{E_\alpha}^{-1} \langle\phi\rangle^2\Big]\cdot Y^{\chi}\langle\chi^\ell\rangle\,,\\
    \mathcal{Y}^{e(g)}_{i3}&= -Y^e\cdot M_{E_\alpha}^{-1}\cdot\Big[Y^{\phi}_2\cdot (Y^\phi_2)^\dag \langle\phi\rangle^2M_{E_\alpha}^{-2}\Big]\cdot Y^{\chi}\langle\chi^\ell\rangle\,,
\end{align}
\end{subequations}
where the corresponding spurion diagrams are shown in Figs.~\ref{fig:massNLOd}, \ref{fig:massNLOe}, \ref{fig:massNLOf}, and~\ref{fig:massNLOg}.

So far, we have analyzed the spurionic structure of the minimal FD model focusing on the lepton sector, without discussing the possible presence of CP-violating phases in the Yukawa couplings. However, the existence of the physical CKM phase in the SM implies that the corresponding Yukawa Lagrangian in the quark sector of the FD framework must contain complex couplings. In Appendix~\ref{sec:$CP$ violation phase}, we use the spurion analysis to reproduce the SM Jarlskog invariant~\cite{Jarlskog:1985ht,Bernabeu:1986fc,Botella:2004ks} in terms of the ultraviolet parameters of the model. This analysis assumes that the quark sector features vector-like fermions and scalar fields analogous to those introduced in the lepton sector, following the construction proposed in Ref.~\cite{Barbieri:2023qpf}. As shown in Appendix~\ref{sec:$CP$ violation phase}, reproducing the observed value of the Jarlskog invariant generically requires $\mathcal{O}(1)$ complex phases in the high-energy quark Yukawa couplings.
Once complex phases are present in the quark sector, there is no compelling reason to assume real couplings in the lepton sector. In the following, we therefore allow for complex leptonic Yukawa interactions and investigate their implications for CP-violating observables, with particular emphasis on EDMs.

\subsection{Lepton Yukawa matrix diagonalization}
\label{sec:lepyukdiag}
In this subsection, we reproduce the lepton Yukawa matrix shown in Eq. \eqref{eq:Yukawa electron after diag} by eliminating the heavy-light fermion mixing originating from the SSB chain in Eq. \eqref{eq:SSB}. We then diagonalize this matrix by applying rotations to the chiral SM leptons.

After the breaking of the full gauge symmetry down to the SM gauge group, the lepton mass matrix exhibits heavy–light fermion mixing of the form
\begin{equation} \label{eq:massmatrix1}
    \mathcal{L}_{Y}^e\supset \begin{pmatrix} \bar{\ell} & \bar{E}_{L} \end{pmatrix} \begin{pmatrix}
        0 & 0 \\ X_E & M_E
    \end{pmatrix} \begin{pmatrix} e \\ E_{R} \end{pmatrix}+\text{h.c.}\,,
\end{equation}
where each block of the matrix is understood to be a $3 \times 3$ matrix in flavor space, whose entries can be straightforwardly derived from Eq.~\eqref{eq:Yukawa}. Explicitly, they read
\begin{equation}
    X_E=\begin{pmatrix}
        0 & Y^\phi_{2,1}\langle\phi\rangle & Y^\chi_1\langle\chi^\ell\rangle \\ 0 & Y^\phi_{2,2}\langle\phi\rangle & Y^\chi_2\langle\chi^\ell\rangle \\ 
        Y^\sigma & 0 & 0
    \end{pmatrix}\,,\qquad M_E=\begin{pmatrix}
        M_{E_1} & 0 & \hat{Y}^\phi_{3,1}\langle\phi\rangle \\ 0 & M_{E_2} & \hat{Y}^\phi_{3,2}\langle\phi\rangle \\ (Y^\phi_{3,1})^*\langle\phi\rangle & (Y^\phi_{3,2})^*\langle\phi\rangle & M_{E_3}
    \end{pmatrix}\,.
\end{equation}
In order to eliminate the heavy–light mixing and isolate the physical degrees of freedom of the theory, we perform a rotation of the original right-handed fermions according to
\begin{equation}
\label{eq:mathcalUe rotation}
    \begin{pmatrix}
        e_R \\ E_R
    \end{pmatrix}\rightarrow \mathcal{W}_e\begin{pmatrix}
        e_R \\ E_R
    \end{pmatrix}\,,\qquad \mathcal{W}_e=\begin{pmatrix}
        \mathbb{I}-\frac{1}{2}\mathcal{E}_e^\dag\mathcal{E}_e & \mathcal{E}^\dag_e \\ -\mathcal{E}_e & \mathbb{I}-\frac{1}{2}\mathcal{E}_e\mathcal{E}^\dag_e
        \end{pmatrix}\,,
\end{equation}
such that the massless eigenvectors correspond to the SM fermions. Explicitly, we obtain
\begin{equation}\label{eq:mathcalEe def}
  \mathcal{E}_e =  M_E^{-1}X_E=\begin{pmatrix}
-Y^\sigma \hat{Y}_{3,1}^{\phi} \varepsilon_\phi \varepsilon_\sigma & Y_{2,1}^{\phi} \varepsilon_\phi & Y_{1}^{\chi} \varepsilon_{\chi} \\
-Y^\sigma \hat{Y}_{3,2}^{\phi } \varepsilon_\phi \varepsilon_\sigma & Y_{2,2}^{\phi} \varepsilon_\phi & Y_2^{\chi }  \varepsilon_\chi\\
Y^\sigma \varepsilon_\sigma & \approx 0 & \approx 0
\end{pmatrix}\,.
\end{equation}
After this rotation, the SM Yukawa Lagrangian takes the form
\begin{equation}
    \mathcal{L}_Y^e\supset\sum_{i,j=1,2,3}\overline{\ell}_L^i \mathcal{Y}^e_{ij}H e_R^j+\mathrm{h.c.}\,,
\end{equation}
where the charged lepton Yukawa matrix correctly reproduces the texture obtained from the LO \emph{spurion analysis} in Eq. \eqref{eq:Yukawa electron after diag}.

Although at this stage all SM leptons are massless, it is convenient to work in the mass-eigenstate basis by diagonalising the Yukawa matrix. This procedure allows a direct matching of the SMEFT Wilson coefficients (WCs) to physical observables. The diagonalization is performed as follows
\begin{equation}
\label{eq:ULUR rotation def}
    (U_L^\ell)^\dagger \mathcal{Y}^e U_R^e=\hat{\mathcal{Y}}^e=\text{diag}(y_e,y_\mu,y_\tau)\sim y_{\tau}\,\text{diag}(\varepsilon_\phi\varepsilon_\sigma,\varepsilon_\phi,1)\,,
\end{equation}
where, as usual, the unitary rotation matrices can be generically written as
\begin{equation}
    U[\theta_1,\theta_2,\theta_3]=\begin{pmatrix}
            1 & 0 & 0 \\ 0 & \cos{\theta_1} & \sin{\theta_1} e^{i\delta_{1}} \\ 0 & -\sin{\theta_1} e^{-i\delta_{1}} & \cos{\theta_1}
        \end{pmatrix}\begin{pmatrix}
          \cos{\theta_2} & 0 & \sin{\theta_2} e^{i\delta_{2}} \\ 0 & 1 & 0 \\ -\sin{\theta_2} e^{-i\delta_{2}} & 0 & \cos{\theta_2}
        \end{pmatrix}\begin{pmatrix}
            \cos{\theta_3} & \sin{\theta_3} e^{i\delta_{3}} & 0 \\ -\sin{\theta_3} e^{-i\delta_{3}} & \cos{\theta_3} & 0 \\ 0 & 0 & 1
        \end{pmatrix}\,.
\end{equation}
After expanding around small angles\footnote{This assumption does not necessarily hold for $\theta_{3}^{L}$ due to the large mixing between the first two generations.}, we find 
\begin{subequations}
    \begin{align}
    &(\theta_1^L)^*\approx -(y^e_3)^{-1} (Y^e\cdot Y^\chi)_{2}\,\varepsilon_\chi\,,\\
    &(\theta_2^L)^*\approx -(y^e_3)^{-1} (Y^e\cdot Y^\chi)_{1}\,\varepsilon_\chi\,,\\
    &e^{-i\delta_3}\tan{\theta_3^L}\approx \frac{(Y^e\cdot Y^\phi_2)_{1}}{(Y^e\cdot Y_2^\phi)_{2}} \,,
\end{align}
\end{subequations}
and
\begin{subequations}
    \begin{align}
    &\theta_1^R\approx |y^e_3|^{-2}(Y^e\cdot Y^\phi_2)^\dag (Y^e\cdot Y^\chi)\varepsilon_\chi\varepsilon_\phi\,,\\
    &\theta_2^R=\mathcal{O}(\varepsilon^3)\,, \\
    &\theta_3^R\approx -Y^\sigma\frac{(Y^e\cdot Y_2^\phi)^\dag (Y^e\cdot \hat{Y}_3^\phi)}{(Y^e\cdot Y_2^\phi)^\dag(Y^e\cdot Y_2^\phi)}\varepsilon_\sigma\,. 
\end{align}
\end{subequations}
This expansion allows us to express the rotation matrix up to 
$\mathcal{O}(\varepsilon^2)$ as
\begin{equation}\label{eq:rotationmatrix}
    U_{L}^\ell=\mathbb{I}+\Delta_{\ell}\,,\qquad U_{R}^e=\mathbb{I}+\Delta_e\,,
\end{equation}
where
\begin{subequations}\label{eq: Deltaell def}
\begin{gather}
    \Delta_\ell\approx\begin{pmatrix}
        \cos\theta^{L}_{3}-1 & \sin{\theta^{L}_3} e^{i\delta_{3}} & (\theta_2^L)^* \\ -\sin{\theta^{L}_3} e^{-i\delta_{3}} & \cos\theta^{L}_{3}-1 & (\theta_1^L)^* \\ -\theta_2^L & -\theta_1^L & -\frac{1}{2}|\theta_1^L|^2- \frac{1}{2}|\theta_2^L|^2
    \end{pmatrix},\\
    \Delta_e\approx\begin{pmatrix}
        -\frac{1}{2}|\theta_3^R|^2 & (\theta_3^R)^* & (\theta_2^R)^* \\ -\theta_3^R & -\frac{1}{2}|\theta_3^R|^2 & (\theta_1^R)^* \\ -\theta_2^R+\theta_1^R\theta_3^R & -\theta_1^R & - \frac{1}{2}|\theta_1^R|^2
    \end{pmatrix},
\end{gather}
\end{subequations}
leading to the following scaling behaviour for the lepton Yukawa eigenvalues
\begin{equation}\label{eq:eigenvalues Ye}
    y_\tau^2\approx |y^e_3|^2\,,\qquad y_\mu^2\approx (Y^e\cdot Y_2^\phi)^\dag (Y^e\cdot Y_2^\phi)\varepsilon_\phi^2\,,\qquad y_e^2\sim\varepsilon_\sigma^2\varepsilon_\phi^2\, .
\end{equation}

After eliminating the heavy–light fermion mixing and diagonalising 
the SM Yukawa matrices, the full Lagrangian can be written as
\begin{equation}
\begin{split}
    &\mathcal{L}_{UV}=\mathcal{L}_{\text{gauge}}+\sum_{\psi=q,\ell}\overline{\psi}_L(U_L^\psi)^\dag i\slashed D(U_L^\psi)\psi_{L}+\sum_{\psi=u,d,e}\begin{pmatrix}
        \overline{\psi}_R (U_R^\psi)^\dag & \overline{F}_R
    \end{pmatrix}\mathcal{W}_\psi^\dag i\slashed D \mathcal{W}_\psi\begin{pmatrix}
       (U_R^\psi) \psi_R \\ F_R
    \end{pmatrix}\\
    &+\sum_{F=U,D,E}\overline{F}_Li\slashed D F_L+\left(\sum_{F=U,D,E} \overline{F}_L M_F F_R+\text{h.c.}\right)+\mathcal{L}_{\text{scalars}}+\sum_{\psi=u,d,e}\mathcal{L}_{Y}^\psi\,.
\end{split}
\end{equation}
In particular, the lepton Yukawa Lagrangian is modified as follows
\begin{equation}
    -\mathcal{L}_Y^e=\sum_{i=1,2,3}\overline{\ell}_L^i \hat{\mathcal{Y}}^e_{ii}H e_R^i+\sum_{i,\alpha=1,2}\overline{\ell}_L^i H Y^e_{i\alpha}E_R^\alpha+\overline{\ell}_L^3 H y^e_{e} (\mathcal{E})^{*}_{\alpha3}E_R^\alpha+\mathcal{L}_Y^{\text{res}}+\mathrm{h.c.}\,,
\end{equation}
where $\mathcal{L}_Y^{\text{res}}$ contains interactions involving heavy scalar fields together with either one SM fermion and one heavy fermion, 
or two SM fermions with suppressed couplings.

\section{Effective field theory for lepton flavor deconstruction}\label{sec:SMEFT}

The model presented in the previous sections features a non-trivial flavor structure, obtained by augmenting the SM particles and gauge content with additional new heavy dynamics. As a result, the proliferation of free parameters and new particles renders a systematic study of the phenomenology of the model challenging.
In this respect, it is convenient to adopt an effective field theory (EFT) approach where the SM Lagrangian is extended by an appropriate set of gauge invariant operators depending on the SM fields, the so-called SMEFT. The phenomenological impact of the heavy gauge bosons, heavy vector-like fermions and new scalars arising in flavor-deconstructed models can be then conveniently addressed by matching their effects onto the coefficients of SMEFT operators~\cite{Buchmuller:1985jz,Grzadkowski:2010es}
\begin{equation}
\mathcal{L} = \mathcal{L}_{\rm SM} + \sum_i C_i^{(6)} \mathcal{O}_i^{(6)}\,,
\end{equation}
where we restricted to dimension-six operators and we use a convention where the $C_i$ are dimensionfull. In Tab.~\ref{table:SMEFT}, we list the complete set of  operators relevant to our analysis which depend on lepton and quark fields, the gauge bosons and on the scalar electroweak doublet.

\begin{table}[h!]
\centering

\begin{minipage}[t]{0.47\textwidth}
\centering
\renewcommand{\arraystretch}{1.2}
\setlength{\tabcolsep}{5pt}

\begin{tabular}{c c c}
\toprule
\textbf{Class} & \textbf{Operator} & \textbf{Structure} \\
\midrule

\multirow{2}{*}{$H^4D^2$}
& $\mathcal{O}_{H\Box}$
& $(H^\dagger H)\Box(H^\dagger H)$ \\

& $\mathcal{O}_{HD}$
& $(H^\dagger D_\mu H)^\star(H^\dagger D^\mu H)$ \\

\midrule

\multirow{4}{*}{$\psi^2H^3,\psi^2H^2D$}

& $\mathcal{O}_{H\ell}^{(1)}$
& $(H^\dagger i\!\stackrel{\leftrightarrow}{D_\mu}\!H)
(\bar\ell_p\gamma^\mu\ell_r)$ \\
& $\mathcal{O}_{H\ell}^{(3)}$
& $(H^\dagger i\!\stackrel{\leftrightarrow}{D_\mu^I}\!H)
(\bar\ell_p\gamma^\mu\tau_I\ell_r)$ \\
& $\mathcal{O}_{He}$
& $(H^\dagger i\!\stackrel{\leftrightarrow}{D_\mu}\!H)
(\bar e_p\gamma^\mu e_r)$ \\
& $\mathcal{O}_{eH}$
& $(H^\dagger H)(\bar\ell_p e_r H)$ \\

\midrule

\multirow{2}{*}{$\psi^2XH$}
& $\mathcal{O}_{eB}$
& $(\bar\ell_p\sigma_{\mu\nu}e_r)HB^{\mu\nu}$ \\

& $\mathcal{O}_{eW}$
& $(\bar\ell_p\sigma_{\mu\nu}e_r)\tau^a HW^{a,\mu\nu}$ \\

\bottomrule
\end{tabular}
\end{minipage}
\hfill
\begin{minipage}[t]{0.47\textwidth}
\centering
\renewcommand{\arraystretch}{1.2}
\setlength{\tabcolsep}{5pt}

\begin{tabular}{c c c}
\toprule
\textbf{Class} & \textbf{Operator} & \textbf{Structure} \\
\midrule

\multirow{9}{*}{$\psi^4$}
& $\mathcal{O}_{\ell\ell}$
& $(\bar\ell_p\gamma_\mu\ell_r)
(\bar\ell_s\gamma^\mu\ell_t)$ \\

& $\mathcal{O}_{ee}$
& $(\bar e_p\gamma_\mu e_r)
(\bar e_s\gamma^\mu e_t)$ \\

& $\mathcal{O}_{\ell e}$
& $(\bar\ell_p\gamma_\mu\ell_r)
(\bar e_s\gamma^\mu e_t)$ \\

& $\mathcal{O}_{qe}$
& $(\bar q_p\gamma_\mu q_r)
(\bar e_s\gamma^\mu e_t)$ \\

& $\mathcal{O}_{\ell u}$
& $(\bar\ell_p\gamma_\mu\ell_r)
(\bar u_s\gamma^\mu u_t)$ \\

& $\mathcal{O}_{\ell q}^{(1)}$
& $(\bar\ell_p\gamma_\mu\ell_r)
(\bar q_s\gamma^\mu q_t)$ \\

& $\mathcal{O}_{ed}$
& $(\bar e_p\gamma_\mu e_r)
(\bar d_s\gamma^\mu d_t)$ \\

& $\mathcal{O}_{eu}$
& $(\bar e_p\gamma_\mu e_r)
(\bar u_s\gamma^\mu u_t)$ \\

& $\mathcal{O}_{\ell d}$
& $(\bar\ell_p\gamma_\mu\ell_r)
(\bar d_s\gamma^\mu d_t)$ \\

\bottomrule
\end{tabular}
\end{minipage}

\caption{List of SMEFT operators divided by classes according to the particle content of the operators. Here $\tau_a$ refers to the Pauli matrices while $H^\dagger i\!\!\stackrel{\leftrightarrow}{D_\mu}\! H= i H^{\dag}(D_{\mu}- \overleftarrow{D}_{\mu})H$.}
\label{table:SMEFT}
\end{table}

\subsection{Tree-level matching with the SMEFT} \label{SMEFT_tree}

The construction of the SMEFT Lagrangian has been achieved by systematically applying the equation of motion of the heavy degrees of freedom. This procedure consists of replacing the heavy degrees of freedom in the Lagrangian with operators terms composed only by light fields and suppressed by the heavy mass scale. In particular, we integrate out the heavy vector-like fermions, the heavy scalars and the heavy gauge bosons without considering RGE effects so as to neglect the impact of the running and mixing of WCs across the various energy scales. According to the previous sections, we retain only the contribution up $\varepsilon_{i}^{2}$ in the parameter expansion. As mentioned in Sec.~\ref{sec:review}, we neglect effects stemming from the integration of $Z_{H}$, since the corresponding WCs will resemble the ones generated by the integration of $Z_{23}$ and $Z_{23}^{\prime}$ but with a larger mass suppression. In addition, we generally refer to $V$ in order to indicate $Z_{23}$ or $Z_{23}^{\prime}$, and, for phenomenological purposes, we will assume a mass hierarchy $M_{Z_{23}} < M_{Z_{23}^{\prime}}$. The results derived in this section by applying the equation of motions has been cross checked with Ref. \cite{deBlas:2017xtg} and the \texttt{Mathematica} package \texttt{Matchete} \cite{Fuentes-Martin:2022jrf}.

\paragraph{First class: \texorpdfstring{$\bm{H^{4}D^{2}}$}{x}.} This first class of operators involves the Higgs doublet and the SM gauge bosons and are generated by integrating out the vector bosons $Z_{23}$ and $Z_{23}^{\prime}$ at tree level. The corresponding WCs then read
\begin{subequations}
    \begin{align}
    &C_{H\Box}=-\frac{1}{2M_{V}^2}g_{V}^2(Q_{V}^H)^{2}\,,\\
    &C_{HD}=-\frac{2}{M_{V}^2}g_{V}^2(Q_{V}^H)^{2}=4C_{H\Box}\,,
\end{align}
\end{subequations}
where $Q_{V}^{H}$ denotes the coupling of the Higgs field to the vector boson $V$. Moreover, similarly to the Higgs mass parameter in Eq.~\eqref{eq: scalar lagrangian broken symmetry}, integrating out the heavy scalar fields induces a shift in the renormalizable quartic interaction of the Higgs potential according to
\begin{equation}
    \mathcal{L}_{\text{scalar}}\supset -\left[\lambda + \sum_i\frac{\langle \phi_i\rangle^2}{2M_{\phi_i}^2}\lambda_{iH}^2\right] (H^\dag H)^2\,.
\end{equation}

\paragraph{Second class: \texorpdfstring{$\bm{\psi^2 H^2 D}$}{x} and \texorpdfstring{$\bm{\psi^2 H^3}$}{x}.}\label{secondclacss} 
In the second class, we consider dimension-six operators containing a single fermion bilinear, restricting our analysis to the lepton sector. To parametrize the flavor structure of the vector bilinears, we introduce the following quantities in flavor space:
\begin{subequations}
\begin{align}
    &\begin{alignedat}{2}
    (P^e_{V})_{pr}=&\,Q_{V}^{e_p}\delta_{pr}+\sum_{\alpha= \text{heavy}}Q_{V}^{E_\alpha}(\mathcal{E}_e)^*_{\alpha p}(\mathcal{E}_e)_{\alpha r}
    -\frac{1}{2}\Big[Q_{V}^{e_p}+Q_{V}^{e_r}\Big](\mathcal{E}_e^\dag\mathcal{E}_e)_{pr}\\
    &+Q_{V}^{e_p}(\Delta_e)_{pr}+Q_{V}^{e_r}(\Delta_e)^*_{rp}+\sum_{k=\text{light}}Q_{V}^{e_k}(\Delta_e)_{kp}^*(\Delta_e)_{kr}\,,
\end{alignedat}\\
&(P^\ell_{V})_{pr}=Q_{V}^{\ell_p}\delta_{pr}+Q_{V}^{\ell_p}(\Delta_\ell)_{pr}+Q_{V}^{\ell_r}(\Delta_\ell)^*_{rp}+\sum_{k=\text{light}}Q_{V}^{\ell_k}(\Delta_\ell)_{kp}^*(\Delta_\ell)_{kr}\,,
\end{align}
\label{eq:Pe and Pell def}
\end{subequations}
where $Q_{V}^{f}$ parametrizes the coupling between the fermion $f$ and the vector boson $V$, while the matrices $\mathcal{E}_{e}$ and $\Delta_{e/\ell}$ are defined in Eqs. \eqref{eq:mathcalEe def} and \eqref{eq: Deltaell def}, respectively.
The expressions in Eq.~\eqref{eq:Pe and Pell def} encode the effect of the diagonalization of the Yukawa matrices done in Sec. \ref{sec:lepyukdiag}.

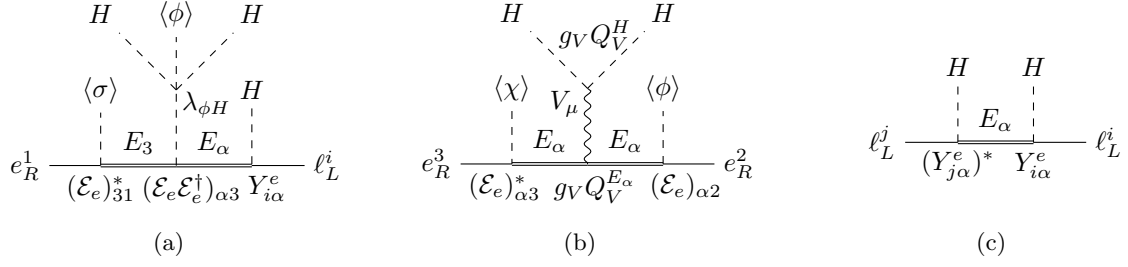
\begin{figure}
    \centering
    \begin{subfigure}[b]{0.3\textwidth}
\centering
\scalebox{1}{
    \begin{tikzpicture}
        \begin{feynman}
            \vertex(a) at (0,0) {$e_R^1$};
            \vertex(b) at (1,0);
            \vertex(c) at (2,0);
            \vertex(b1) at (1,1) {$\langle\sigma\rangle$};
            \vertex(b2) at (1,2) {$H$};
            \vertex(bbb) at (1,-0.3) {$(\mathcal{E}_e)_{31}^*$};
            \vertex(ccc) at (2.2,-0.3) {$(\mathcal{E}_e\mathcal{E}_e^\dag)_{\alpha3}$};
            \vertex(cccc) at (3.2,-0.3) {$Y^e_{i\alpha}$};
            \vertex(c1) at (2,2) {$\langle\phi\rangle$};
            \vertex(d) at (3,0);
            \vertex(d2) at (3,1) {$H$};
            \vertex(cd) at (2,1);
            \vertex(d1) at (3,2) {$H$};
            \vertex(e) at (4,0) {$\ell_L^i$};
            \vertex(E1) at (1.5,0.3) {$E_3$};
            \vertex(E2) at (2.5,0.3) {$E_\alpha$};
            \vertex(lambda) at (2.4,0.8){$\lambda_{\phi H}$};
            \diagram{
            (a) -- [black] (b) -- [double, black] (c) -- [double,black] (d) -- [black] (e),
            (b) -- [scalar] (b1),
            (c) -- [scalar] (cd),
            (cd) -- [scalar] (d1),
            (cd) -- [scalar] (c1),
            (d2) -- [scalar] (d),
            (cd) -- [scalar] (b2),
            };
        \end{feynman}
    \end{tikzpicture}}
    \caption{}
\end{subfigure}
\hfill
\begin{subfigure}[b]{0.3\textwidth}
\centering
\scalebox{1}{
    \begin{tikzpicture}
        \begin{feynman}
            \vertex(a) at (0,0) {$e_R^3$};
            \vertex(b) at (1,0);
            \vertex(c) at (2,0);
            \vertex(b1) at (2,1);
            \vertex(b2) at (1,1) {$\langle\chi\rangle$};
            \vertex(bbb) at (0.9,-0.3) {$(\mathcal{E}_e)_{\alpha 3}^*$};
            \vertex(h) at (1,2) {$H$};
            \vertex(c1) at (3,2) {$H$};
            \vertex(d) at (3,0);
            \vertex(ccc) at (3.3,-0.3) {$(\mathcal{E}_e)_{\alpha 2}$};
            \vertex(cccc) at (2.1,-0.3) {$g_V Q_V^{E_\alpha}$};
            \vertex(ccccc) at (2.1,1.7) {$g_V Q_V^{H}$};
            \vertex(d2) at (3,1) {$\langle \phi\rangle$};
            \vertex(e) at (4,0) {$e_R^2$};
            \vertex(E1) at (1.5,0.3) {$E_\alpha$};
            \vertex(E2) at (2.5,0.3) {$E_\alpha$};
            \vertex(Z) at (1.7,0.8) {$V_\mu$};
            \diagram{
            (a) -- [black] (b) -- [double,black] (c) -- [double,black] (d) -- [black] (e),
            (c) -- [photon] (b1),
            (b1) -- [scalar] (h),
            (b1) -- [scalar] (c1),
            (d) -- [scalar] (d2),
            (b2) -- [scalar] (b),
            };
        \end{feynman}
    \end{tikzpicture}}
    \caption{}
\end{subfigure}
\hfill
\begin{subfigure}[b]{0.3\textwidth}
\centering
\scalebox{1}{
    \begin{tikzpicture}
        \begin{feynman}
            \vertex(a) at (0,0) {$\ell_L^j$};
            \vertex(b) at (1,0);
            \vertex(b2) at (1,1) {$H$};
            \vertex(d) at (2,0);
            \vertex(d2) at (2,1) {$H$};
            \vertex(e) at (3,0) {$\ell_L^i$};
            \vertex(E1) at (1.5,0.3) {$E_\alpha$};
            \vertex(ccccc) at (1,-0.3) {$(Y^e_{j\alpha})^*$};
            \vertex(cccc) at (2,-0.3) {$Y^e_{i\alpha }$};
            \vertex(label) at (1.5,-0.8) {};
            \diagram{
            (a) -- [black] (b) -- [double, black] (d) -- [black] (e),
            (b) -- [scalar] (b2),
            (d) -- [scalar] (d2),
            };
        \end{feynman}
    \end{tikzpicture}}
    \caption{}
    \end{subfigure}

    \caption{Examples of Feynman diagrams contributing to the second class of SMEFT operators. Diagram $(a)$ generates the operator $\mathcal{O}_{eH}$, diagram $(b)$ contributes to $\mathcal{O}_{He}$ and, upon the replacement $e_R \leftrightarrow e_L$, to $\mathcal{O}_{H\ell}^{(1,3)}$, while diagram $(c)$ generates $\mathcal{O}_{H\ell}^{(1,3)}$.}
    \label{fig:second class feynman diagrams}
\end{figure}

With this definition, we derive the matching conditions for the operators in the class $\psi^2 H^2 D$. These operators arise either from integrating out a heavy vector boson, as in the case of $\mathcal{O}_{He}$, or from integrating out a heavy vector-like fermion, as in the case of $\mathcal{O}_{H\ell}^{(3)}$, see Fig.~\ref{fig:second class feynman diagrams}. The operator $\mathcal{O}_{H\ell}^{(1)}$, instead, receives contributions from both types of heavy fields.\footnote{For the terms proportional to $1/M_{E_\alpha}^2$, we neglect possible effects arising from flavor rotations, since they already contribute at $\mathcal{O}(\varepsilon^2)$ according to Eq.~\eqref{eq:varepsilon definition}.
}
\begin{subequations}
    \begin{align}
    &(C_{He})_{pr} =-\sum_{V=Z_{23},Z_{23}^\prime}\frac{1}{M_{V}^2}g_{V}^2Q_{V}^{H}(P^e_V)_{pr}\,,
        \\
    &(C^{(1)}_{H\ell})_{pr}=-\sum_{V=Z_{23},Z_{23}^\prime}\frac{1}{M_{V}^2}g_{V}^2Q_{V}^{H}(P^\ell_V)_{pr}-\frac{1}{4M^2_{E_\alpha}}Y^e_{p\alpha}(Y^e_{r\alpha})^*\,,
        \\
    &(C^{(3)}_{H\ell})_{pr}=-\frac{1}{4M^2_{E_\alpha}}Y^e_{p\alpha}(Y^e_{r\alpha})^*\,.
\end{align}\label{eq:C_He and C_Hell}
\end{subequations}

Regarding the class $\psi^2H^3$, only one operator is generated in our setup, $\mathcal{O}_{eH}$. 
This is obtained by the integration of a heavy fermion, followed by the subsequent integration of a heavy scalar, see Fig.~\ref{fig:second class feynman diagrams}. 
As a result, these interactions generate WCs that are approximately aligned with the Yukawa matrix $\mathcal{Y}^e$ in Eq.~\eqref{eq:Yukawa electron after diag}. The only exception is that interactions of the type $i3$ involve the quartic coupling $\lambda_{\chi H}$, whereas in the remaining terms the field $\chi$ is replaced by $\phi$. This generically induces an effective misalignment with the Yukawa matrix whenever $\lambda_{\phi H} \neq \lambda_{\chi H}$. 
Diagrammatically, this effect can be understood by attaching the Higgs bilinear $H^\dagger H$ to the scalar vacuum expectation values appearing in Fig.~\ref{fig:mssLO}.
In the flavor basis, we have 
\begin{equation}
    C_{e H}=\frac{\lambda_{H\phi}}{M_{\phi}^2} \left[\mathcal{Y}^e+\begin{pmatrix}
        0 & 0 & \delta_{\phi\chi}\mathcal{Y}^e_{13} \\ 0 & 0 & \delta_{\phi\chi}\mathcal{Y}^e_{23} \\ 0 & 0 & -\mathcal{Y}^e_{33}
    \end{pmatrix}\right]\,,\qquad \delta_{\phi\chi}\equiv \frac{\lambda_{\chi H}}{\lambda_{\phi H}}\frac{M_\phi^2}{M_\chi^2}-1\,.
\end{equation}
By rotating the light fields to the mass basis as in Eq.~\eqref{eq:ULUR rotation def}, we find that the approximate alignment with the Yukawa matrix renders the corresponding WCs nearly diagonal. The only off-diagonal contributions arise from the rotation of the terms misaligned with the Yukawa structure, which are formally NLO in the spurion expansion
\begin{equation}\label{eq: delta CeH}
    \Delta C^{\rm LO}_{eH}=\frac{\lambda_{\phi H}}{M_{\phi}^2}(\mathbb{I}+\Delta_\ell)^\dag \begin{pmatrix}
        0 & 0 & \delta_{\phi\chi}\mathcal{Y}^e_{13} \\ 0 & 0 & \delta_{\phi\chi}\mathcal{Y}^e_{23} \\ 0 & 0 & -\mathcal{Y}^e_{33}
    \end{pmatrix}(\mathbb{I}+\Delta_e)\,.
\end{equation}

In addition, a second NLO contribution arises from the same spurionic structures responsible for the NLO corrections to the mass matrix (see Fig.~\ref{fig:massNLO}), analogously to the LO case. A complete parametrization of this contribution would require the inclusion of multiple spurion insertions along the corresponding lines and is therefore beyond the scope of the present analysis. In the following, we collectively denote these effects by $C_{eH}^{\text{NLO}}$. 
Overall, after diagonalizing the Yukawa matrix $\mathcal{Y}^e$, the WC of the operator $\mathcal{O}_{eH}$ takes the form\footnote{As in the case of $C_{\ell H}^{(1,3)}$, the NLO contribution is unaffected by the diagonalization matrices at this order in the perturbative expansion, since it already scales as $C_{eH}^{\text{NLO}} \sim 1/M_{E_\alpha}^2$.}
\begin{equation}
    (C_{eH})_{pr}=\frac{\lambda_{\phi H}}{M_\phi^2}y_p\delta_{pr}+(\Delta C^{\rm LO}_{eH})_{pr}+(C_{eH}^{\text{NLO}})_{pr}\,,
\end{equation}
where $y_p$ are the eigenvalues of $\mathcal{Y}^e$ in Eq. \eqref{eq:eigenvalues Ye}.

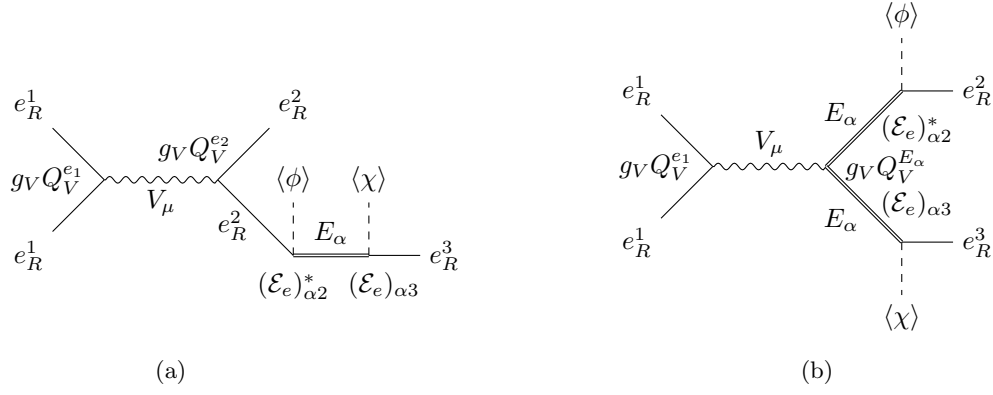
\begin{figure}
    \centering
    \begin{subfigure}[b]{0.45\textwidth}
\centering
        \begin{tikzpicture}
        \begin{feynman}
        \vertex(phantom) at (-2.5,0) {};
            \vertex(a) at (0,0);
            \vertex(b) at (-1,1) {$e_R^1$};
            \vertex(c) at (-1,-1) {$e_R^1$};
            \vertex(d) at (1.5,0);
            \vertex(e) at (2.5,1) {$e_R^2$};
            \vertex(f) at (2.5,-1); \vertex(ff) at (1.7,-0.6) {$e_R^2$};
            \vertex(h) at (3.5,-1);
            \vertex(i) at (4.5,-1) {$e_R^3$};
            \vertex(phi) at (2.5,0) {$\langle\phi\rangle$};
            \vertex(EE) at (3,-0.7) {$E_\alpha$};
            \vertex(Ea1) at (2.5,-1.4) {$(\mathcal{E}_e)^*_{\alpha 2}$};
            \vertex(Ea2) at (3.7,-1.4) {$(\mathcal{E}_e)_{\alpha 3}$};
            \vertex(chi) at (3.5,0) {$\langle\chi\rangle$};
            \vertex(zz) at (1.2,0.4) {$g_V Q_V^{e_2}$};
            \vertex(zz) at (-0.75,0) {$g_V Q_V^{e_1}$};
            \vertex(Z) at (0.75,-0.3) {$V_\mu$};
            \vertex(phantom) at (0,-2) {};
            \diagram{
            (b) -- [black] (a) -- [black] (c),
            (a) -- [photon] (d),
            (e) -- [black] (d) -- [black] (f) -- [double,black] (h) -- [black] (i),
            (f) -- [scalar] (phi),
            (h) -- [scalar] (chi),
            };
        \end{feynman}
    \end{tikzpicture}
    \caption{}
    \end{subfigure}
    \hfill \begin{subfigure}[b]{0.45\textwidth}
\centering
        \begin{tikzpicture}
        \begin{feynman}
        \vertex(phantom) at (4,0) {};
            \vertex(a) at (0,0);
            \vertex(b) at (-1,1) {$e_R^1$};
            \vertex(c) at (-1,-1) {$e_R^1$};
            \vertex(d) at (1.5,0);
            \vertex(e) at (2.5,1);
            \vertex(ee) at (2.5,2) {$\langle\phi\rangle$};
            \vertex(f) at (2.5,-1);
            \vertex(ff) at (2.5,-2) {$\langle\chi\rangle$};
            \vertex(eee) at (3.5,1) {$e_R^2$};
            \vertex(fff) at (3.5,-1) {$e_R^3$};
            \vertex(Z) at (0.75,0.3) {$V_\mu$};
            \vertex(zz) at (2.3,0) {$g_V Q_V^{E_\alpha}$};
            \vertex(zz) at (-0.75,0) {$g_V Q_V^{e_1}$};
            \vertex(lab1) at (2.7,0.5) {$(\mathcal{E}_e)_{\alpha2}^*$};
            \vertex(lab2) at (2.7,-0.5) {$(\mathcal{E}_e)_{\alpha3}$};
            \vertex(Ea1) at (1.7,0.7) {$E_\alpha$};
            \vertex(Ea2) at (1.7,-0.7) {$E_\alpha$};
            \diagram{
            (b) -- [black] (a) -- [black] (c),
            (a) -- [photon] (d),
            (eee) -- [black] (e) -- [double,black] (d) -- [double,black] (f) -- [black] (fff),
            (f) -- [scalar] (ff),
            (e) -- [scalar] (ee),
            };
        \end{feynman}
    \end{tikzpicture}
    \caption{}
    \end{subfigure}
    \caption{Examples of Feynman diagrams contributing to the third class of SMEFT operator $\mathcal{O}_{ee}$.}
    \label{fig:third class feynman diagrams}
\end{figure}

\paragraph{Third class: \texorpdfstring{$\bm{\psi^4}$}{x}.}\label{Thirdclass} 
The third class consists of four-fermion operators with vector-vector interactions generated by integrating out heavy vector bosons; see e.g.~Fig.~\ref{fig:third class feynman diagrams} for the representative 
case of $\mathcal{O}_{ee}$.

Unlike the previous sections, we also include operators involving quark fields, since certain semileptonic interactions play an important role in the phenomenology of lepton flavor physics.
The corresponding matching conditions read
\begin{subequations}\label{eq: 4fermion operators}
    \begin{align}
    &(C_{\ell\ell})_{prst} =-\frac{1}{2}\sum_{V=Z_{23},Z_{23}^\prime}\frac{1}{M_{V}^2}g_{V}^2(P^\ell_V)_{pr}(P^\ell_V)_{st}\,,\\
    &(C_{ee})_{prst} =-\frac{1}{2}\sum_{V=Z_{23},Z_{23}^\prime}\frac{1}{M_{V}^2}g_{V}^2(P^e_V)_{pr}(P^e_V)_{st}\,,\\
    &(C_{\ell e})_{prst} =-\sum_{V=Z_{23},Z_{23}^\prime}\frac{1}{M_{V}^2}g_{V}^2(P^\ell_V)_{pr}(P^e_V)_{st}\,,\\
    &(C_{\ell u})_{prst} =-\sum_{V=Z_{23},Z_{23}^\prime}\frac{1}{M_{V}^2}g_{V}^2(P^\ell_V)_{pr}(P^u_V)_{st}\,,\\
    &(C_{\ell d})_{prst} =-\sum_{V=Z_{23},Z_{23}^\prime}\frac{1}{M_{V}^2}g_{V}^2(P^\ell_V)_{pr}(P^d_V)_{st}\,,\\
    &(C_{qe})_{prst} =-\sum_{V=Z_{23},Z_{23}^\prime}\frac{1}{M_{V}^2}g_{V}^2(P^q_V)_{pr}(P^e_V)_{st}\,,\\
    &(C_{\ell q}^{(1)})_{prst} =-\sum_{V=Z_{23},Z_{23}^\prime}\frac{1}{M_{V}^2}g_{V}^2(P^{\ell}_V)_{pr}(P^{q}_V)_{st}\,,\\
    &(C_{ed})_{prst} =-\sum_{V=Z_{23},Z_{23}^\prime}\frac{1}{M_{V}^2}g_{V}^2(P^e_V)_{pr}(P^d_V)_{st}\,,\\
    &(C_{eu})_{prst} =-\sum_{V=Z_{23},Z_{23}^\prime}\frac{1}{M_{V}^2}g_{V}^2(P^e_V)_{pr}(P^{u}_V)_{st}\,,
\end{align}
\end{subequations}
where $P_V^{q,u,d}$ mirrors the structure of $P_V^{\ell,e}$, with the matrices defined in Eq.~\eqref{eq:Pe and Pell def} replaced by their analogues in the quark sector. However, since our analysis is not focused on flavor-violating processes involving quarks, we restrict our attention to the diagonal components only. For simplicity, we further assume the quark charges $Q_V^{q,u,d}$ to coincide with the corresponding SM hypercharges.

\subsection{Loop effects}
\label{sec:dipole}
In the following subsection, we examine the contributions to the SMEFT dipole operators arising from loop diagrams involving heavy states. On general grounds, dipole operators are generated at the one-loop level by considering the spurionic structures shown in Figs.~\ref{fig:mssLO} and~\ref{fig:massNLO}, closing the corresponding diagrams either through quartic scalar interactions or gauge-boson exchange, and attaching an electroweak gauge boson to one of the fermion lines. A complete computation of the dipole WCs would require a systematic separation between short-distance UV contributions and long-distance infrared effects in the one-loop amplitudes of the UV theory. However, for the purposes of the present work, we restrict ourselves to a dimensional estimate of the size and flavor structure of the resulting coefficients. Since anomalous magnetic moments (AMMs) and electric dipole moments (EDMs) are defined in the mass eigenstate basis, it is essential to perform the field rotations introduced in Eq.~\eqref{eq:rotationmatrix}.

At leading order, AMM contributions arise from loop diagrams constructed from the spurionic structures in Fig.~\ref{fig:mssLO}, yielding contributions proportional to the lepton masses once the Yukawa matrix is diagonalized. In contrast, the leading contributions to EDMs and LFV observables require, respectively, complex and flavor off-diagonal WCs. As discussed above, generating these structures crucially relies on NLO corrections to the Yukawa sector, which induce both physical CP phases and flavor misalignment relative to the leading-order contribution. Consequently, the complex and flavor off-diagonal components of the dipole WCs are controlled by the same combinations of couplings entering the NLO corrections to the Yukawa matrix.

\begin{figure}[htp]
\centering

\begin{subfigure}[b]{0.45\textwidth}
\centering
\scalebox{1}{\begin{tikzpicture}[baseline]
        \begin{feynman}
            \vertex(a) at (0,0) {$e_R^1$};
            \vertex(b) at (1,0);
            \vertex(c) at (1.6,0);
            \vertex(b1) at (1,1) {$\langle\sigma\rangle$};
            \vertex(c1) at (1.5,1.5) {$\langle\phi\rangle$};
            \vertex(d) at (3.4,0);
            \vertex(cd) at (2.5,1);
            \vertex(d1) at (3.5,1.5) {$H$};
            \vertex(e) at (4.4,0){$\ell_L^i$};
             \vertex(f) at (3.4,0.7);
            \vertex(g) at (4.7,1.5) {$\gamma$};
            \diagram{
            (a) -- [black] (b) -- [double, black] (c) -- [double,black] (d) -- [black] (e),
            (b) -- [scalar] (b1),
            (c) -- [scalar,quarter left] (cd) -- [scalar,quarter left] (d),
            (cd) -- [scalar] (d1),
            (cd) -- [scalar] (c1),
            (f) -- [boson] (g)
            };
        \end{feynman}
    \end{tikzpicture}}
\caption{}
\end{subfigure}
\hfill
\begin{subfigure}[b]{0.45\textwidth}
\centering
\scalebox{1}{\begin{tikzpicture}[baseline]
        \begin{feynman}
            \vertex(a) at (-1,0) {$e_R^1$};
            \vertex(a1) at (0,0);
            \vertex(b) at (1,0);
            \vertex(c) at (2,0);
            \vertex(b1) at (1,1) {$\langle\sigma\rangle$};
            \vertex(c1) at (2,1) {$\langle\phi\rangle$};
            \vertex(d) at (3,0);
            \vertex(d1) at (3,1) {$H$};
            \vertex(e) at (5,0) {$\ell_L^i$};
            \vertex(e1) at (4,0);
            \vertex(f) at (4.1,0.7);
            \vertex(g) at (5,1.7) {$\gamma$};
            \diagram{
            (a) -- [black] (b) -- [double, black] (c) -- [double,black] (d) -- [black] (e),
            (b) -- [scalar] (b1),
            (c) -- [scalar] (c1),
            (d) -- [scalar] (d1),
            (a1) -- [boson,half left] (e1),
            (f) -- [boson] (g) 
            };
        \end{feynman}
    \end{tikzpicture}}
\caption{}
\end{subfigure}
\caption{Feynman diagrams for dipoles in minimal flavor deconstruction at the leading order in the \emph{spurion} expansion. The photon field has to be attached to all possible internal and external lines.}
\label{fig:diploop}
\end{figure}
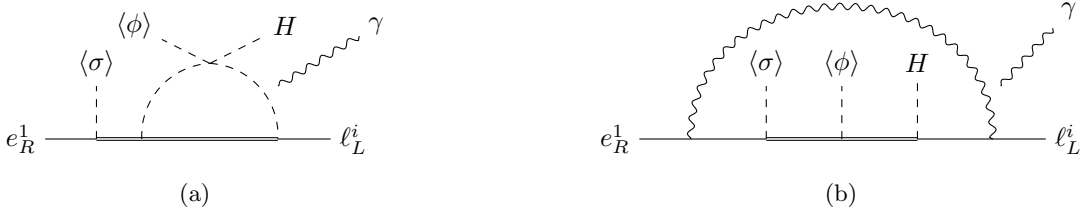

Similarly to the case of $\mathcal{O}_{eH}$, by considering the LO spurionic structures we obtain
\begin{equation}
    C_{eX}=\frac{g_X}{16\pi^2}D_{12}^X \left[\mathcal{Y}^e+\begin{pmatrix}
        0 & 0 & \delta^X_D\mathcal{Y}^e_{13} \\ 0 & 0 & \delta^X_D \mathcal{Y}^e_{23} \\ 0 & 0 & -\mathcal{Y}^e_{33}
    \end{pmatrix}\right]\,,\qquad \delta^X_{D}\equiv \frac{D_3^X}{D_{12}^X}-1\,,
\end{equation}
where $\mathcal{Y}^e$ is the electron Yukawa matrix in Eq. \eqref{eq:Yukawa electron after diag}, $X= B,W^{I}$, and 
the loop functions $D^X_{12(3)}$ read
\begin{subequations}
    \begin{align}
        &D^X_{12}=\frac{\lambda_{H\phi}}{M_\phi^2}F_\phi^X+\sum_V\frac{g_V^2}{M_V^2}F_{V}^X(Q_V^{\psi_{1,2}})^2\,,\\
    &D^X_{3}=\frac{\lambda_{H\chi}}{M_\chi^2}F_\chi^X+\sum_V\frac{g_V^2}{M_V^2}F_{V}^XQ_V^{\psi_{1,2}}Q_V^{\psi_{3}}\,.
    \end{align}
    \label{eq:dipole LO loop functions}
\end{subequations}
Notice that the loop functions associated with the the first two fermion generations, $D^X_{12}$, are identical. 
This reflects the fact that, at the TeV scale, the flavor-deconstructed framework still treats the first two fermion generations as charged under the same gauge symmetry, unlike the third-generation leptons. As a consequence, already at leading order in the spurion expansion, the WC $C_{eX}$ is not exactly proportional to the Yukawa matrix $\mathcal{Y}^e$, with the misalignment originating from the couplings involving third-generation right-handed leptons. 
On the other hand, the entries $(C_{eX})_{3i}$ (with $i=1,2$) remain aligned with the corresponding Yukawa-matrix elements, reflecting the universal nature of the $\mathrm{SU}(2)_L$ interactions in the underlying ultraviolet theory. After diagonalizing the Yukawa matrix $\mathcal{Y}^e$ as in Eq. \eqref{eq:ULUR rotation def}, the LO dipole WC takes the form
\begin{equation}
    C_{eX}=\frac{g_X}{16\pi^2} D_{12}^X\left[\hat{\mathcal{Y}}^e+U_L^\dag \begin{pmatrix}
        0 & 0 & \delta^X_D\mathcal{Y}^e_{13} \\ 0 & 0 & \delta^X_D \mathcal{Y}^e_{23} \\ 0 & 0 & -\mathcal{Y}^e_{33}
    \end{pmatrix} U_R\right]\,.
\end{equation} 

As already emphasized, it is necessary to estimate the NLO contribution as well, since it provides effects comparable in size to the leading contributions to LFV and CP-violating dipole observables. The relevant spurionic structures are illustrated in Fig.~\ref{fig:massNLO}. Restricting ourselves to naive dimensional analysis, we can generically write
\begin{equation}
    (C_{eX}^{\text{NLO}})_{pr}=\frac{g_X}{16\pi^2}\sum_{i}\mathcal{Y}^{e(i)}_{pr}D^{X(i)}\,,
\end{equation}
where the sum runs over all possible NLO spurionic structures connecting a left-handed field of flavor $p$ to a right-handed field of flavor $r$, as given in Eqs.~\eqref{eq: spurions NLO i1} and~\eqref{eq: spurions NLO i2}. The loop functions $D^{X(i)}$ encode, for each spurionic structure, the different possible loop contractions of the diagrams in Fig.~\ref{fig:massNLO}. Combining LO and NLO contributions, we obtain
\begin{equation} 
\label{dipolegeneral}
    (C_{eX})_{pr}=\frac{g_X}{16\pi^2}D_{12}^X y_p\delta_{pr}+(\Delta C^{\rm LO}_{eX})_{pr}+(C_{eX}^{\text{NLO}})_{pr}\,,
\end{equation}
where $y_p$ denote the eigenvalues of the Yukawa matrix $\mathcal{Y}^e$ defined in Eq.~\eqref{eq:eigenvalues Ye}, while
\begin{equation}\label{eq: delta CeX}
    \Delta C^{\rm LO}_{eX}=\frac{g_X}{16\pi^2}D_{12}^X(\mathbb{I}+\Delta_\ell)^\dag \begin{pmatrix}
        0 & 0 & \delta_{D}^X\mathcal{Y}^e_{13} \\ 0 & 0 & \delta_{D}^X\mathcal{Y}^e_{23} \\ 0 & 0 & -\mathcal{Y}^e_{33}
    \end{pmatrix}(\mathbb{I}+\Delta_e)\,.
\end{equation}

\subsection{Flavor structure of the SMEFT operators}
\label{sec:flav structure to smeft}
As a preliminary step before discussing the phenomenological implications of the model, we summarize the flavor suppression factors associated with the fermionic bilinears entering the SMEFT WCs. We begin with chirality-conserving bilinears, which appear in $\psi^2 H^2 D$ and four-fermion ($\psi^4$) operators.
For left-left bilinears, namely $\overline{\ell}_p \Gamma \ell_r$, the flavor suppression is entirely controlled by the quantities $(P_V^\ell)_{pr}$ defined in Eq.~\eqref{eq:Pe and Pell def}. Similarly, for right-right bilinears, $\overline{e}_p \Gamma e_r$, the dominant flavor suppression factors are determined by $(P_V^e)_{pr}$.
The resulting suppression patterns are summarized in Tab.~\ref{tab:ChiChi fermionic bilienars suppression factors}, while the explicit spurion expansions of the flavor structures are reported in Appendix~\ref{app:Flavor structure relations of fermionic bilinears}.

\begin{table}[]
    \centering
    \begin{tabular}{|c|c|c|c|c|}
    \toprule
        $pr$ & $LL$ & $RR$ & $LR$ & $RL$\\
        \hline
        $33$ &  $1+\varepsilon_\chi^2$ & $1+\varepsilon_\chi^2+\varepsilon_\phi^2$ & $1+\varepsilon_{\chi,\phi}^2$ & $1+\varepsilon_{\chi,\phi}^2$\\
        $23$ & $\varepsilon_\chi$ & $\varepsilon_\chi\varepsilon_\phi$ & $\varepsilon_\chi$ & $\varepsilon_\chi\varepsilon_\phi$\\
        $13$ & $\varepsilon_\chi$ & $\varepsilon_\chi\varepsilon_\phi\varepsilon_\sigma$ & $\varepsilon_\chi$ & $\varepsilon_\sigma\varepsilon_\chi\varepsilon_\phi$ \\
        $22$ & $1+\varepsilon_\chi^2$ & $1+\varepsilon_\chi^2+\varepsilon_\phi^2$ & $\varepsilon_\phi+\varepsilon_\phi\varepsilon_{\chi,\phi}^2$ & $\varepsilon_\phi+\varepsilon_\phi\varepsilon_{\chi,\phi}^2$\\
        $12$ & $\varepsilon_\chi^2$ & $\varepsilon_\phi^2\varepsilon_\sigma$ & $\varepsilon_\chi^2\varepsilon_\phi$ & $\varepsilon_\sigma\varepsilon_\chi\varepsilon_\phi$\\
        $11$ & $1+\varepsilon_\chi^2$ & $1+\varepsilon_\sigma^2$ & $\varepsilon_\sigma\varepsilon_\phi+\varepsilon_\sigma\varepsilon_\phi\varepsilon_{\chi,\phi}^2$ & $\varepsilon_\sigma\varepsilon_\phi+\varepsilon_\sigma\varepsilon_\phi\varepsilon_{\chi,\phi}^2$\\
        \bottomrule
    \end{tabular}
    \caption{Flavor suppression factors associated with fermionic bilinears of the form $\overline{\psi}_p \Gamma \psi_r$ for a fixed chirality structure. For the diagonal components of $LR$ and $RL$ operators, the LO contribution is real, while NLO effects can induce complex phases.}
    \label{tab:ChiChi fermionic bilienars suppression factors}
\end{table}
The situation is more involved for chirality-violating fermionic bilinears, which enter $\psi^2 H^3$ and dipole operators, as in this case the leading complex contributions are of particular phenomenological relevance, especially for dipole observables.
In particular, from Eqs.~\eqref{eq: delta CeH} and~\eqref{eq: delta CeX}, we find
\begin{equation} 
\label{eq:dipolematrixdelta}
    \Delta C^{\rm LO}_{eH}\frac{M_\phi^2}{\lambda_{\phi H}},~ \Delta C^{\rm LO}_{eX}\frac{16\pi^2}{g_X D_{12}^X}~\sim y_{\tau} \begin{pmatrix}
        \varepsilon_\phi\varepsilon_\sigma\varepsilon_\chi^2 & \varepsilon_\phi\varepsilon_\chi^2 & \varepsilon_\chi \\ \varepsilon_\sigma\varepsilon_\phi\varepsilon_\chi^{2}& \varepsilon_\phi\varepsilon_\chi^2 & \varepsilon_\chi\\\varepsilon_\sigma\varepsilon_\phi\varepsilon_\chi& \varepsilon_\chi\varepsilon_\phi & 1+\varepsilon_\chi^2
    \end{pmatrix}\,.
\end{equation}
Note that both classes of operators exhibit the same flavor suppression pattern, since they originate from the same underlying spurionic structures. Instead, the genuine NLO contributions satisfy
\begin{equation}\label{eq:dipolematrixNLO}
    C_{eH}^{\text{NLO}}\frac{M_\phi^2}{\lambda_{\phi H}},~ C_{eX}^{\text{NLO}}\frac{16\pi^2}{g_X D_{12}^X}~\sim y_{\tau} \begin{pmatrix}
        \varepsilon_\phi\varepsilon_\sigma\left(\varepsilon_\chi^2+\varepsilon_\phi^2\right) & \varepsilon_\phi\varepsilon_\chi^2 & \varepsilon_\chi \varepsilon_\phi^2\\ \varepsilon_\phi\varepsilon_\sigma\left(\varepsilon_\chi^2+\varepsilon_\phi^2\right)& \varepsilon_\phi\varepsilon_\chi^2 & \varepsilon_\chi \varepsilon_\phi^2 \\\approx 0& \approx 0 & \varepsilon_{\phi,\chi}^2
    \end{pmatrix}\,.
\end{equation}
Notice that, for the light generations, the genuine NLO contributions 
($C^{\rm NLO}_{eX}$) and the residual LO terms ($\Delta C^{\rm LO}_{eX}$) arise 
at the same parametric order.
The resulting flavor suppression patterns are summarized in Tab.~\ref{tab:ChiChi fermionic bilienars suppression factors}, while the explicit spurion expansions are reported in Appendix~\ref{app:Flavor structure relations of fermionic bilinears}. Interestingly, the resulting flavor hierarchies are consistent with an approximate global $\mathrm{U}(2)^5$ flavor symmetry structure, as expected~\cite{Allwicher:2023shc}.

\subsection{LEFT-SMEFT matching} \label{section:LEFT}

Following a top-down approach, we consistently match the SMEFT operators onto the Low-Energy Effective Theory (LEFT) at tree level, following Ref.~\cite{Jenkins:2017jig}. Restricting our attention to the lepton sector, the higher-dimensional operators primarily induce modifications of the SM gauge currents 
\begin{subequations}
    \begin{align}
    J_{Z}^{\mu} &= [Z_{e_{L}}]_{ij} \,\bar{e}_{L}^{i}\gamma^{\mu}e^{j}_{L}+ [Z_{e_{R}}]_{ij} \, \bar{e}^{i}_{R}\gamma^{\mu}e^{j}_{R} + [Z_{\nu}]_{ij} \bar{\nu}_{L}^{i} \gamma^{\mu}\nu_{L}^{j}\,, \\
    J_{W}^{\mu} &= [W_{l}]_{ij} \,\bar{\nu}_{L}^{i}\gamma^{\mu}e^{j}_{L}\,,
\end{align}
\end{subequations}
where 
\begin{subequations}
    \begin{align}
    [Z_{e_{L}}]_{ij} & = \delta_{ij} \left(-\frac{1}{2} + s_W^{2}\right) - \frac{v^{2}}{2} (C_{H\ell}^{(1)})_{ij} -\frac{v^{2}}{2} (C_{H\ell}^{(3)})_{ij}\,,\\
    [Z_{e_{R}}]_{ij} & = s_W^{2} \delta_{ij} -\frac{v^{2}}{2} (C_{He})_{ij} \,,\\
    [Z_{\nu_{L}}]_{ij} & =  \frac{1}{2} \delta_{ij} - \frac{v^{2}}{2} (C_{H\ell}^{(1)})_{ij} +\frac{v^{2}}{2} (C_{H\ell}^{(3)})_{ij}\,,\\
    [W_{l}]_{ij} & = \delta_{ij} + v^{2} (C_{H\ell}^{(3)})_{ij}\,.
\end{align}
    \label{eq:Z_LR}
\end{subequations}
Similarly, both the mass terms and the Yukawa interactions receive corrections such that, whenever the Yukawa matrix and the WC $C_{eH}$ are misaligned in flavor space, the Higgs couplings to leptons are no longer proportional to the corresponding fermion masses. As a consequence, both flavor-violating and CP-violating Higgs interactions can arise. However, the additional rephasing and diagonalization of the mass matrix induced by the misaligned contributions 
$\Delta C^{\rm LO}_{eH}$ and $C_{eH}^{\mathrm{NLO}}$ generate effects that effectively correspond to dimension-eight corrections in the dimension-six WCs and can therefore be consistently neglected within our working accuracy. Observable effects may instead arise from the modification of the dimension-four Yukawa interactions at the matching scale, leading to
\begin{equation}
    \mathcal{L}_{Y} \supset -[Y_{H}]_{ij} h\, \bar{e}_{L} e_{R} \mathrm{ \, + \, h.c.}\,, \quad \quad [Y_{H}]_{ij} \simeq \frac{\sqrt{2}\,m_{i}}{v} \, \delta^{ij} - \frac{v^{2}}{\sqrt{2}} C_{eH}^{ij}\,.
\end{equation}

In addition to inducing observable effects in LFV and LFUV decays of the Higgs, $W$, and $Z$ bosons, the modified structure of the SM couplings also generates corrections to low-energy four-fermion interactions once the electroweak gauge bosons are integrated out. These effects lead to deviations from the standard low-energy SM predictions.
Such contributions combine with the four-fermion operators discussed in Sec.~\ref{SMEFT_tree}, giving rise to the following effective Lagrangian terms
\begin{equation}
    \begin{split}
        \mathcal{L} &\,\supset (C^{V,XY}_{ee})_{prst}\, (\bar{e}^{p}\gamma_{\mu}P_{X}\,e^{r})\, (\bar{e}^{s}\gamma_{\mu} P_{Y}\,e^{t}) + 
     (C^{S,XY}_{ee})_{prst}\, (\bar{e}^{p} P_{X}\,e^{r})\, (\bar{e}^{s}P_{Y}\,e^{t}) \\
     & \quad + (C^{V,LX}_{\nu e})_{prst}\, (\bar{\nu}^{p}\gamma_{\mu}P_{L}\,\nu^{r})\, (\bar{e}^{s}\gamma_{\mu} P_{X}\,e^{t}) + (C^{V,XY}_{ eu})_{pr}) (\bar{e}^{p}\gamma_{\mu}P_{X}\,e^{r})\, (\bar{u}\gamma_{\mu} P_{Y}\,u) \\
     & \quad + (C^{V,XY}_{ ed})_{pr}) (\bar{e}^{p}\gamma_{\mu}P_{X}\,e^{r})\, (\bar{d}\gamma_{\mu} P_{Y}\,d)\,,
    \end{split}
\end{equation}
where $u$ and $d$ denote the up- and down-type quark fields, respectively. The corresponding LEFT WCs are summarized in Tab.~\ref{table:SMEFT_LEFT_matching}. 
Compared to the interactions mediated by gauge bosons, scalar four-fermion operators are subject to an additional suppression proportional to $Y_H^2 \sim (m/v)^2$, where $m$ denotes the mass of a light SM fermion. Consequently, these contributions are numerically subleading and will be neglected in the following analysis.
\begin{table}[h!]
\centering
\renewcommand{\arraystretch}{1.25}
\setlength{\tabcolsep}{8pt}
\begin{tabular}{@{}c c@{}}
\toprule
\textbf{LEFT Coefficient} & \textbf{Matching} \\
\midrule
$(C^{V,LL}_{ee})_{prst}$ & $(C_{ll})_{prst} - \frac{g_{Z}^{2}}{4M_{Z}^{2}} [Z_{e_{L}}]_{pr}[Z_{e_{L}}]_{st} - \frac{g_{Z}^{2}}{4M_{Z}^{2}} [Z_{e_{L}}]_{pt}[Z_{e_{L}}]_{sr}$  \\
$(C^{V,RR}_{ee})_{prst}$ &$(C_{ee})_{prst} - \frac{g_{Z}^{2}}{4M_{Z}^{2}} [Z_{e_{R}}]_{pr}[Z_{e_{R}}]_{st} - \frac{g_{Z}^{2}}{4M_{Z}^{2}} [Z_{e_{R}}]_{pt}[Z_{e_{R}}]_{sr}$ \\
$(C^{V,LR}_{ee})_{prst}$ & $(C_{le})_{prst} - \frac{g_{Z}^{2}}{M_{Z}^{2}} [Z_{e_{L}}]_{pr}[Z_{e_{R}}]_{st}$ \\
$(C^{V,LL}_{\nu e})_{prst}$ & $(C_{ll})_{prst} + (C_{ll})_{stpr} - \frac{g_{2}^{2}}{2M_{W}^{2}} [W_{{l}}]_{pt}[W_{l}]^{*}_{rs} - \frac{g_{Z}^{2}}{M_{Z}^{2}} [Z_{\nu_{L}}]_{pr}[Z_{e_{L}}]_{st}$ \\
$(C^{V,LR}_{\nu e})_{prst}$ & $(C_{le})_{prst}  - \frac{g_{Z}^{2}}{M_{Z}^{2}} [Z_{\nu_{L}}]_{pr}[Z_{e_{R}}]_{st}$ \\[0.1cm]
\hline
\addlinespace
$(C^{S,RR}_{ee})_{prst}$ & $\frac{1}{M_{h}^{2}}[Y_{H}]_{pr}[Y_{H}]_{st}$ \\
$(C^{S,RL}_{ee})_{prst}$ & $\frac{1}{2M_{h}^{2}}[Y_{H}]_{pr}[Y_{H}]^{*}_{ts}$ \\[0.1cm]
\hline
\addlinespace
$(C^{V,LR}_{eu(d)})_{pr}$ & $(C_{lu(d)})_{pr11} - \frac{g_{Z}^{2}}{M_{Z}^{2}} [Z_{e_{L}}]_{pr} \, \rho_{u(d)}^{R}$ \\
$(C^{V,RL}_{eu(d)})_{pr}$ & $(C_{qe})_{11pr} - \frac{g_{Z}^{2}}{M_{Z}^{2}} [Z_{e_{R}}]_{pr} \, \rho_{u(d)}^{L}$ \\
$(C^{V,RR}_{eu(d)})_{pr}$ & $(C_{eu(d)})_{pr11} - \frac{g_{Z}^{2}}{M_{Z}^{2}} [Z_{e_{R}}]_{pr} \, \rho_{u(d)}^{R}$\\
$(C^{V,LL}_{eu(d)})_{pr}$ & $(C_{lq}^{(1)})_{pr11} - \frac{g_{Z}^{2}}{M_{Z}^{2}} [Z_{e_{L}}]_{pr} \, \rho_{u(d)}^{L}$\\

\bottomrule
\end{tabular}
\caption{Tree-level matching between LEFT four-fermion operators and SMEFT operators. Here $g_Z = g_2/c_W = e/(c_W s_W)$, with $s_W$ and $c_W$ denoting the sine and cosine of the Weinberg angle, respectively. The quantities $\rho_{u,d}^{L/R}$ parametrize the SM couplings of up- and down-type quarks to the $Z$ boson and are given by
$\rho_{u}^{L} = 1/2 - (2/3)s_{W}^{2}$, $\rho_{u}^{R} =-(2/3)s_{W}^{2}$, $\rho_{d}^{L} = -1/2 + (1/3)s_{W}^{2}$, $\rho_{d}^{R} = (1/3)s_{W}^{2}$.}
\label{table:SMEFT_LEFT_matching}
\end{table}

Finally, the dimension-six SMEFT dipole operators are matched onto the corresponding dimension-five operators in the LEFT according to
\begin{equation}
    \mathcal{L}\supset \frac{e}{8\pi^2}\, (C_{D})_{ij} \,\bar{e}^{i}_L\, \sigma_{\mu \nu} \,e^j_R\, F^{\mu \nu} + \mathrm{h.c.}\,,
\end{equation}
where 
\begin{equation}
    (C_{D})_{ij} = \frac{8 \pi^2}{e} \frac{v}{\sqrt{2}} \Big[c_{W} (C_{eB})_{ij} - s_{W}(C_{eW})_{ij}\Big]\,,
    \label{eq:dipole_LEFT}
\end{equation}
being $s_{W}$ and $c_{W}$ the sine and cosine of the Weinberg angle.

\section{Phenomenology of flavor deconstruction in the lepton sector}
\label{sec: pheno FD lepton sector}

In the following section, we analyze the phenomenology of the minimal flavor deconstruction model \cite{Barbieri:2023qpf,Barbieri:2024zkh} in the leptonic sector, assessing the sensitivity of present and future experiments to the underlying NP scale. 
Among the various observables, we focus on lepton flavor violating (LFV), lepton flavor universality violating (LFUV), and CP-violating observables, motivated by the remarkable current experimental sensitivities and the significant improvements expected in future facilities.
Tab.~\ref{table:pheno} summarizes the current bounds and measurements, together with the projected sensitivities, for the observables relevant to our analysis.

The main goal of this section is to investigate the interplay between LFV observables, such as $\mu \rightarrow e\gamma$, $\mu \rightarrow 3e$, and $\mu-e$ conversion in nuclei, and CP-violating observables, with particular emphasis on the electron EDM. In particular, we will show that, in the presence of sizable CP phases and natural couplings, the electron EDM provides a highly competitive probe of the framework.

To this end, most of the UV parameters are assumed to be $\mathcal{O}(1)$ in the numerical estimates presented below. The only couplings allowed to vary over a broader range are the quartic scalar couplings $\lambda_{ij}$ appearing in the scalar potential (see Eq.~\eqref{eq: scalar lagrangian broken symmetry} and the related discussion), the physical CP phase $\sin\varphi_{\rm CP}$ (see Appendix~\ref{sec:$CP$ violation phase}), and the charges of the SM leptons $Q_V^{e,\ell}$ under the heavy gauge bosons $V = Z_{23}, Z'_{23}$ (see Eq.~\eqref{eq:universality between first 2 generations} and the surrounding discussion). 
Also, since $M_V \sim \langle\chi\rangle,\langle\phi\rangle$, and using Eq.~\eqref{eq:varepsilon definition}, we assume $M_V/M_{E_\alpha} \sim \varepsilon_\chi$. Finally, we adopt $\theta_3^L \sim \lambda \approx 0.2$ as a benchmark value, motivated by the analogy with the quark sector, where such a relation is expected to hold.
Imposing that the model correctly reproduces the observed fermion masses and mixing pattern, we normalize the expansion parameters $\varepsilon_i$ as \cite{Barbieri:2023qpf,Barbieri:2024zkh}
\begin{equation}\label{eq:espilon}
    \varepsilon_\chi=0.04=|V_{cb}|\,,\qquad \varepsilon_\phi=0.06=\frac{m_\mu}{m_\tau}\,,\qquad \varepsilon_\sigma=0.005=\frac{m_e}{m_\mu}\,.
\end{equation}
In particular, while the values of $\varepsilon_{\phi,\sigma}$ are directly fixed by the Yukawa matrix in Eq.~\eqref{eq:Yukawa electron after diag}, the parameter $\varepsilon_\chi$ does not directly control the charged-lepton mass spectrum. Nevertheless, since $\varepsilon_\chi$ governs the mixing between the third lepton generation and the lighter ones, an upper bound can be inferred from SMEFT analyses of scenarios with an approximate $\mathrm{U}(2)^5$ flavor symmetry, leading to the estimate reported in Eq.~\eqref{eq:espilon} for NP scales around the TeV scale~\cite{Covone:2025lee}.

In the following, we present the predictions for the observables listed in Tab.~\ref{table:pheno}.

\begin{table}[p]
\centering
\renewcommand{\arraystretch}{1.25}
\setlength{\tabcolsep}{8pt}
\begin{tabular}{@{}c c c c@{}}
\toprule
\textbf{Class} & \textbf{Observables} & \textbf{Present bound} & \textbf{Future Sensitivities} \\
\midrule

\multirow{3}{*}{LFV Z decays}
 & $\mathrm{Br}(Z \rightarrow e \mu)$ 
 & $< 4.2 \times 10^{-7}$ \cite{ATLAS:2021bdj}
 & $\mathcal{O}(10^{-9})$ \cite{Altmann:2025feg}\\
 & $\mathrm{Br}(Z \rightarrow e \tau)$  
 & $< 4.1 \times 10^{-6}$ \cite{ATLAS:2021bdj}
 & $\mathcal{O}(10^{-8}\div10^{-10})$ \cite{Altmann:2025feg}\\
  & $\mathrm{Br}(Z \rightarrow \mu \tau)$ 
 & $< 5.3 \times 10^{-6}$ \cite{ATLAS:2021bdj}
 & $\mathcal{O}(10^{-9})$ \cite{Altmann:2025feg}\\[0.1cm]
\hline
\addlinespace
\multirow{3}{*}{LFV H decays}
 & $\mathrm{Br}(h \rightarrow e \mu)$  
 & $< 6.1 \times 10^{-5}$ \cite{ATLAS:2019xlq}
 & $\mathcal{O}(10^{-5})$ \cite{Altmann:2025feg}\\
 & $\mathrm{Br}(h\rightarrow e \tau)$  
 & $< 2.2 \times 10^{-3}$ \cite{CMS:2017onh}
 & $\mathcal{O}(10^{-4})$ \cite{Altmann:2025feg}\\
 & $\mathrm{Br}(h \rightarrow \mu \tau)$ 
 & $< 1.5 \times 10^{-3}$ \cite{CMS:2017onh}
 & $\mathcal{O}(10^{-4})$ \cite{Altmann:2025feg}\\[0.1cm]
\hline
\addlinespace
\multirow{6}{*}{LFV lepton decays}
 & $\mathrm{Br}(\mu \rightarrow e \gamma)$ 
 & $< 1.5 \times 10^{-13}$ \cite{MEGII:2025gzr}
 & $6 \times 10^{-14}$ \cite{MEGII:2025gzr}\\
 & $\mathrm{Br}(\tau \rightarrow e \gamma)$
 & $< 3.3 \times 10^{-8}$ \cite{BaBar:2009hkt}
 & $3 \times 10^{-9}$ \cite{Belle-II:2018jsg} \\
 &  $\mathrm{Br}(\tau \rightarrow \mu\gamma)$ 
 &  $< 4.2 \times 10^{-8}$ \cite{Belle:2025bpu}
 & $1 \times 10^{-9}$ \cite{Belle-II:2018jsg} \\
 &  $\mathrm{Br}(\mu \rightarrow 3e)$ 
 &  $< 1.0 \times 10^{-12}$ \cite{SINDRUM:1987nra}
 & $\mathcal{O}(10^{-16})$ \cite{Blondel:2013ia} \\
 &  $\mathrm{Br}(\tau \rightarrow 3e)$ 
 &  $< 2.7 \times 10^{-8}$ \cite{Hayasaka:2010np}
 & $5.0 \times 10^{-10}$  \cite{Belle-II:2018jsg} \\
 &  $\mathrm{Br}(\tau \rightarrow 3\mu)$ 
 &  $< 3.3 \times 10^{-8}$ \cite{Hayasaka:2010np}
 & $4 \times 10^{-10}$ \cite{Belle-II:2018jsg} \\[0.1cm]
\hline
\addlinespace
\multirow{3}{*}{$\mu- e$ in nuclei}
 &  $\mathrm{Br}(\mu -e,\mathrm{Au})$ 
 &  $< 7.0 \times 10^{-13}$ \cite{SINDRUMII:2006dvw}
 & - \\
 &  $\mathrm{Br}(\mu-e,\mathrm{Ti})$ 
 &  $< 4.2 \times 10^{-12}$ \cite{SINDRUMII:1993gxf}
 & - \\
 &  $\mathrm{Br}(\mu-e,\mathrm{Al})$ 
 & -
 & $10^{-16} \div 10^{-17}$ \cite{COMET,Mu2e:2014fns}\\[0.1cm]
\hline
\addlinespace
\multirow{6}{*}{AMMs and EDMs}
 & $\Delta a_{e}$ 
 & $< 8 \times 10^{-13}$\cite{AMMe2020} 
 & - \\
 & $\Delta a_{\mu}$ 
 & $ <1.60 \times 10^{-9}$ \cite{Muong-2:2025xyk}
 & -\\ 
 & $\vert d_{e}/(eQ)\vert $ 
 & $ < 4.1 \times 10^{-30}$ cm \cite{Roussy:2022cmp}
 &  
 \begin{tabular}[c]{@{}l@{}}
        $ 0.3 \, \times\, 10^{-30} $ cm \cite{Hiramoto:2022fyg} \  \\
        $10^{-31} \div 10^{-32} $ cm  \cite{Fitch:2020jil}
        \end{tabular}\\
 & $\vert d_{\mu}/(eQ)\vert $ 
 & $ < 1.8 \times 10^{-19}$ cm \cite{Muong-2:2008ebm} 
 & $ 6 \times 10^{-23}$ cm \cite{Adelmann:2025nev} \\
 &$\vert d_{\tau}/(eQ)\vert $
 & $< 10^{-18}$ cm \cite{Belle:2021ybo}  
 & $ \mathcal{O}(10^{-19})$ cm \cite{aggarwal2022snowmasswhitepaperbelle} \\[0.1cm]
 \hline
\addlinespace
 \multirow{6}{*}{LFUV}
 & $g^\mu_A/g^e_A$
 & $1.0002(13)$ \cite{ParticleDataGroup:2024cfk} & $ \sim\,1.5\times10^{-4}$ \cite{FCC:2025lpp,Belloni:2022due}\\
 & $g^\tau_A/g^e_A$
 & $1.0019(15)$ \cite{ParticleDataGroup:2024cfk} & $\sim \,1.6\times10^{-4}$ \cite{FCC:2025lpp,Belloni:2022due}\\
 & $g^\mu_V/g^e_V$
 & $0.962(63)$ \cite{ParticleDataGroup:2024cfk} & $ \sim\,3\times10^{-4}$ \cite{FCC:2025lpp,Belloni:2022due}\\
 & $g^\tau_V/g^e_V$
 & $0.958(29)$ \cite{ParticleDataGroup:2024cfk} & $ \sim \,1.4\times10^{-4}$ \cite{FCC:2025lpp,Belloni:2022due}\\
 & $(g_{\tau}/g_\mu)_\tau$ 
 & $1.0009(14)$ \cite{HFLAV:2022esi}& -\\
 & $(g_{\tau}/g_e)_\tau$ 
 & $1.0027(14)$ \cite{HFLAV:2022esi}& -\\
\bottomrule
\end{tabular}
\caption{Current and future projections for relevant observables in the leptonic sector. $\Delta a_{i}$ has been estimated by subtracting the SM theoretical prediction \cite{Aliberti:2025beg} from the experimental determination and allowing for a 2 $\sigma$ uncertainty. Regarding the future projections for the $Z$-boson coupling ratios, we report only the expected experimental sensitivities, following Refs.~\cite{FCC:2025lpp,Belloni:2022due}.
}
\label{table:pheno}
\end{table}

\subsection{Lepton flavor violation} \label{sec: LFV}

LFV is primarily controlled by the flavor structure of the model in the lepton sector. In general, this involves a large number of parameters encoded in the left-handed ($P_V^\ell$) and right-handed ($P_V^e$) mixing matrices, as well as in the Yukawa couplings $Y^{e}_{i\alpha}$ associated with the light generations. Restricting to tree-level vector-vector four-fermion interactions, LFV effects can arise both from processes mediated by heavy $Z'$ bosons, encoded in the WCs discussed in Sec.~\ref{Thirdclass}, and from modifications of the SM $Z$-boson couplings. The latter originate, in the EFT description, from operators of the $\psi^2 H^2 D$ class introduced in Sec.~\ref{secondclacss}, whose low-energy structure is discussed in Sec.~\ref{section:LEFT}.

Although the precise predictions depend on the assumptions made on the underlying parameters of the model, several general features emerge rather robustly. First, flavor-violating processes involving the first two generations are subject to much stronger experimental constraints than those involving $\tau$ leptons, especially in view of future experimental sensitivities. In particular, the strong suppression of the right-handed flavor mixing, encoded in the flavor entry $(P_V^e)_{12}$ in Eq.~\eqref{eq:Pe12}, implies that processes such as $\mu \rightarrow 3e$ and $\mu-e$ conversion in nuclei are dominantly mediated by left-handed currents.\footnote{This feature is a direct consequence of the flavor-deconstructed framework under consideration, where $\mathrm{SU}(2)_L$ remains universal. As a result, left-handed SM leptons can exhibit sizable mixing angles, whereas right-handed leptons experience a much stronger suppression due to the deconstruction of the $\mathrm{U}(1)_Y$ gauge symmetry.} Consequently, experiments employing polarized muons could provide a useful handle to test the chiral structure of the model, especially in scenarios where the final-state electrons are predominantly right-handed~\cite{Bolton:2022lrg,Kuno:1999jp}.

As a first class of observables, we consider LFV $Z$ decays, characterized by the branching ratios
\begin{equation}
    \mathrm{Br}(Z\rightarrow\ell_i\overline{\ell}_j) \approx
    \frac{M_Z}{6\pi\,\Gamma_Z }
    \frac{e^{2}}{c_{W}^{2}s_{W}^{2}}
    \vert [Z_{eL}]_{ij}\vert^2\,,
\end{equation}
where we neglected the loop-suppressed dipole contributions as well as 
right-handed vector currents proportional to $[Z_{eR}]_{ij}$ since they 
are subdominant in our framework. Exploiting Eqs.~\eqref{eq:Pe and Pell def}, \eqref{eq:C_He and C_Hell}, and \eqref{eq:Z_LR}, the coefficients $[Z_{e_{L}}]_{ij}$ can be written as
\begin{align}
        [Z_{e_{L}}]_{i\neq j} &= \sum_{V=Z_{23},Z_{23}^\prime}
        \frac{v^{2}}{2M_{V}^2}g_{V}^2Q_{V}^{H}(P^\ell_V)_{ij}+\frac{v^{2}}{4M^2_{E_\alpha}}
        Y^e_{i\alpha}(Y^e_{j\alpha})^*\,,
\end{align}
where, according to Appendix~\ref{app:Flavor structure relations of fermionic bilinears}, the quantities $(P_V^{\ell})_{ij}$ are well approximated by
\begin{align}
    (P_V^{\ell})_{12} \sim (Q_{V}^{\ell_{3}}-Q_{V}^{\ell_2}) \varepsilon_{\chi}^{2}\,, \qquad\qquad
    (P_V^{\ell})_{13} \approx (P_V^{\ell})_{23} \sim (Q_{V}^{\ell_{3}}-Q_{V}^{\ell_2}) \varepsilon_{\chi}\,,
\end{align}
in the limit where $\theta_3^L$ is sufficiently small (e.g. $\theta_3^L\sim\lambda\sim0.2$). As a result, the following estimates are obtained
\begin{subequations}
    \begin{align}
    &\frac{\mathrm{Br}(Z\to \tau \mu(e))}{5\cdot 10^{-6}}\approx g_V^{4} Q_{H}^{2}\left( Q^{\ell_3}_{V}-Q^{\ell_2}_{V}\right)^2\left(\frac{\varepsilon_\chi}{0.04}\right)^2\left(\frac{\mathrm{0.7 \,TeV}}{M_V}\right)^4\,,\\
    &\frac{\mathrm{Br}(Z\to \mu e)}{4\cdot 10^{-7}}\approx \left(Y^{e}_{1\alpha} (Y^{e}_{2\alpha})^{*} + 2\, g_{V}^{2}Q_{H}^{2} (Q^{\ell_3}_{V}-Q^{\ell_2}_{V})\right)^4 \left(\frac{\varepsilon_\chi}{0.04}\right)^4\left(\frac{0.2\,\mathrm{TeV}}{M_V}\right)^4\,.
\end{align}
\end{subequations}

Purely leptonic decay processes, such as $\ell_j \rightarrow 3\ell_i$, represent highly sensitive probes of LFV. In our framework, these observables are dominantly mediated by four-fermion operators, and the corresponding branching ratios are well approximated by
\begin{equation}
    \begin{split}
    \frac{\mathrm{Br}(\ell_{j} \rightarrow 3\ell_{i})}{\mathrm{Br}(\ell_{j} \rightarrow \ell_{i} \nu\bar\nu)}
         \approx \frac{1}{G_{F}^2}
         \Bigg(\vert (C^{V,LL}_{ee})^{ijii}\vert^{2} + \vert C^{V,LR}_{ee})^{ijii}\vert^{2}\Bigg) \,,
    \end{split}\label{eq:lto3l}
\end{equation}
where
\begin{subequations}
    \begin{align}
(C^{V,LL}_{ee})_{ijii} &=  
-\frac{1}{2}\sum_{V=Z_{23},Z_{23}^\prime}\frac{g_{V}^2}{M_{V}^2}(P^\ell_V)_{ij}Q^{\ell_i}_V
- \frac{g_{Z}^{2}}{2M_{Z}^{2}} [Z_{e_{L}}]_{ij}
\left(-\frac{1}{2}+s^2_W\right) \,,\\
(C^{V,LR}_{ee})_{ijii} &=  
-\sum_{V=Z_{23},Z_{23}^\prime}\frac{g_{V}^2}{M_{V}^2}(P^\ell_V)_{ij}Q^{e_i}_V
- \frac{g_{Z}^{2}}{2M_{Z}^{2}} [Z_{e_{L}}]_{ij}s_{W}^{2}\,.
\end{align}
\end{subequations}
This, in turn, leads to the following predictions:
\begin{subequations}
    \begin{align}
    &\frac{\mathrm{Br}(\mu\to 3e)}{1\cdot 10^{-12}} \approx  \left( (Y^{e}_{2\alpha} Y^{e}_{1\alpha} )^2 + 2.4\,g_VY^{e}_{2\alpha} Y^{e}_{1\alpha}(Q_{V}^{\ell_{2}}-Q_{V}^{\ell_{3}})+ 17\, g_V^{2}(Q_{V}^{\ell_{3}}-Q_{V}^{\ell_{2}})^2\right) \left(\frac{\varepsilon_\chi}{0.04}\right)^4\left(\frac{7\,\mathrm{TeV}}{M_V}\right)^4 \\
    &\frac{\mathrm{Br}(\tau\to 3e(\mu))}{4\cdot 10^{-8}} \approx  g_{V}^{4}\left( 0.1 + (Q_{V}^{\ell_{2}}-Q_{V}^{\ell_{3}})\right)^{2}\left(\frac{\varepsilon_\chi}{0.04}\right)^4\left(\frac{2\,\mathrm{TeV}}{M_V}\right)^4\,,
\end{align}
\end{subequations}
where we assumed the benchmark value \(Q_{V}^{\psi} = Q_{Y}^{\psi}\) for all SM fields $\psi$ not explicitly shown, with $Q_{Y}$ denoting the SM hypercharge.

$\mu-e$ conversion in nuclei depends on the UV parameters of the lepton flavor sector, as well as on the couplings of light quarks to the heavy vector bosons, rendering the corresponding expressions rather cumbersome. Moreover, one must specify the nuclear form factors, which depend on the experimental target under consideration. 
In the following, we assign to the quark charges $Q_V^{q,u,d}$ the same values as the corresponding SM hypercharges, as done for the lepton couplings above. 
In our numerical analysis, we consider the experimental bound on $\mu-e$ conversion in Gold, which currently provides the most stringent constraint, see Tab.~\ref{table:pheno}.
The corresponding branching ratio in our framework can be expressed as
\begin{equation}
    \mathrm{Br}(\mu -e,\mathrm{N}) \approx \frac{B_{N}}{2 G_{F}^{2}}\vert  C_{N,L}\vert^{2}\,,
\end{equation}
where the coefficient $C_{N,L}$ depends on the semileptonic WCs as follows
\begin{equation}
    \begin{split}
        C_{N,L} =& F_{V}^{u}\,(C^{V,LL}_{eu} +C^{V,LR}_{eu})_{e\mu uu} +  F_{V}^{d}\,(C^{V,LL}_{ed} + C^{V,LR}_{ed})_{e\mu dd}\,,
    \end{split}
\end{equation}
where 
\begin{subequations}
    \begin{align}
    (C^{V,LL}_{eu(d)})_{ij11} &=  
-\sum_{V=Z_{23},Z_{23}^\prime}\frac{g_{V}^2}{M_{V}^2}(P^\ell_V)_{ij}Q^{q}_V
- \frac{g_{Z}^{2}}{M_{Z}^{2}} [Z_{e_{L}}]_{ij} \rho_{u(d)}^{L}\,, \\
(C^{V,LR}_{eu(d)})_{ij11} &=  
-\sum_{V=Z_{23},Z_{23}^\prime}\frac{g_{V}^2}{M_{V}^2}(P^\ell_V)_{ij}Q^{u(d)}_V
- \frac{g_{Z}^{2}}{M_{Z}^{2}} [Z_{e_{L}}]_{ij} \rho_{u(d)}^{R}\,,
\end{align}
\end{subequations}
and where $\rho_{u(d)}^{L/R}$ parametrizes the SM interaction between the $Z$ boson and up- and down-type quarks, as defined in Tab.~\ref{table:SMEFT_LEFT_matching}. The nucleus-dependent coefficients $F_X^i$ and $B_N$ are given in Appendix~A of Ref.~\cite{Ardu:2024bua}. As a result, we find the following estimate
\begin{equation}
    \frac{\mathrm{Br}(\mu-e,\mathrm{N})}{7\cdot 10^{-13}}\approx \left(Y^{e}_{2\alpha} Y^{e}_{1\alpha}  + 4.3\, g_{V}^{2}(Q_{V}^{\ell_2}-Q_{V}^{\ell_3})\right)^2\left(\frac{\varepsilon_\chi}{0.04}\right)^4\left(\frac{15\,\mathrm{TeV}}{M_V}\right)^4\,.
\end{equation}

Finally, we discuss the radiative LFV processes $\ell_j \to \ell_i \gamma$ which arise at the one-loop level. Unlike tree-level observables, the corresponding branching ratios depend not only on the charges and Yukawa parameters, but also on the full set of loop contributions. In the flavor basis, these correspond to the different possible ways of closing the loop diagrams shown in Figs.~\ref{fig:mssLO} and~\ref{fig:massNLO} (see Refs.~\cite{Calibbi:2008qt,Calibbi:2020emz,Lopez-Ibanez:2021yzu} for a discussion of this effect in the context of other flavor symmetries). At the level of expansion parameters, the structure of the dipole matrices is reported in Eqs.~\eqref{eq:dipolematrixdelta} and~\eqref{eq:dipolematrixNLO}. The relevant branching fraction reads
\begin{equation}\label{BRLFV}
    \frac{Br(\ell_{j}\, \rightarrow \ell_{i} \,\gamma)}{\mathrm{Br}(\ell_{j} \rightarrow \ell_{i} \nu\bar\nu)} = \frac{3 \,\alpha}{\pi G_{F}^{2} m_{j}^{2}} \Big( \vert (C_{D})_{ij} \vert^2 + \vert (C_{D})_{ji} \vert^2\Big)\,,
\end{equation}
 where
 \begin{equation}
    (C_{D})_{ij} = \frac{8 \pi^2}{e} \frac{v}{\sqrt{2}} \Big[c_{W} (C_{eB})_{ij} - s_{W}(C_{eW})_{ij}\Big]\,,
\end{equation}
and the dependence of the dipole matrices $(C_{eB})_{ij}$ and $(C_{eW})_{ij}$ on the expansion parameters $\varepsilon_i$ was already given in Eqs.~\eqref{eq:dipolematrixdelta} and~\eqref{eq:dipolematrixNLO}.
As a result, the general structure of ${\rm Br}(\ell_{j} \rightarrow \ell_{i} \gamma)$ can be written as
\begin{subequations}
    \begin{align}
    &\frac{\mathrm{Br}(\mu\to e\gamma)}{1.5\cdot 10^{-13}}\approx \left[g_V^2 F_V\left(\frac{3\,\mathrm{TeV}}{M_V}\right)^2+\lambda_{ij}
    F_\phi
    \left(\frac{3\,\mathrm{TeV}}{M_\phi}\right)^2 \right]^2\left(\frac{\varepsilon_\phi}{0.06}\right)^2\left(\frac{\varepsilon_\chi}{0.04}\right)^4\,,\\
    &\frac{\mathrm{Br}(\tau\to \mu(e)\gamma)}{4.2\cdot 10^{-8}}\approx \left[g_V^2 F_V\left(\frac{0.5\,\mathrm{TeV}}{M_V}\right)^2+\lambda_{\phi H}F_\phi\left(\frac{0.5\,\mathrm{TeV}}{M_\phi}\right)^2\right]^2\left(\frac{\varepsilon_\chi}{0.04}\right)^2\,,
\end{align}
\end{subequations}
where $F_{V}$ and $F_{\phi}$ denote generic loop functions. 
According to the dipole matrix structure in Eqs.~\eqref{eq:dipolematrixdelta} and~\eqref{eq:dipolematrixNLO}, the process $\mu \to e \gamma$ receives comparable contributions from $\Delta C^{\rm LO}_{eX}$ and $C_{eX}^{\mathrm{NLO}}$, where the former originates from the off-diagonal entries of the dipole matrix at LO, while the latter corresponds to the genuine NLO contribution. By contrast, the $\tau \to \ell_i \gamma$ processes are dominated by the LO off-diagonal dipole entries.

Additional LFV effects arise from modified Higgs--lepton Yukawa couplings induced by the misalignment between the dimension-four Yukawa interaction and \(C_{eH}\). Nevertheless, the experimental bounds reported in Tab.~\ref{table:pheno} imply a negligible impact on $h \to \ell_i \ell_j$ for boson masses at the TeV scale.

\subsection{Lepton flavor universality violation} \label{sec:LFUV}
The search for lepton flavor universality violation (LFUV) represents one of the most powerful probes of NP phenomena, 
since the SM predicts only negligible LFUV effects.

LEP and SLD measurements provide stringent constraints on LFUV in the effective couplings of the $Z$ boson to charged leptons. Adopting the standard parametrization
\begin{equation}
\mathcal{L}_Z =
\frac{g}{2 c_W}\,
Z_\mu \,
\bar e^i \gamma^\mu
\left(g_V^{\ell_i}-g_A^{\ell_i}\gamma_5\right)e^i \, ,
\end{equation}
the current global electroweak fits reported by the PDG~\cite{ParticleDataGroup:2024cfk}, based in particular on the $Z$ decay widths,
left-right and forward-backward asymmetries, confirm the universality of the axial couplings at the per-mille level, while weaker bounds are obtained for the vector couplings due to their accidental suppression in the SM. In particular, the experimental determinations are reported in Tab. \ref{table:pheno} and show that sizeable LFUV effects in $Z$-pole observables are strongly constrained.

According to Eq.~\eqref{eq:Z_LR}, $g_V^i$ and $g_A^i$ can be written as
\begin{equation}\label{eq: def of g_V,g_A}
    g_V^i=[Z_{e_{L}}]_{ii}+[Z_{e_{R}}]_{ii}\,,\qquad g_A^i=[Z_{e_{L}}]_{ii}-[Z_{e_{R}}]_{ii}\,.
\end{equation}
In our setup, the flavor symmetry enforces an $\varepsilon^2$ suppression of LFUV effects between the first two generations. By contrast, sizeable LFUV effects involving the third generation are generically expected.
To quantify these effects, we calculate the quantities
$g_V^\tau/g_V^e$ and $g_A^\tau/g_A^e$ in our setup
\begin{subequations}
    \begin{align}
\frac{g_V^\tau}{g_V^e} &= 1 + \frac{g_{V}^{2}}{4 s_{W}^{2}-1} \frac{v^2}{M_{V}^{2}} Q_{V}^{H} \left[ (Q_V^{\ell_{3}}-Q_V^{\ell_{2}} )+ (Q_V^{e_{3}}-Q_V^{e_{2}})\right],
\\
\frac{g_A^\tau}{g_A^e} &=1 + {g_{V}^{2}} \frac{v^2}{M_{V}^{2}} Q_{V}^{H} \left[ (Q_V^{\ell_{3}}-Q_V^{\ell_{2}} ) - (Q_V^{e_{3}}-Q_V^{e_{2}})\right],
\end{align}
\end{subequations}
which yields the numerical estimates
\begin{subequations}
    \begin{align}
\frac{g_V^\tau}{g_V^e} &= 1 + 2.9\cdot10^{-2}g_{V}^{2} Q_{V}^{H}\left[ (Q_V^{\ell_{3}}-Q_V^{\ell_{2}} )+ (Q_V^{e_{3}}-Q_V^{e_{2}})\right]\left(\frac{4 \, \mathrm{TeV}}{M_{V}}\right)^{2},
\\
\frac{g_A^\tau}{g_A^e} &=1 + 1.5\cdot10^{-3}g_{V}^{2} Q_{V}^{H}\left[ (Q_V^{\ell_{3}}-Q_V^{\ell_{2}} )- (Q_V^{e_{3}}-Q_V^{e_{2}})\right]\left(\frac{6\, \mathrm{TeV}}{M_{V}}\right)^{2}.
\end{align}
\end{subequations}
Furthermore, the first phase of the future FCC-ee collider at CERN is planned to operate at the $Z$-pole, with the goal of probing LFU at the $10^{-4}$ level~\cite{FCC:2025lpp}. Such a precision would allow sensitivity to NP scales of several tens of TeV. 

Moreover, also low-energy measurements at $B$ factories reached a precision on LFUV effects at the per-mille level by comparing different $\tau$-decay rates with those of muons or mesons. Focusing on purely leptonic decays, these observables are governed by modified $W$-boson couplings, as well as by flavor off-diagonal interactions of the $Z$ boson, leading to contributions suppressed by $\varepsilon_\chi^2$. Such effects are subdominant with respect to those arising from LFUV observables at the $Z$-pole.

Modifications of the $Z$-boson couplings to neutrinos affect the extraction of the number of neutrino species, $N_\nu$, from the invisible $Z$-decay width. 
In the SMEFT framework, $N_\nu$ is given by
\begin{equation}\label{eq: N_nu def}
    N_\nu\approx3 -\frac{v^2}{\Lambda^2}\sum_i\text{Re}\left[(C_{H\ell}^
    {(1)})_{ii}-(C_{H\ell}^
    {(3)})_{ii}\right]\,.
\end{equation}
Using Eq.~\eqref{eq:C_He and C_Hell}, we obtain
\begin{equation}
    \frac{N_\nu-3}{5\cdot 10^{-3}}\approx g_V^2Q_V^H(Q_V^{\ell_3}+2Q_V^{\ell_2})\left(\frac{3\,\mathrm{TeV}}{M_V}\right)^2\,.
\end{equation}
Thus, $N_\nu$ provides a probe of NP effects that is complementary to, and comparable in sensitivity with, the LFUV tests at the $Z$-pole discussed above.

\subsection{Electric and magnetic dipole moments}
\label{PhenoLoop}
Lepton dipole moments provide highly sensitive probes of the SM and its possible NP extensions.  The anomalous magnetic moments (AMMs) of the electron and muon, $a_{\ell}\equiv \frac{(g-2)_\ell}{2}$ ($\ell=e,\mu$), are measured with remarkable precision~\cite{Fan:2022eto,Muong-2:2023cdq,Muong-2:2025xyk}.
In order to fully exploit their sensitivity to NP, a comparable level of theoretical control is required. In the case of $a_e$, the dominant uncertainty is associated with the determination of the fine-structure constant~\cite{Morel:2020dww,Parker:2018vye}, whereas for $a_\mu$
the main limitation originates from hadronic vacuum-polarization contributions, for which substantial recent progress has been achieved using lattice QCD techniques~\cite{Aliberti:2025beg}. At present, the comparison between SM predictions and experimental measurements does not show a statistically significant discrepancy~\cite{Aliberti:2025beg}, and a continuous effort is ongoing to further reduce the theoretical uncertainties~\cite{Hertzog:2025ssc,Aliberti:2025beg} and fully capitalize on the impressive experimental precision.
By contrast, the electric dipole moments (EDMs) of the electron and muon, $d_{\ell}$, are negligibly small within the SM. Consequently, the observation of a non-zero signal below the current experimental bounds~\cite{Roussy:2022cmp,Muong-2:2008ebm} would provide unambiguous evidence for CP-violating NP. 

Leptonic EDMs are generated by the imaginary flavor-diagonal components of the dipole matrix,
\begin{equation}
      d_{\ell_i} = -\frac{e}{4 \pi^2} \mathrm{Im}[(C_{D})_{ii}]
      = -\frac{e}{4 \pi^2}|(C_{D})_{ii}|\,\sin\varphi_{\rm CP}\,,
\end{equation}
where $\varphi_{\rm CP}$ denotes the relevant CP-violating phase. According to Eq.~\eqref{eq:dipole_LEFT}, $(C_D)_{11}$ can be estimated by naive dimensional analysis as
\begin{equation}
    (C_D)_{11}\approx \frac{8\pi^2}{e} m_e
    \times \frac{C_{V,\phi}\, F_{V,\phi}}{16\pi^2 M_{V,\phi}^2}\,
    e^{i\varphi_{\rm CP}}\,.
\end{equation}
where $C_{V,\phi} = g_{V}^{2},\lambda_{ij}$.

As already discussed in Sec.~\ref{sec:flav structure to smeft}, in minimal FD models the leading contributions to the ``flavored'' CP-violating phase $\varphi_{\rm CP}$ arise only at NLO in the spurion expansion ~\cite{Calibbi:2008qt,Calibbi:2020emz}, resulting in an effective suppression $\sin\varphi_{\rm CP}\sim\varepsilon_\chi^2$, see Tab.~\ref{tab:ChiChi fermionic bilienars suppression factors}.
This built-in CP-protection mechanism allows 
the typical NP bounds from leptonic EDMs, which in scenarios with $\mathcal{O}(1)$ phases would probe scales of hundreds of TeV, to be relaxed down to the $\mathcal{O}(10)\,\mathrm{TeV}$ regime.
Focusing on the electron EDM, we obtain the prediction
\begin{align} 
\label{eq:EDM_prediction}
    \!\frac{d_e}{6\cdot 10^{-30}e \,\mathrm{cm}}
    \approx 
    \Bigg[g_V^2 F_V
    \left(\frac{10\,\mathrm{TeV}}{M_V}\right)^2 
+\lambda_{ij} F_\phi
\left(\frac{10\,\mathrm{TeV}}{M_\phi} \right)^2 
\Bigg]
\left(\frac{\varepsilon_\chi}{0.04}\right)^2,
\end{align}
where $d_e$ is normalized to a value close to the current experimental upper bound reported in Tab.~\ref{table:pheno}.

In FD models, lepton EDMs obey the naive scaling relation, 
$d_{\ell_i}/d_{\ell_j}=m_{\ell_i}/m_{\ell_j}$~\cite{Giudice:2012ms}, as shown in Tab.~\ref{tab:ChiChi fermionic bilienars suppression factors}. Consequently, after accounting for the current theoretical predictions and experimental sensitivities, the muon and tau EDMs are found to be several orders of magnitude less sensitive to NP effects than the electron EDM; see, e.g., ref.~\cite{Valori:2025hlp}.

NP contributions to AMMs, defined as $\Delta a_{\ell_i}\equiv a_{\ell_i}^{\rm exp}-a_{\ell_i}^{\rm SM}\equiv a_{\ell_i}^{\rm NP}$, are generated by the real flavor-diagonal entries of the dipole matrix. Under the naive scaling hypothesis~\cite{Giudice:2012ms}, the AMMs scale quadratically with the lepton masses, $\Delta a_{\ell_i}/\Delta a_{\ell_j}=m_{\ell_i}^2/m_{\ell_j}^2$, thereby strongly suppressing NP effects for the lighter leptons. 
As follows from Eq.~\eqref{dipolegeneral}, FD models obey the naive scaling relation, implying that the present constraint from the muon AMM probes NP scales of at most a few hundred GeV.
Moreover, once the current theoretical predictions and experimental sensitivities are taken into account, the AMMs of the electron and the tau are found to provide even weaker constraints on NP effects than the muon AMM (see \cite{Valori:2025hlp} and references therein).

\subsection{Renormalization group evolution effects}
In the phenomenological analysis presented above we have neglected the renormalization group evolution (RGE) of the dimension-six operators between the matching scale and the electroweak scale. In this subsection, we estimate the size of these effects and discuss under which conditions they can become phenomenologically relevant. The running of the Wilson coefficients is governed by
\begin{equation}
 16\pi^2\,\frac{d C_i(\mu)}{d\log\mu}
 =
 \gamma_{ji}\,C_j(\mu) \, ,
\label{eq:RGE_general}
\end{equation}
where $\gamma_{ji}$ denotes the anomalous-dimension matrix. The full
one-loop SMEFT anomalous dimensions are known~\cite{Alonso:2013hga,Jenkins:2013wua,Jenkins:2013zja}, and allow one to consistently evolve the coefficients generated at the heavy scale down to the electroweak scale. The diagonal entries of $\gamma_{ji}$ induce a self-renormalization of the Wilson coefficients. In the purely leptonic sector this running is dominated, below the electroweak scale, by QED effects and typically amounts to corrections at the level of a few tens of percent in low-energy observables. These effects do not modify the flavor structure discussed in the previous sections and can be included as an overall refinement of the numerical predictions.

On the contrary, by extending the model to the quark sector, operators featuring heavy quarks may induce sizeable mixing to operators relevant for the phenomenology in the lepton sector. This mechanism was shown in Refs.~\cite{Feruglio:2016gvd,Feruglio:2017rjo,Cornella:2018tfd} to play a crucial role in scenarios with
semileptonic NP above the electroweak scale: even if the NP is introduced only through semileptonic operators, electroweak running generates corrections to the leptonic $W$- and $Z$-boson couplings, as well as purely leptonic effective interactions. These loop-induced effects are therefore probed by precision observables such as $Z$-pole measurements, $\tau$ decays and charged-lepton flavor violating processes.
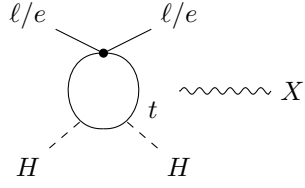
\begin{figure}[t]
\centering
\scalebox{1.0}{\begin{tikzpicture}[baseline]
        \begin{feynman}
        \vertex(c1) at (1.5,1.5) {$\ell/e$};
        \vertex(d1) at (3.5,1.5) {$\ell/e$};
        \vertex(cd) [dot,minimum size= 3pt] at (2.5,1) {};
        \vertex(dc) at (2.5,0);
        \vertex(ab) at (2.2,0.1);
        \vertex(ba) at (2.8,0.1);
        \vertex(o) at (1.5,-0.5) {$H$};
        \vertex(f) at (3.5,-0.5) {$H$};
        \vertex(z) at (2.5,-0.1) {};
        \vertex(w) at (5,0.5) {$X$};
        \vertex(ww) at (3.5,0.5);
        \diagram{
            (c1) -- [black] (cd) -- [black] (d1),
            (cd) -- [black, half left,edge label=$t$] (dc),
            (cd) -- [black, half right] (dc),
            (o) -- [scalar,black] (ab), (f) -- [scalar,black] (ba), (ww) -- [boson, black] (w)
            };
        \end{feynman}
    \end{tikzpicture}}
\caption{Diagrammatic representation of the mixing of a semileptonic operator into operators of the $\psi^{2}H^{2}D$ class. Here, $X=B,W^{I}$ denotes a SM electroweak gauge boson attached to the fermion loop.}
\label{fig:RGE}
\end{figure}

In the present FD model the tree-level exchange of the heavy vector bosons generates, among others, semileptonic operators involving top-quark currents. Keeping only the top-Yukawa enhanced terms, the relevant SMEFT mixing is
\begin{subequations}
    \begin{align}
16\pi^2\frac{d(C_{H\ell}^{(1)})_{pr}}{d\log\mu}
 &=
 6y_t^2 (C_{\ell q}^{(1)})_{pr33}
 -
 6y_t^2 (C_{\ell u})_{pr33}\, ,
\label{eq:RGE_CHl}
\\
16\pi^2\frac{d(C_{He})_{pr}}{d\log\mu}
 &=
 6y_t^2 (C_{qe})_{33pr}
 -
 6y_t^2 (C_{eu})_{pr33}\, .
\label{eq:RGE_CHe}
\end{align}
\end{subequations}
At the diagrammatic level, the origin of these effects can be directly understood from the one-loop topology shown in Fig.~\ref{fig:RGE}.
Solving these equations at leading-log accuracy between the heavy-vector scale $M_V$ and the electroweak scale $m_{\rm EW}$, one obtains
\begin{subequations}
    \begin{align}
(\delta C_{H\ell}^{(1)})_{pr}
 &\approx
 \frac{3y_t^2}{8\pi^2}
 \log\left[\frac{M_V}{m_{\rm EW}}\right]
 \Big[
 (C_{\ell q}^{(1)})_{pr33}
 -
 (C_{\ell u})_{pr33}
 \Big],
\label{eq:delta_CHl}
\\
(\delta C_{He})_{pr}
 &\approx
 \frac{3y_t^2}{8\pi^2}
 \log\left[\frac{M_V}{m_{\rm EW}}\right]
 \Big[
 (C_{qe})_{33pr}
 -
 (C_{eu})_{pr33}
 \Big].
\label{eq:delta_CHe}
\end{align}
\end{subequations}

After electroweak symmetry breaking these contributions shift the leptonic
$Z$ couplings. 
Adopting the notation introduced in Sec.~\ref{sec:SMEFT}, and using the corrections to the modified $Z$ couplings defined in Eq.~\eqref{eq:Z_LR}, the LFU observables discussed in Sec.~\ref{sec:LFUV} can be expressed, at leading-order accuracy, as
\begin{subequations}
    \begin{align}
\delta\!\left(\frac{g_V^\tau}{g_V^e}\right)
&\approx
\frac{g_V^2}{4s_W^2-1}
\frac{v^2}{M_V^2}
\frac{3y_t^2}{8\pi^2}
\log\left[\frac{M_V}{m_{\rm EW}}\right]
(Q_V^{q_3}-Q_V^{u_3})
\left[
 (Q_V^{\ell_3}-Q_V^{\ell_2})
 +(Q_V^{e_3}-Q_V^{e_2})
\right] ,
\\
\delta\!\left(\frac{g_A^\tau}{g_A^e}\right)
&\approx
g_V^2
\frac{v^2}{M_V^2}
\frac{3y_t^2}{8\pi^2}
\log\left[\frac{M_V}{m_{\rm EW}}\right]
(Q_V^{q_3}-Q_V^{u_3})
\left[
 (Q_V^{\ell_3}-Q_V^{\ell_2})
 -(Q_V^{e_3}-Q_V^{e_2})
\right] ,
\\
\delta N_\nu
&\approx
g_V^2
\frac{v^2}{M_V^2}
\frac{3y_t^2}{8\pi^2}
\log\left[\frac{M_V}{m_{\rm EW}}\right]
(Q_V^{q_3}-Q_V^{u_3})
(Q_V^{\ell_3}+2Q_V^{\ell_2}) \,.
\label{eq:RGE_LFU}
\end{align}
\end{subequations}
Numerically, for TeV-scale vectors these corrections correspond to few percent effects in the ratios of charged-lepton $Z$ couplings and to corrections at the ten-percent level in $N_\nu$, translating into an $\mathcal{O}(10\%)$ shift of the corresponding bounds on $M_V$. 
Moreover, LFV observables such as $\mu\!-\!e$ conversion and $\mu\to3e$ exhibit a quadratic dependence on the LEFT Wilson coefficients. As a consequence, including the interference between the $Z$- and $Z_V$-mediated amplitudes induces corrections of order $\mathcal{O}(10\%)$ with respect to the tree-level predictions.
Thus, the RGE effects do not change the qualitative pattern of the bounds
derived above, but they represent a non-negligible correction to precision
observables. 

\subsection{Discussion and generalization to other flavor-deconstructed models} 

The phenomenological analysis presented in this section highlights several distinctive features of the minimal FD framework in the lepton sector. 
The setup naturally predicts correlated effects in charged LFV, LFUV, and 
CP-violating observables, all originating from the same underlying spurionic structures responsible for the generation of the fermion mass hierarchies. 
In particular, the interplay between flavor and CP violation emerges as one 
of the most distinctive features of the framework.

Our main results can be summarized as follows:
\begin{itemize}
    \item Among LFV observables, $\mu-e$ conversion in nuclei provides the strongest constraints on the FD model over a broad region of the parameter space, probing NP scales up to $\mathcal{O}(20\div30)\,\mathrm{TeV}$ 
    for $\mathcal{O}(1)$ couplings and charges, namely $Y_{i\alpha}^{e}=\mathcal{O}(1)$ 
    and $g_V(Q_V^{\ell_2}-Q_V^{\ell_3})=\mathcal{O}(1)$. This sensitivity is mainly driven by tree-level contributions mediated by the heavy gauge bosons $Z_{23}$ and $Z'_{23}$, which induce semileptonic four-fermion operators 
    with a characteristic flavor structure. Thanks to the excellent projected experimental sensitivities, future $\mu-e$ conversion searches are 
    expected to improve the current reach on NP scales by approximately one 
    order of magnitude.    
    \item The decay $\mu \to 3e$ is governed by the same source of flavor violation. The corresponding bound on the NP scale is of $\mathcal{O}(10)\,\mathrm{TeV}$, approximately a factor of three weaker than that derived from $\mu-e$ conversion in nuclei. As a result, this channel is currently subleading in light of existing experimental limits~\cite{Davidson:2020hkf}, although it still provides complementary information. Future experimental improvements are expected to extend the sensitivity to NP scales up to $100\,\mathrm{TeV}$.
    \item Despite the approximate alignment between dipole operators and Yukawa matrices, which significantly suppresses both LFV transitions 
    and CP-violating effects, the electron EDM currently provides one of 
    the most sensitive probes of the framework, testing NP scales up to $\mathcal{O}(10)\,\mathrm{TeV}$. Future experiments employing ThO~\cite{Hiramoto:2022fyg} and YbF~\cite{Fitch:2020jil} molecules are expected to improve the present sensitivity by approximately one and two orders of magnitude, respectively, thereby extending the reach in the multi-10~TeV regime. By contrast, the radiative decay $\mu\to e\gamma$
    currently probes NP scales up to $\mathcal{O}(3)\,\mathrm{TeV}$, and the projected experimental improvements are not expected to substantially strengthen this sensitivity.
    \item Tests of LFUV at the $Z$ pole performed at LEP probe NP scales in the range $\mathcal{O}(5\div10)\,\mathrm{TeV}$. Although this sensitivity is subleading compared to that of $\mu-e$ conversion in nuclei and the electron EDM, it nevertheless provides a particularly robust probe of FD models, being comparatively less sensitive to the detailed flavor and CP structure of the framework. Most notably, future Tera-$Z$ facilities, such as circular $e^+e^-$ colliders operating at the $Z$ pole, could improve the current experimental precision on LFUV observables by roughly two orders of magnitude, corresponding to sensitivities to NP scales of $\mathcal{O}(50\div100)\,\mathrm{TeV}$.
    \item Processes involving $\tau\to\ell$ transitions (with $\ell=e,\mu$), such as $\tau \to 3\ell$, $\tau \to \ell\gamma$, and $Z \to \tau \ell$, constitute significantly less sensitive probes of FD models. Indeed, although the associated flavor-mixing angles are generally larger than those relevant for $\mu \to e$ transitions, this enhancement is more than compensated by the comparatively weaker experimental bounds.
\end{itemize}

\begin{figure}
    \centering
    \includegraphics[width=0.8\linewidth]{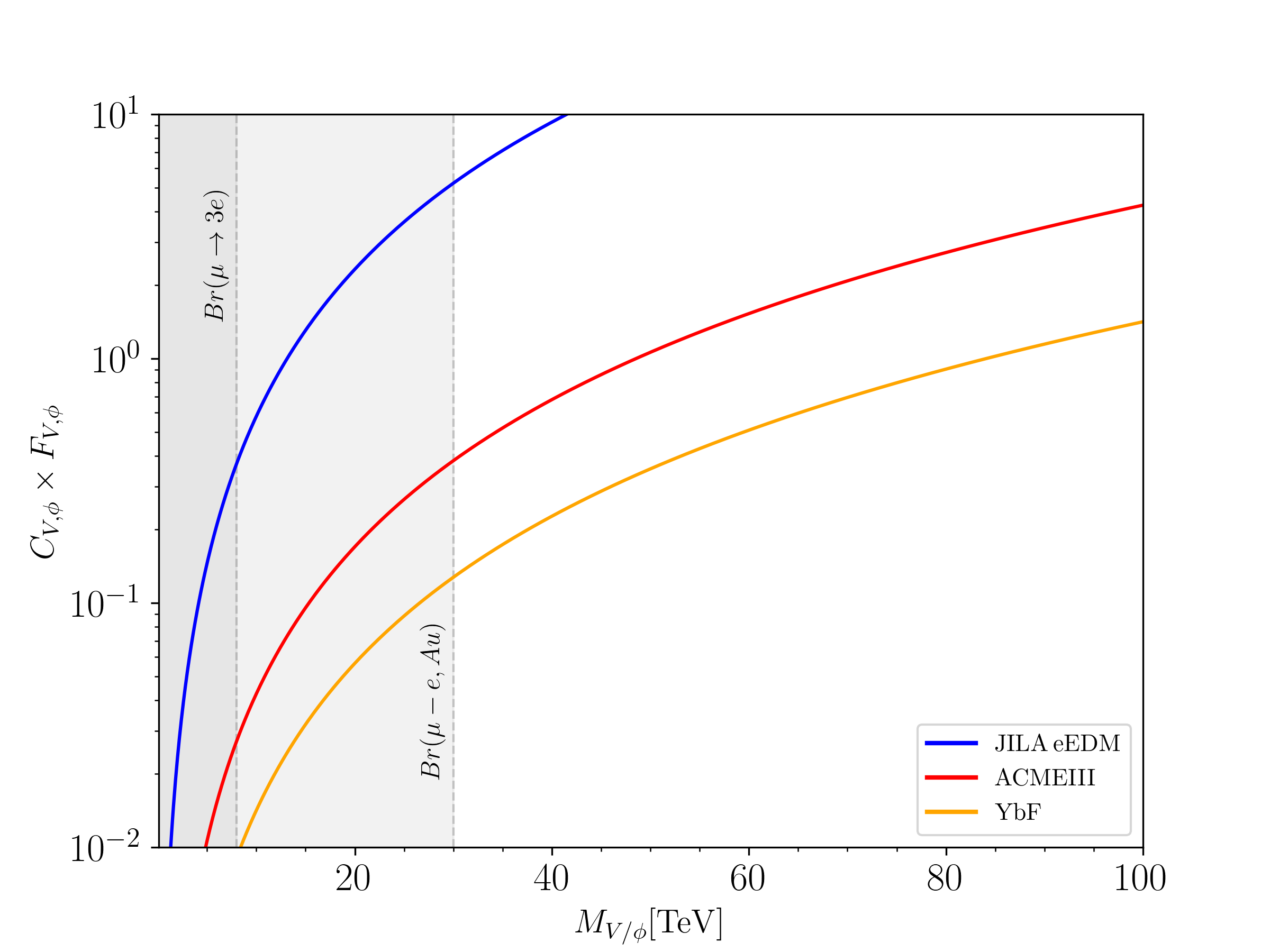}
    \caption{Plot of $C_{V,\phi} \times F_{V,\phi}$ and the NP masses assuming that the current upper limit (JILA eEDM \cite{Roussy:2022cmp}) and future projections (ACMEIII \cite{Hiramoto:2022fyg} and YbF \cite{Fitch:2020jil}) for the electron EDM experiments are saturated. 
    Vertical dashed lines represent the energy scales currently probed by the present bounds on $\mathrm{Br}(\mu \rightarrow 3e)$ \cite{SINDRUM:1987nra} and $\mathrm{Br}(\mu-e, \mathrm{Au})$ \cite{SINDRUMII:2006dvw} in the limit $g_{V}= O(1)$. Benchmark values for UV parameters are chosen as described in the text. 
    }
    \label{fig:FigureEDMvsLFV}
\end{figure}

\begin{figure}
    \centering
    \includegraphics[width=0.8\linewidth]{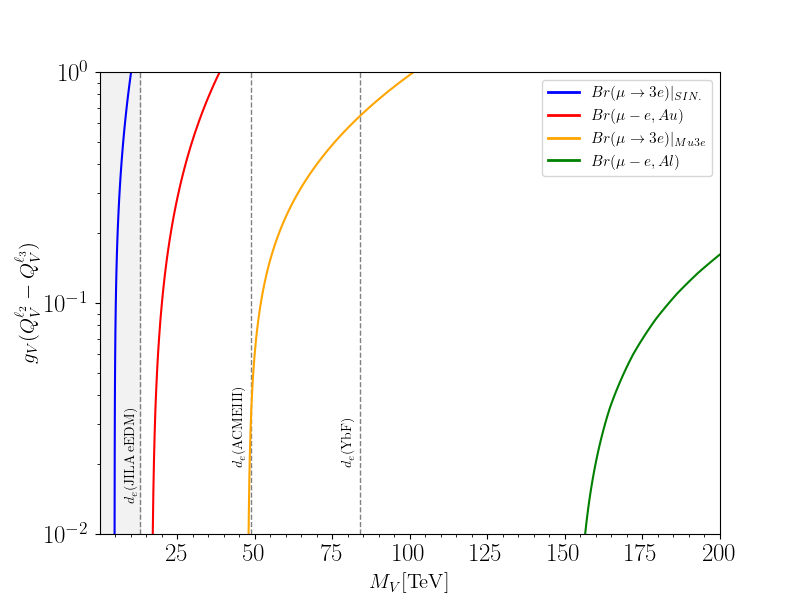}
    \caption{Plot of the charge difference $g_V(Q_{V}^{\ell_{2}}-Q_{V}^{\ell_{3}})$ and the NP masses assuming that the current upper limit for $\mu-e$ conversion in nuclei \cite{SINDRUMII:2006dvw} and $\mu \rightarrow 3e$ \cite{SINDRUM:1987nra} as well as future projections \cite{Mu2e:2014fns,COMET,Blondel:2013ia} are saturated. 
    Vertical dashed lines represent the energy scales probed by the present bounds (JILA eEDM \cite{Roussy:2022cmp}) and future projections on $d_{e}$ (ACMEIII \cite{Hiramoto:2022fyg} and YbF \cite{Fitch:2020jil}) in the limit of single diagram contribution, assuming $\mathcal{O}(1)$ parameters for the phase and couplings.}
    \label{fig:mutoe vs edm}
\end{figure}

To illustrate the above points, Fig.~\ref{fig:FigureEDMvsLFV} displays the sensitivity of the electron EDM to the NP scale $M_{V/\phi}$ as a function of the product of NP couplings $C_{V,\phi}$ and the loop function $F_{V,\phi}$ generating the dipole amplitude relevant for the electron EDM. The figure includes the current upper bound from the JILA eEDM experiment~\cite{Roussy:2022cmp}, together with the projected sensitivities 
of future experiments, namely ACMEIII~\cite{Hiramoto:2022fyg} and YbF~\cite{Fitch:2020jil}. Assuming $g_V\sim\lambda_{ij}=\mathcal{O}(1)$, 
these bounds translate into a sensitivity to heavy gauge-boson masses in 
the $50\div80\,\mathrm{TeV}$ range.
The vertical dashed lines denote the NP scales currently probed by the 
present bounds on $\mathrm{Br}(\mu \to 3e)$~\cite{SINDRUM:1987nra} and 
on $\mathrm{Br}(\mu-e,\mathrm{Au})$~\cite{SINDRUMII:2006dvw}.
Similarly, by varying the charge difference $Q_V^{\ell_2}-Q_V^{\ell_3}$, 
it is possible to illustrate the interplay between LFV observables and $d_e$. Fig.~\ref{fig:mutoe vs edm} shows the NP scales probed by $\mathrm{Br}(\mu-e,\mathrm{N})$ and $\mathrm{Br}(\mu \to 3e)$ as a function of the heavy 
vector-boson mass, while varying the difference between the third- and 
second-generation (or first-generation) charges, $Q_V^{\ell_2}-Q_V^{\ell_3}$. As can be inferred from the figure, future experiments will be sensitive to NP scales up to approximately $100\,\mathrm{TeV}$ even in the presence of flavor alignment at the percent level.

Although the phenomenological analysis presented above has been carried out in the minimal FD realization of Refs.~\cite{Barbieri:2023qpf,Barbieri:2024zkh}, most of our conclusions rely only on a set of assumptions which are common to a broader class of FD models. 
The first key ingredient is the separation of the third generation from the first two at the lowest flavor-breaking scale. This is a generic feature of FD in which the large third-generation Yukawa couplings are generated at leading order, while the lighter-generation masses and inter-generational mixings arise from insertions of link fields or from integrating out heavy mediator states. If the Higgs couples dominantly to the third-generation, the low-energy theory inherits an approximate $\mathrm{U}(2)^5$ flavor symmetry acting on the first two generations~\cite{Davighi:2023iks}. As a consequence, the flavor off-diagonal couplings of heavy neutral vectors to charged leptons exhibit the parametric pattern
\begin{equation}
  (P^{e,\ell}_{V})_{23} \sim (P^{e,\ell}_{V})_{13}\,, \qquad
  (P^{e,\ell}_{V})_{12} \ll (P^{e,\ell}_{V})_{13,23}\,,
  \label{eq:generic_PV_pattern}
\end{equation}
up to model-dependent charge factors and order-one coefficients. In the minimal model analysed here, this expectation is realized explicitly, with
$(P^\ell_V)_{13,23}\sim \varepsilon_\chi$ and $(P^\ell_V)_{12}\sim \varepsilon_\chi^2$, while the right-handed sector can display additional suppressions controlled by the spurions entering the charged-lepton Yukawa matrix. 

\paragraph{Abelian flavor deconstruction.} The same logic extends to other abelian deconstructions of hypercharge, $\mathrm{B-L}$, or $T_{3R}$, provided that the light generations remain approximately aligned before the final breaking to the SM gauge group. This observation implies that several phenomenological results derived in the previous sections are robust. In particular, LFV transitions involving a tau lepton, such as $Z\to \tau\ell$ and \(\tau\to 3\ell\), are expected to be controlled by a single third-to-light mixing insertion. By contrast, purely light-generation transitions such as $\mu\to 3e$ and $\mu-e$ conversion require two such insertions, or an additional misalignment between the first two generations. Therefore, while the detailed numerical bounds depend on the gauge charges and on the relative alignment of the lepton rotations, the hierarchy between $\tau$-LFV and $\mu-e$ observables follows directly from the approximate $\mathrm{U}(2)$ structure. The exceptional experimental sensitivity of $\mu-e$ observables can nevertheless compensate for their stronger flavor suppression, making $\mu-e$ conversion and $\mu\to 3e$ among the most powerful probes of FD models in the lepton sector.

A second robust feature concerns chirality-violating observables. 
Dipole operators are approximately aligned with the charged-lepton Yukawa matrix at LO in the spurion expansion \cite{Calibbi:2020emz,Calibbi:2021qto,Lopez-Ibanez:2021yzu}. However, once NLO spurion structures 
are included, this alignment is no longer exact. In a generic FD completion 
the electron EDM is therefore expected to scale as
\begin{equation}
  d_e \sim
  \frac{e}{16\pi^2}\,
  \frac{m_e}{\Lambda_{[23]}^2}\,
  \sin\varphi_{\rm CP}\,,
  \label{eq:generic_de_scaling}
\end{equation}
where $\Lambda_{[23]}$ denotes the scale associated with the deconstruction of the third generation, $\varphi_{\rm CP} \sim \varepsilon^2$ 
is a flavored CP-violating phase, generated by a double first--third flavor mixing, e.g. $(P^{e,\ell}_{V})_{13}\sim \varepsilon$. The precise loop function and the relative importance of vector, scalar, and vector-like-fermion exchange are model dependent. Nevertheless, Eq.~\eqref{eq:generic_de_scaling} captures the main reason why EDMs are particularly sensitive probes of FD dynamics: 
they test the same flavor-breaking sector responsible for light-generation masses, but are also directly sensitive to irreducible CP-violating phases. This conclusion does not depend on the abelian nature of the benchmark model, but only on the coexistence of hierarchical flavor breaking and complex spurions.

\paragraph{Non-abelian flavor deconstruction.} It is useful to compare these conclusions with the non-abelian electroweak deconstruction studied in Ref.~\cite{Davighi:2023xqn}. In that framework the 
gauge group $\mathrm{SU}(2)_{L}^{[1]}\times \mathrm{SU}(2)_{L}^{[2]}\times \mathrm{SU}(2)_{L}^{[3]}$
is broken sequentially to the diagonal SM $\mathrm{SU}(2)_L$, with the Higgs charged under the third-family factor. The resulting spectrum contains two heavy $\mathrm{SU}(2)_L$ triplets: a heavier $W_{12}$, associated with the breaking of the first two generations, and a lighter $W_{23}$, associated with the breaking of the third generation from the light-family sector. The phenomenology is therefore controlled by the same organizing principle as in the present work: an approximate $\mathrm{U}(2)^5$ protection of the light generations combined with sizable third-generation non-universality.

There are, however, important differences. In the abelian model considered here, the leading tree-level effects arise from heavy neutral vectors and generate neutral-current four-fermion operators and modified $Z$-boson couplings. In the deconstructed $\mathrm{SU}(2)_L$ model of
Ref.~\cite{Davighi:2023xqn}, the heavy vectors form electroweak triplets and necessarily generate both neutral- and charged-current operators. This leads to a richer set of correlations among semileptonic flavor observables, electroweak precision tests including LFUV in charged currents, and high-$p_T$ Drell--Yan searches. In particular, Ref.~\cite{Davighi:2023xqn} finds that the lighter triplet $W_{23}$ is constrained by a highly complementary set of probes, including electroweak precision data, $B_s\to\mu^+\mu^-$,
Drell--Yan tails, and LFV searches, with present bounds excluding masses below roughly $9$ TeV in the most favorable region of parameter space. The heavier $W_{12}$, instead, is pushed to scales of hundreds of TeV by meson-mixing constraints, reflecting the extreme sensitivity of first--second generation flavor violation.

The comparison highlights both the universality and the model dependence of FD
phenomenology. The universality lies in the flavor structure: whenever the third generation is separated from the first two at the lowest scale, the dominant new effects are governed by third-to-light mixing, while purely light-generation transitions are protected by the residual $\mathrm{U}(2)^5$ symmetry. This explains why $\mu\to 3e$, $\mu-e$ conversion, LFUV tests, and
$\tau$-LFV observables provide complementary probes in both abelian and non-abelian FD realizations. The model dependence lies instead in the gauge representation of the heavy vectors and in the presence or absence of charged currents. In the $\mathrm{SU}(2)_L$-deconstructed case, electroweak precision observables and high-$p_T$ searches are intrinsically tied to the same heavy triplet that controls flavor. In the abelian case, the phenomenology is more
directly controlled by the flavor-dependent charges and by the scalar/vector-like sector entering the Yukawa spurion expansion.

A further distinction concerns CP violation. The analysis of Ref.~\cite{Davighi:2023xqn} focuses primarily on tree-level flavor, collider, and electroweak constraints, whereas the present work emphasizes the impact of CP-violating spurions on loop-induced dipole observables. This is a particularly important aspect of abelian FD models with vector-like fermions, where complex Yukawa structures are required to reproduce the observed CKM phase and naturally induce physical phases in the lepton sector. In such models, the electron EDM can probe scales in the multi-TeV to multi-10-TeV regime, even when LFV dipole amplitudes remain suppressed
by approximate alignment. Thus, EDMs provide a qualitatively different handle on FD dynamics, complementary to the tree-level probes emphasized in non-abelian electroweak deconstruction.

Finally, the comparison also clarifies the role of future experiments. In the electroweak deconstruction of Ref.~\cite{Davighi:2023xqn}, FCC-ee electroweak precision measurements are expected to cover a large fraction of the natural parameter space, while HL-LHC Drell--Yan searches and Mu3e will provide important shorter-term tests. In the present abelian framework, future $\mu-e$ conversion experiments and improved EDM searches play an analogous role, probing
flavor-breaking scales well beyond the direct reach of colliders. These two classes of models therefore illustrate complementary experimental strategies for testing the same underlying idea: flavor hierarchies emerging from gauge deconstruction near the TeV scale.\\

In summary, the minimal model studied in this paper captures the essential low-energy features of a broad class of FD scenarios. The detailed numerical bounds are not universal, since they depend on the gauge group, the heavy spectrum, and the charge assignments. Nevertheless, the parametric organization of flavor violation, the emergence of approximate $\mathrm{U}(2)^5$ protection, and the sensitivity of EDMs to complex higher-order spurions are generic.
For this reason, the combined study of LFV, LFUV, EDMs, electroweak precision observables, and high-energy searches provides a coherent and broad strategy for testing flavor deconstruction.

\section{Conclusions}\label{sec:conclusions}

Flavor deconstruction provides an attractive framework to address the origin of fermion mass hierarchies through the spontaneous breaking of flavor-dependent gauge symmetries at comparatively low scales. In this work, we have investigated the anatomy and phenomenology of a minimal flavor-deconstructed scenario \cite{Barbieri:2023qpf,Barbieri:2024zkh} in the lepton sector, with particular emphasis on the interplay between flavor violation and CP violation.

Starting from the ultraviolet completion of the model, we derived the effective structure of the lepton Yukawa sector by means of a systematic spurion analysis, extending the expansion beyond leading order. This allowed us to identify the dominant sources of flavor and CP violation and to clarify how complex phases naturally emerge once the framework is required to reproduce the observed CKM phase. In particular, we have shown that, while leading-order dipole structures are approximately aligned with the Yukawa matrices, next-to-leading order effects generically induce non-trivial flavor misalignment and physical CP-violating phases. As a result, electric dipole moments and charged lepton flavor violating observables emerge as especially sensitive probes of the framework.

We then constructed the corresponding SMEFT description obtained after integrating out the heavy vector bosons, vector-like fermions, and scalar states predicted by the model. The resulting EFT framework provides a systematic and model-independent characterization of the low-energy phenomenology, including tree-level matching for four-fermion and Higgs-current operators as well as loop-induced dipole interactions. The subsequent matching onto LEFT enables a direct connection between the ultraviolet structure of the theory and low-energy observables.

From a phenomenological perspective, we find that $\mu-e$ conversion in nuclei, $\mu \to 3e$, tests of lepton flavor universality, and the electron EDM provide the most stringent constraints on the parameter space of the model, probing new-physics scales in the $\mathcal{O}(10)\,\mathrm{TeV}$ range. Remarkably, the projected sensitivities of future experiments are expected to extend the reach of all these observables up to scales of $\mathcal{O}(100)\,\mathrm{TeV}$.
An important outcome of our analysis is the strong complementarity between flavor-violating and CP-violating observables, whose relative sensitivity depends on the underlying flavor charges, CP phases, and details of the ultraviolet dynamics. 
More generally, our results indicate that precision observables in the lepton sector provide a remarkably powerful window into flavor-deconstructed theories, potentially probing energy scales far beyond the direct reach of present and future colliders. 

The framework developed in this work can be naturally extended to more general realizations of flavor deconstruction, including setups based on non-abelian flavor symmetries or extended neutrino sectors~\cite{Davighi:2023evx,Greljo:2024ovt,Isidori:2025rci}. 
We have also argued that the key phenomenological features identified in this analysis are not peculiar to the minimal flavor deconstruction setup considered here, but rather represent generic predictions of a broader class of flavor-deconstructed theories.

Overall, the present study highlights how the combined study of lepton flavor violation, CP violation, and effective field theory techniques can provide a powerful and systematic strategy to uncover the underlying dynamics responsible for the flavor structure of the Standard Model.

\section{Acknowledgments}

The authors acknowledge Marco Ardu for useful discussions and for his collaboration in the early stages of this work. NV  and OV acknowledge the support of the grant PID2023-151418NB-I00, which is funded by MCIU/AEI/10.13039/501100011033/ FEDER, UE. OV acknowledges funding from Generalitat Valenciana CIPROM/2021/054. 
AM and PP received funding by the INFN Iniziative Specifiche AMPLITUDES, APINE, and TASP and from the European Union’s Horizon 2020 research and innovation programme under the Marie Sklodowska-Curie grant agreements n. 860881 — HIDDeN, n. 101086085 — ASYMMETRY. This work was also partially supported by the Italian MUR Departments of Excellence grant 2023-2027 “Quantum
Frontiers”, the European Union — Next Generation EU, and the Italian Ministry of University and Research (MUR) via the PRIN 2022 project n. 2022K4B58X — AxionOrigins. DQ acknowledges support from Generalitat Valenciana CIDEGENT/2019/024 and CIESGT2024-021.
AM and NV acknowledge the Munich Institute for Astro-, Particle and BioPhysics (MIAPbP) which is funded by the Deutsche Forschungsgemeinschaft (DFG, German Research Foundation) under Germany´s Excellence Strategy – EXC-2094 – 390783311 for the hospitality during the final stage of this project. AS and NV thank the Galileo Galilei Institute for Theoretical Physics for the hospitality and the INFN for partial support during the completion of this work.

\appendix

\section{Model independent correction to the mass terms} \label{Appendix:massterms}

In the following, we show, in a model independent way, how pure mass terms and wave function renormalization corrections affect the final expression for the mass term. While this procedure has been exploited in the context of the \emph{spurion analysis} in FD models, the conclusions obtained can be applied to any situation where an expansion parameter governs the Yukawa texture of the model.
Following Ref. \cite{Bellazzini:2010gn}, let us consider the kinetic and mass terms of the lepton fields
\begin{eqnarray}
{\cal L}
&=&\overline{\ell}_{jL} \! \left( Z_{L}^{\ell} \right)^{ji} \! i\!\not\!\partial\, \ell_{iL} +
\overline{\ell}_{jR} \! \left( Z^{\ell}_{R} \right)^{ji} \!i\! \not\!\partial \,\ell_{iR} 
- \left( \overline{\ell}_{jR}\left(m^{\circ}_{\ell}+\eta^{\ell}_{m}\right)^{ji} \!\ell_{iL} + \mathrm{h.c.}\right)\,, 
\end{eqnarray}
where $m^{\circ}_{\ell}$ and $\eta^{\ell}_{m}$ denote the LO and NLO contributions to the lepton mass matrix, respectively, while $Z^{\ell}_{L(R)}$ are Hermitian matrices encoding wave-function renormalisation effects, defined as
\begin{equation}
\label{eq:WFZ}
\left(Z^{\ell}_{L}\right)_{ij}= \delta_{ij} + \left(\eta^{\ell}_{L}\right)_{ij}\,,\qquad\qquad
\left(Z^{\ell}_{R}\right)_{ij}= \delta_{ij} + \left(\eta^{\ell}_{R}\right)_{ij}\,,
\end{equation}
where $(\eta^{\ell})^\dagger=\eta^{\ell}$. We define
\begin{equation}
\hat{Z}^{\ell}_L = L^\dagger_{\ell} Z^{\ell}_L L_{\ell}~,
\qquad\qquad
\hat{Z}^{\ell}_R = R^\dagger_{\ell} Z^{\ell}_{R} R_{\ell}~,
\end{equation}
such that $\hat{Z}^a_{L,R}$ are diagonal matrices with positive entries, 
while $L_\ell$ and $R_\ell$ are unitary matrices. We then rescale the 
lepton fields in order to render their kinetic terms canonical,
\begin{eqnarray}
\ell_{L}
\rightarrow
L_{\ell}\left(\hat{Z}^\ell_L\right)^{-\frac{1}{2}}~\ell_{L}\,,
\qquad\qquad
\ell_{R}
\rightarrow
R_{\ell}\left(\hat{Z}^{\ell}_{R}\right)^{-\frac{1}{2}}~
\ell_{R}\,,
\label{eqn:rescaling}
\end{eqnarray}
where $(\hat{Z}^a)^{-1/2}$ denotes the diagonal matrix with entries $(\hat{Z}^a)^{-1/2}_{ii}$. The charged lepton mass terms are then given by
\begin{eqnarray}
&{\cal L}_{\rm mass}=
-\overline{\ell}_{jR}
\left(\hat{Z}^{\ell}_{R}\right)^{-\frac{1}{2}}R^{\dagger}_{\ell}
\bigg(m^{\circ}_\ell + \eta^{\ell}_{m} \bigg)
L_{\ell}\left(\hat{Z}^{\ell}_{L}\right)^{-\frac{1}{2}}
\ell_{iL}+\mbox{h.c.}
\label{eqn:effmassterm}
\end{eqnarray}
and can be diagonalized by two independent unitary rotations
\begin{eqnarray}
\ell_{L}\rightarrow L_m \ell_L\,,
\qquad\qquad
\ell_{R}\rightarrow R_m \ell_R\,,
\label{eqn:rotations}
\end{eqnarray}
where $L_m$ and $R_m$ are unitary matrices. We obtain
\begin{equation}
R_m^\dagger\left(\hat{Z}^{\ell}_{R}\right)^{-\frac{1}{2}}R^{\dagger}_{\ell}
\bigg(m^{\circ}_\ell + \eta^{\ell}_{m} \bigg)
L_{\ell}\left(\hat{Z}^{\ell}_{L}\right)^{-\frac{1}{2}}L_{m}= m_{\ell}\,,
\end{equation}
where $m_{\ell}$ is diagonal.
After performing the rescaling in Eq.~\eqref{eqn:rescaling} and the field rotations in Eq.~\eqref{eqn:rotations}, the kinetic terms are brought 
to canonical form and the charged lepton mass matrix is diagonalised.
Since we are interested in retaining effects up to NLO, it is sufficient 
to perform the field redefinition at LO as follows
\begin{eqnarray}
\ell_{L}
\rightarrow
\left(1 - {1\over2}\eta_{L}^{\ell}\right)
\ell_{L}\,,
\qquad\qquad
\ell_{R} 
\rightarrow
\left(1 - {1\over2}\eta_{R}^{\ell}\right) \ell_{R}\,,
\label{eqn:rescaling_1loop}
\end{eqnarray}
and we obtain
\begin{equation}
R_m^\dagger
\bigg(m^{\circ}_\ell 
+ \eta^{\ell}_{m} -{1\over2}\eta_{R}^{\ell} m^{\circ}_{\ell}-{1\over2}m^{\circ}_{\ell} \eta_{L}^{\ell}\bigg)
L_{m}= m_{\ell}\,.
\end{equation}

We stress that the above expression systematically captures all contributions 
to the lepton mass matrix at NLO. In particular, in addition to the LO term $m^{\circ}_{\ell}$, there is a genuine NLO mass contribution $\eta^{\ell}_{m}$, as well as effective NLO terms arising from the interplay between the LO mass matrix and LO wave-function renormalisation effects.

\section{CP-violating couplings in flavor-deconstructed models}\label{sec:$CP$ violation phase}
In this Appendix, we show that reproducing the observed CP violation in the quark sector within the FD framework requires the presence of natural $\mathcal{O}(1)$ CP-violating phases in the underlying UV parameters. 
This result will play a key role in motivating our subsequent phenomenological analysis.

In order to make this connection explicit, we express the SM quark Yukawa matrices in terms of the underlying UV parameters, in close analogy with the lepton sector, as follows
\begin{align}
\mathcal{Y}^a =& Y^a_3 + Y^a\cdot\varepsilon_\chi\cdot Y^{\chi_a}_{3} +Y^a \cdot \varepsilon_\phi \cdot \hat Y^{\phi_a}_{2}   + Y^a \cdot  \varepsilon_\phi \cdot \hat Y^{\phi_a}_{3} \cdot \varepsilon_\sigma\cdot Y^{\sigma_a}\,,
\end{align}
As discussed in the previous subsections, fermion masses and mixings are obtained via a singular value decomposition of $\mathcal{Y}^a$. From this, and neglecting $\mathcal{O}(1)$ coefficients, one can derive the parametric structure of the eigenvalue ratios required to reproduce the observed quark masses and mixings, namely
\begin{equation}
    \varepsilon_\phi^d,\varepsilon_\chi^d\sim \lambda^2\,,\qquad \varepsilon_\phi^u,\varepsilon_\chi^u\sim \lambda^3\,,\qquad \varepsilon_\sigma^d\sim\lambda\,,\qquad \varepsilon_\sigma^u\sim\lambda^2\,, 
\end{equation} 
where $\lambda\approx0.2$ is the Cabibbo angle.

The SM expression for the Jarlskog invariant is given by~\cite{Jarlskog:1985ht,Bernabeu:1986fc,Botella:2004ks}
\begin{align}
J = \text{Im}\left\{\text{Tr}\left[\HH_u \HH_d \HH_u^2 \HH_d^2 \right]\right\}\,,
\end{align}
with $\HH_a = \mathcal{Y}^a {\mathcal{Y}^a}^\dagger$. 
Expressing $J$ in terms of quark masses and mixing angles, and factoring out a common $(y_b y_t)^6$ term, it is found that $J\sim \lambda^{16}$. Using the results derived above, one can then determine the leading contribution to $J$ in terms of the corresponding high-energy invariants, namely
\begin{equation}
   \begin{split}
       \frac{J}{|y_3^d|^2|y_3^u|^2 } \sim&  ~  \text{Im}\Big(|y_3^u|^2~(\varepsilon_{\phi}^d)^2 ~(\varepsilon_{\phi}^u)^2 ~(\varepsilon_{\chi}^d)^2\,\text{Tr} \{{Y^d}^\dagger \, Y^d \, \HH^{\phi_d} \, {Y^d}^\dagger \, Y^u\, {\HH^{\phi_u}} \, {Y^u}^\dagger \, Y^d \, \HH^{\chi_d} \} \, 
\cr& ~+ 
(\varepsilon_{\chi}^d)^3 (\varepsilon_{\chi}^u)^3 y_3^d ~{y_3^u}^*\left( ~\text{Tr} \{{Y^d}^\dagger \, Y^d \, \HH^{\chi_d} \, {Y^d}^\dagger \, Y^u\, {\HH^{\chi_u}} \, {Y^u}^\dagger \, Y^u \, Y^{\chi_u}\, {Y^{\chi_d}}^\dagger \} \right. \cr & \left. ~- ~\text{Tr} \{{Y^d}^\dagger \, Y^d \, Y^{\chi_d} \, { Y^{\chi_u}}^\dagger  \, {Y^u}^\dagger \, Y^u\, {Y^{\chi_u}} \, {Y^{\chi_d}}^\dagger \, {Y^d}^\dagger \, Y^u \, Y^{\chi_u}\, {Y^{\chi_d}}^\dagger \} \right)  \Big)\,.
   \end{split} 
\end{equation}
Notice that the combination $(\varepsilon_{\phi}^d)^2 ~(\varepsilon_{\phi}^u)^2 ~(\varepsilon_{\chi}^d)^2\sim \mathcal{O}(\lambda^{14})$, while the associated trace can be written as
\begin{equation}
    \text{Tr} \left\{ G^{\phi_d}~ G^{\phi_u} G^{\chi_d}\right\}\,,\qquad\text{where}\qquad G^x = Y^x \HH^{a_x} {Y^x}^\dagger\,,
\end{equation}
which are Hermitian matrices with indices $i,j=1,2$, corresponding to the SM left-handed quarks. 
This implies that the imaginary component of the trace necessarily involves first-generation Yukawa couplings, schematically of the form Im~$(G^{\phi_d}_{12}  G^{\phi_u}_{22} G^{\chi_d}_{21})$. Taking into account that $Y^d_{1 \alpha} \sim \lambda$, as required to reproduce the Cabibbo angle in the CKM matrix, one finds that the leading non-vanishing imaginary invariant arises at $\mathcal{O}(\lambda^{16})$.
One can similarly verify that the difference between the contributions in the second and third lines is of $\mathcal{O}(\lambda^{17})$.

Therefore, requiring the FD framework to reproduce the observed value of $J$ implies the presence of non-vanishing phases in the underlying high-energy couplings. In principle, CP violation could originate from complex vacuum expectation values; however, given the sequential breaking of the symmetry at separated energy scales, the scalar potential is too constrained to naturally accommodate spontaneous CP violation with $\mathcal{O}(1)$ phases. We therefore assume that the relevant Yukawa couplings are intrinsically complex at the symmetry-breaking scale, with phases of $\mathcal{O}(1)$, or at least $\mathcal{O}(\lambda^2)\sim \mathcal{O}(10^{-1})$. 
Once complex phases are required in the quark sector, there is no compelling reason to assume real couplings in the lepton sector. In the following, we thus allow for complex leptonic Yukawa interactions and investigate their implications for CP-violating observables, in particular EDMs.

\section{Flavor structure relations of fermionic bilinears}\label{app:Flavor structure relations of fermionic bilinears}
In Sec. \ref{sec:flav structure to smeft} we listed the flavor entries of fermionic bilinears of same and different chiralities. Flavor-violating entries turned out to be suppressed. Here we explicitly compute the flavor suppression factors. For $LL$ bilinears, namely $\overline{\ell}_p\Gamma\ell_r$, we have that the flavor suppression factors are all dominated by $(P_V^\ell)_{pr}$ defined in Eq. \eqref{eq:Pe and Pell def}. Explicitly, we find
\begin{subequations}
    \begin{align}
    &(P_V^\ell)_{33}=Q_V^{\ell_3}+\sum_{i=1,2}\Big[Q_V^{\ell_i}-Q_V^{\ell_3}\Big]|(\Delta_\ell)_{3i}|^2\,,\\
    &(P_V^\ell)_{23}=\Big[Q_V^{\ell_2}-Q_V^{\ell_3}\Big](\Delta_\ell)_{23}+\sum_{i=1,2}Q_V^{\ell_i}(\Delta_\ell)_{i3}(\Delta_\ell^*)_{i2}\,,\\
    &(P_V^\ell)_{32}=(P_V^\ell)_{23}^*\,,\\
    &(P_V^\ell)_{13}=\Big[Q_{V}^{\ell_1}-Q_V^{\ell_3}\Big](\Delta_\ell)_{13}+\sum_{i=1,2}Q_V^{\ell_i}(\Delta_\ell)_{i3}(\Delta_\ell^*)_{i1}\,,\\
    &(P_V^\ell)_{31}=(P_V^\ell)_{13}^*\,,\\
    &(P_V^\ell)_{22}=Q_V^{\ell_2}+\Big[Q_V^{\ell_3}-Q_V^{\ell_2}\Big]|(\Delta_\ell)_{23}|^2\,,\\
    &(P_V^\ell)_{12}=\Big[Q_V^{\ell_3}-Q_V^{\ell_2}\Big](\Delta_\ell)_{31}(\Delta_\ell^*)_{32} \label{Pell12}\,,\\
    &(P_V^\ell)_{21}=(P_V^\ell)_{12}^*\,,\\
    &(P_V^\ell)_{11}=Q_V^{\ell_1}+\Big[Q_V^{\ell_3}-Q_V^{\ell_1}\Big]|(\Delta_\ell)_{13}|^2\,,
\end{align}
\end{subequations}
where we have used the fact that $Q_{V}^{\ell_1}=Q_V^{\ell_2}$ and the unitarity of $U_L^\ell=\mathbb{I}+\Delta_\ell$. For $RR$ bilinears, namely $\overline{e}_p\Gamma e_r$, we have that the flavor suppression factors are all dominated by $(P_V^e)_{pr}$ defined in Eq. \eqref{eq:Pe and Pell def}. Explicitly, we find
\begin{subequations}
\begin{align}
    &(P_V^e)_{33}=Q_V^{e_3}+\sum_{i=1,2}\Big[Q_V^{e_i}-Q_V^{e_3}\Big]|(\Delta_e)_{3i}|^2+\sum_{\alpha=1,2,3}\Big[Q_V^{E_\alpha}-Q_V^{e_3}\Big] |(\mathcal{E}_e)_{\alpha 3}|^2\,,\\
    &(P_V^e)_{23}=\Big[Q_V^{e_2}-Q_V^{e_3}\Big](\Delta_e)_{23}+\sum_{i=1,2}Q_V^{e_i}(\Delta_e)_{i3}(\Delta_e^*)_{i2}+\sum_{\alpha=1,2,3} (\mathcal{E}_e^*)_{\alpha 2}(\mathcal{E}_e)_{\alpha3}\left[Q_V^{E_\alpha}-\frac{1}{2}Q_V^{e_2}-\frac{1}{2}Q_V^{e_3}\right]\,,\\
    &(P_V^e)_{32}=(P_V^e)_{23}^*\,,\\
    &(P_V^e)_{13}=\Big[Q_V^{e_1}-Q_V^{e_3}\Big](\Delta_e)_{13}+\sum_{i=1,2}Q_V^{e_i}(\Delta_e)_{i3}(\Delta_e^*)_{i1}+\sum_{\alpha=1,2,3} (\mathcal{E}_e^*)_{\alpha 1}(\mathcal{E}_e)_{\alpha3}\left[Q_V^{E_\alpha}-\frac{1}{2}Q_V^{e_1}-\frac{1}{2}Q_V^{e_3}\right]\,,\\
    &(P_V^e)_{31}=(P_V^e)_{13}^*\,,\\
    &(P_V^e)_{22}=Q_V^{e_2}+\Big[Q_V^{e_3}-Q_V^{e_2}\Big]|(\Delta_e)_{23}|^2+\sum_{\alpha=1,2,3}|(\mathcal{E}_e)_{\alpha 2}|^2\Big[Q_V^{E_\alpha}-Q_V^{e_2}\Big] \,,\\
    &(P_V^e)_{12}=\Big[Q_V^{e_3}-Q_V^{e_2}\Big](\Delta_e)_{31}(\Delta_e^*)_{32}+\sum_{\alpha=1,2,3} (\mathcal{E}_e^*)_{\alpha 1}(\mathcal{E}_e)_{\alpha2}\Big[Q_V^{E_\alpha}-Q_V^{e_2}\Big] \label{eq:Pe12}\,,\\
    &(P_V^e)_{21}=(P_V^e)_{12}^*\,,\\
    &(P_V^e)_{11}=Q_V^{e_1}+\Big[Q_V^{e_3}-Q_V^{e_1}\Big]|(\Delta_e)_{13}|^2+\sum_{\alpha=1,2,3}|(\mathcal{E}_e)_{\alpha 1}|^2\Big[Q_V^{E_\alpha}-Q_V^{e_1}\Big]\,,
\end{align}
\end{subequations}
where we have used the fact that $Q_V^{e_1}=Q_V^{e_2}$ and the unitarity of $U_R^e=\mathbb{I}+\Delta_e$. 

Regarding chirality-violating bilinears, we can notice that, to estimate the flavorful suppression factors, in full generality, the off-diagonal LO contribution is of type 
\begin{equation}
    \Delta C_{ij}\equiv ((U_L^\ell)^\dag C U_R^e)_{ij}=(U_L^\ell)^*_{ki} c_k (U_R^e)_{3j}\,,
\end{equation}
where $C_{ij}= \delta_{j3}c_i$. Explicitly, we find that, up to corrections of $\mathcal{O}(\varepsilon^3)$,
\begin{subequations}
    \begin{align}
    &\Delta C_{33}\approx c_1\theta_2^L+c_2\theta_1^L+c_3\left(1-\frac{1}{2}|\theta_1^L|^2-\frac{1}{2}|\theta_2^L|^2\right)\,,\\
    &\Delta C_{23}\approx c_1 \sin{\theta_3}e^{-i\delta_3}+c_2 \cos{\theta_3}-c_3 (\theta_1^L)^*\,,\\
    &\Delta C_{13}\approx c_1 \cos{\theta_3}-c_2 \sin{\theta_3}e^{i\delta_3}-c_3 (\theta_2^L)^*\,,\\
    &\Delta C_{32}\approx -c_3\theta_1^R\,,\\
    &\Delta C_{22}\approx -\left(c_1 \sin{\theta_3}e^{-i\delta_3}+c_2 \cos{\theta_3}\right)\theta_1^R\,,\\
    &\Delta C_{12}\approx\left(-c_1 \cos{\theta_3}+c_2 \sin{\theta_3}e^{i\delta_3}\right)\theta_1^R\,,\\
    &\Delta C_{31}\sim \varepsilon_\sigma\varepsilon_\chi\varepsilon_\phi c_3 \,,\\
    &\Delta C_{21} \sim \varepsilon_\sigma\varepsilon_\chi\varepsilon_\phi \left(c_1 \sin{\theta_3}e^{-i\delta_3}+c_2 \cos{\theta_3}\right)\,,\\
    &\Delta C_{11}\sim \varepsilon_\sigma\varepsilon_\chi\varepsilon_\phi  \left(c_1 \cos{\theta_3}-c_2 \sin{\theta_3}e^{i\delta_3}\right)\,.
\end{align}
\end{subequations}

\section{Formulae for low-energy observables}
\label{setion:obs}
In this section, we collect the general expressions for the relevant observables discussed in Sec.~\ref{sec: pheno FD lepton sector}.

The branching ratios for $Z$-decays can be written as
\begin{equation}
    \mathrm{Br}(Z\rightarrow\ell_i\overline{\ell}_j) =
    \frac{M_Z}{24\pi\,\Gamma_Z }
    \Bigg[
    \frac{M_{Z}^{2}\alpha}{16\pi^{3}}\left( \vert (C_{D}^Z)_{ij}\vert^2 + \vert (C_{D}^Z)_{ji}\vert^2\right)+
    \frac{4 e^{2}}{c_{W}^{2}s_{W}^{2}}\left( \vert [Z_{eR}]_{ij}\vert^2 + \vert [Z_{eL}]_{ij}\vert^2\right)\Bigg],
\end{equation}
where
\begin{equation}
    (C_{D}^Z)_{ij} = \frac{8 \pi^2}{e} \frac{v}{\sqrt{2}} \Big[s_{W} (C_{eB})_{ij} +c_{W}(C_{eW})_{ij}\Big]\,,
\end{equation}
is the WC associated with the dipole contribution mediated by the $Z$ boson.

The branching ratio for Higgs-boson decays is given by
\begin{align}
      \mathrm{Br}(H\rightarrow\ell_i\overline{\ell}_j) &=
    \frac{M_h}{16\pi\,\Gamma_H}\left(
    \vert[Y_{H}]_{ji}\vert^2 +  \vert[Y_{H}]_{ij}\vert^2\right)\,.
\end{align}
The effective number of neutrinos reads
\begin{equation}
    N_\nu\approx3 -\frac{v^2}{\Lambda^2}\sum_i\text{Re}\left[(C_{H\ell}^
    {(1)})_{ii}-(C_{H\ell}^
    {(3)})_{ii}\right]\,.
\end{equation}
LFUV in $\tau$ decays can be probed by defining the following ratio of couplings
\begin{equation}
\label{eq:LFUV:tau}
\left| \frac{g_\tau}{g_{\mu(e)}} \right|
=
\frac{
\mathcal{B}(\tau \to e(\mu)\,\nu\bar{\nu})/
      {\mathcal{B}(\tau \to e(\mu)\,\nu\bar{\nu})_{\mathrm{SM}}}
}{
\mathcal{B}(\mu \to e\,\nu\bar{\nu})/
      {\mathcal{B}(\mu \to e\,\nu\bar{\nu})_{\mathrm{SM}}}
}\,,
\end{equation}
which can be written as
\begin{subequations}
    \begin{align}
    &\bigg\vert \frac{g_{\tau}}{g_{\mu}}\bigg\vert = 1-\frac{v^2}{2}\text{Re}[(C_{\nu e}^{V,LL})_{3113}-(C_{\nu e}^{V,LL})_{2112})]\,,\\
    &\bigg\vert \frac{g_{\tau}}{g_{e}} \bigg\vert = 1-\frac{v^2}{2}\text{Re}[(C_{\nu e}^{V,LL})_{3223}-(C_{\nu e}^{V,LL})_{2112})]\,,
    \end{align}
\end{subequations}
where only left-handed decays are considered.

The branching ratio for the LFV decay $\ell_j \to \ell_i \ell_i \ell_i$ can be written as
\begin{align}
        \frac{\mathrm{Br}(\ell_{j} \rightarrow \ell_{i}\ell_{i}\ell_{i})}{\mathrm{Br}(\ell_{j} \rightarrow \ell_{i}\nu \bar{\nu})} = & \frac{1}{2G_{F}^2}\bigg( \frac{\vert (C^{S,LR}_{ee})_{ijii}\vert^2}{4} + 2 \vert (C^{V,RR}_{ee})^{ijii} + \frac{2 \alpha}{\pi m_{j}} C_{D}^{ji}\vert^{2} + 2 \vert (C^{V,LL}_{ee})^{ijii} + \frac{2 \alpha}{\pi m_{j}} C_{D}^{ij}\vert^{2} \nonumber\\
    & + (64 \log\frac{m_{j}}{m_{i}}-136) \frac{\alpha^{2}}{\pi^{2}m_{j}^{2}}(\vert C_{D}^{ij} \vert^{2} + \vert C_{D}^{ji} \vert^{2}) + \vert (C^{V,LR}_{ee})_{iiij} + \frac{2 \alpha}{\pi m_{j}} C_{D}^{ji} \vert^{2} 
    \nonumber\\
    &+ \vert (C^{V,LR}_{ee})_{ijii} + \frac{2 \alpha}{\pi m_{j}} C_{D}^{ij} \vert^{2}\bigg)\,.
    \label{eq:lto3l}
\end{align}

The $\mu-e$ conversion rate in nuclei receives contributions from several operators. Owing to the coherent enhancement associated with the presence of many nucleons, spin-independent interactions dominate over spin-dependent ones, whose contributions are suppressed by the averaging over nuclear spins. For this reason, we neglect the latter and restrict our analysis to the spin-independent contribution\footnote{For an EFT analysis of spin-dependent contributions, see Ref.~\cite{Cirigliano:2017azj}.}
\begin{equation}
    \mathrm{Br}(\mu -e,\mathrm{N}) = \frac{B_{N}}{2 G_{F}^{2}}\left( \vert d_{N} \frac{e}{8\pi^{2}m_{\mu}}C_{D}^{e\mu} + C_{N,L}\vert^{2}+ \vert d_{N} \frac{e}{8\pi^{2}m_{\mu}}C_{D}^{\mu e} + C_{N,R}\vert^{2}\right)\,,
\end{equation}
where the coefficients $C_{N,X}$ depend on the semileptonic WCs as follows
\begin{equation}
    \begin{split}
        C_{N,X} =& F_{V}^{u}\,(C^{V,XL}_{eu} +C^{V,XR}_{eu})_{e\mu uu} +  F_{V}^{d}\,(C^{V,XL}_{ed} + C^{V,XR}_{ed})_{e\mu dd} +  F_{S}^{u}(C^{S,XL}_{eu}+ C^{S,XR}_{eu})_{e\mu uu} \\  &+ F_{S}^{d}(C^{S,XL}_{ed}+ C^{S,XR}_{ed})_{e\mu dd}
    + F_{S}^{s}(C^{S,XL}_{ed}+ C^{S,XR}_{ed})_{e\mu ss}  \\ &+ F_{S}^{c}(C^{S,XL}_{eu}+ C^{S,XR}_{eu})_{e\mu cc} + F_{S}^{b}(C^{S,XL}_{ed}+ C^{S,XR}_{ed})_{e\mu bb}\,,
    \end{split}
\end{equation}
where $d_N$, $F_X^i$, and $B_N$ are given in Appendix~A of Ref.~\cite{Ardu:2024bua}.

Lepton AMM and EDM can be defined in the mass basis as follows
\begin{equation}
    \Delta a_{\ell_{i}} = - \frac{m_{i}}{2 \pi^{2} Q_{f}} \mathrm{Re}[(C_{D})_{ii}]\,, \quad \quad d_{\ell_{i}} = -\frac{e}{4 \pi^2} \mathrm{Im}[(C_{D})_{ii}]\,.
\end{equation}
Radiative LFV decays, on the other hand, are described by the branching ratio
\begin{equation}
    \frac{Br(\ell_{j}\, \rightarrow \ell_{i} \,\gamma)}{\mathrm{Br}(\ell_{j} \rightarrow \ell_{i} \nu\bar\nu)} = \frac{3 \,\alpha}{\pi G_{F}^{2} m_{j}^{2}} \Big( \vert (C_{D})_{ij} \vert^2 + \vert (C_{D})_{ji} \vert^2\Big)\,.
\end{equation}

\bibliography{references}
\newpage
\bibliographystyle{JHEP}

\end{document}